\documentclass[prb,aps,amssymb,twocolumn,superscriptaddress,notitlepage,longbibliography]{revtex4-1}
\usepackage{amsmath}
\usepackage{amssymb}
\usepackage{amsthm}
\usepackage{amsfonts}
\usepackage{listings}
\usepackage{physics}
\usepackage{enumerate}
\usepackage{latexsym}
\usepackage{comment}
\usepackage{psfrag}
\usepackage{bm}
\usepackage{graphicx}
\usepackage[caption=false]{subfig}
\usepackage{blkarray}
\usepackage{array}
\usepackage{color}
\usepackage[normalem]{ulem}

\usepackage{hyperref}

\newcommand{\stress}[2]{\tau^{#1}_{\hphantom{#1}#2}}
\newcommand{\visc}[5]{#1^{#2\hphantom{#3}#4}_{\hphantom{#2}#3\hphantom{#4}#5}}

\def\beq{\begin{equation}}
\def\eeq{\end{equation}}
\def\bal{\begin{aligned}}
\def\eal{\end{aligned}}
\def\p{\partial}
\begin{document}
\title{Hall viscosity in quantum systems with discrete symmetry: point group and lattice anisotropy}
\author{Pranav Rao}
\author{Barry Bradlyn}
\affiliation{Department of Physics and Institute for Condensed Matter Theory, University of Illinois at Urbana-Champaign, Urbana, IL, 61801-3080, USA}
\date{\today}
\begin{abstract}
Inspired by recent experiments on graphene, we examine the non-dissipative viscoelastic response of anisotropic two-dimensional quantum systems. We pay particular attention to electron fluids with point group symmetries, and those with discrete translational symmetry. We start by extending the Kubo formalism for viscosity to systems with internal degrees of freedom and discrete translational symmetry, highlighting the importance of properly considering the role of internal angular momentum. We analyze the Hall components of the viscoelastic response tensor in systems with discrete point group symmetry, focusing on the hydrodynamic implications of the resulting forces. We show that though there are generally six Hall viscosities, there are only three independent contributions to the viscous force density in the bulk.  To compute these coefficients, we develop a framework to consistently write down the long-wavelength stress tensor and viscosity for multi-component lattice systems. We apply our formalism to lattice and continuum models, including a lattice Chern insulator and anisotropic superfluid. 
\end{abstract}
\maketitle
\section{Introduction}

One of the most peculiar and fascinating manifestations of topology in condensed matter physics is the appearance of nondissipative transport coefficients in insulating systems. The paradigmatic example is the Hall conductivity, which is quantized in a two-dimensional insulator, and proportional to a topological invariant--the Chern number--characterizing the many-body ground state\cite{klitzing1980new,laughlin1981quantized,thouless1982quantized}. Similarly, it has recently been noted that in two-dimensional insulators with broken time-reversal symmetry, there is a non-dissipative \emph{viscosity}\cite{1995-AvronSeilerZograf,2007-TokatlyVignale,read2009non}. In rotationally invariant phases it has been shown that there is a unique Hall viscosity coefficient $\eta^\mathrm{H}$, which for a gapped phase is proportional to the particle density $\bar{n}$ and a quantized invariant of the ground state known as the shift $\mathcal{S}$\cite{read2009non,read2011hall},
\begin{equation}
    \eta^\mathrm{H}=\frac{\hbar}{4}\bar{n}\mathcal{S}.
\end{equation}
The shift quantifies the number of additional magnetic monopoles needed to stabilize the ground state on a sphere\cite{Haldane1983}. In the quantum Hall regime, the Hall viscosity has been proposed as a numerical diagnostic for distinguishing between different competing topological orders\cite{Zaletel-PhysRevLett.110.236801}. When rotational symmetry is broken, there ceases to be a single Hall viscosity coefficient, and the relation between the viscosity and the shift is lost\cite{haldane2015geometry,gromov2017investigating,offertaler2019viscoelastic,souslov2019anisotropic}. 

Outside of insulators, topological considerations can also lead to nondissipative transport coefficients in metallic systems, due to the influence of Berry phase effects in transport. Pioneering work by Karplus and Luttinger\cite{karplus1954hall}, as well as Haldane\cite{haldaneahe} have shown how the Hall conductivity in metallic magnets receives a contribution due to the Berry curvature of the occupied states in a Fermi liquid. Recently, there has been a surge of interest into the non-dissipative viscosity of metallic systems as well, driven in large part by the discovery of hydrodynamic flow in systems like graphene\cite{lucas2018hydrodynamics}. {For electronic fluids with an approximately conserved momentum at long wavelengths,} experiments have been proposed for extracting the Hall viscosity from flow through the width dependence of the Hall conductance in narrow channels\cite{scaffidi2017hydrodynamic,alekseev2016negative}, the flow profile near point contacts\cite{delacretaz2017transport,pellegrino2017nonlocal}, and the semiclassical districution function\cite{holder2019unified}. Cutting-edge experiments in graphene under non-quantizing magnetic fields have started to validate these proposals\cite{berdyugin2019measuring}. Hall viscosity is also an area of active theoretical study, with work on graphene\cite{narozhny2019magnetohydrodynamics,imran2019quantizing,narozhny2019electronic} and the consideration of viscous effects in a variety of other contexts\cite{son2019chiral,pu2020hall,buchel2019holographic,apostolov2019magnetodrag} ongoing.

Despite this progress, the robustness and even the definability of the Hall viscosity in the absence of rotational and translational symmetry have not been systematically treated. For instance, the low-energy Dirac theory of graphene arises as a $\mathbf{k}\cdot\mathbf{p}$ expansion in a highly anisotropic band structure for a system with no translational symmetry. In spite of previous works examining the Hall viscosity in models with broken translational symmetry\cite{shapourian2015viscoelastic,tuegel2015hall,niu2018geometry,hallvisc-disorder-numerics}, the connection between microscopic, low energy descriptions and long-wavelength hydrodynamics relevant to experiment has not been directly addressed. Furthermore, a comprehensive framework for treating momentum transport in systems with broken time reversal symmetry and no external magnetic field (analogous to the formalism for the anomalous Hall conductance) is lacking.

In this work, we will take steps to address these issues by developing a formalism for nondissipative viscosity in nonrotationally invariant systems, both in the continuum and with periodic potentials. With these tools we deduce several conclusions about the viscosity of anisotropic quantum fluids. Our three main conceptual innovations are: 1) a novel detailed analysis of the non-dissipative viscosity tensor, revealing that although there are generally six viscosity coefficients, three are redundant in the bulk (we find a similar redundancy in the dissipative viscosity); 2) a relationship between band topology and Hall viscosity for free fermion systems, showing how the six viscosity coefficients are expressible in terms of quandrupole moments of the Berry curvature of occupied bands, and a correction due to the internal (pseudospin) angular momentum of bands; and 3) the first consistent framework for momentum transport and viscosity on a lattice or tight-binding system, derived only from conservation laws. In formulating these results, we also develop 
an extension the Belinfante-Rosenfeld symmetrization procedure to anisotropic continuum and lattice systems, thus fixing the antisymmetric part of the stress tensor operator. Before moving on, we will give some brief background and establish our notation conventions.

\subsection*{Background and Notation}

To begin, let us establish notation and review how viscosity arises in nonrelativistic quantum systems. We will work throughout in units where $\hbar=c=e=1$. For a quantum system with a single particle type in $d$-dimensions, we can introduce the stress tensor through the conservation law for momentum. We thus start with the momentum density operator $g_\mu(\mathbf{r})$.
Here and throughout this work, Greek indices such as $\mu$ will run over spatial directions $\mu=1,\dots,d$, and we will unless otherwise noted take $d=2$. 
The stress tensor $\tau^\mu_{\hphantom{\mu}\nu}(\mathbf{r})$ can then be defined through the conservation law for momentum density\cite{landau1987fluid},
\begin{equation}
    \partial_t g_\nu(\mathbf{r})+\sum_\mu\partial_\mu\tau^\mu_{\hphantom{\mu}\nu}(\mathbf{r}) = f^\mathrm{ext}_\nu(\mathbf{r}),\label{eq:continuity}
\end{equation}
where $\mathbf{f}^\mathrm{ext}$ is the density of external forces acting on the particles. Throughout this work we will use boldface symbols to refer exclusively to two-dimensional vectors. We also introduce the shorthand $\partial_\mu$ to denote the partial derivative of a function with respect to a component of its (coordinate) vector argument,
\begin{equation}
    \partial_\mu f(\mathbf{x})\equiv\frac{\partial f}{\partial x_\mu},
\end{equation}
and analogously in momentum space
\begin{equation}
    \partial^\mu f(\mathbf{k})\equiv \frac{\partial f}{\partial k^\mu}.
\end{equation}
Furthermore, we will \emph{not} use the Einstein summation convention in this work. Since expressions in lattice systems often involve repeated indices, we will explicitly indicate all summations over indices as above. Because this is an unconventional choice, we will remind the reader periodically that repeated indices are not summed over.

Eq. (\ref{eq:continuity}) defines the stress tensor up to a divergenceless term, which must be fixed from other considerations. In rotationally invariant relativistic systems, it is always possible to choose a stress tensor which is symmetric in flat space (where we can avoid complications due to index raising/lowering)\cite{bradlyn2014low}. It was recently shown how to adapt this symmetrization procedure for rotationally invariant nonrelativstic two-component fermions as well\cite{link2018elastic}. One of the main results of our work is a generalization of this procedure to lattice and continuum systems which lack rotational symmetry.

Having defined the stress tensor, we will be primarily interested in the response of the stress to an applied time-varying strain perturbation $\lambda_\mu^{\phantom{\mu}\nu}\equiv\partial_\mu v^\nu$. To make contact with classical hydrodynamics, we can interpret $v^\nu$ as a spatially dependent velocity field in a fluid. We expand the average of the stress tensor perturbatively in the velocity field to define
\begin{equation}
    \langle\tau^\mu_{\hphantom{\mu}\nu}\rangle= \langle\tau^\mu_{\hphantom{\mu}\nu}\rangle_0-\sum_{\lambda\rho}\left[\kappa^{\mu\hphantom{\nu}\lambda}_{\hphantom{\mu}\nu\hphantom{\lambda}\rho}\int dt \partial_\lambda v^\rho +
    \eta^{\mu\hphantom{\nu}\lambda}_{\hphantom{\mu}\nu\hphantom{\lambda}\rho} \partial_\lambda v^\rho+\dots\right]\label{eq:constitutive}
\end{equation}
Here $\langle \tau^\mu_{\hphantom{\mu}\nu}\rangle_0$ denotes the average stress in the absence of a strain perturbation. For a translation-invariant fluid in $d$-dimensions we have that
\begin{equation}
   \langle \tau^\mu_{\hphantom{\mu}\nu}\rangle_0= \mathcal{p}\delta^{\mu}_{\nu},
\end{equation}
where $\mathcal{p}$ is the hydrostatic pressure.
The tensor $\kappa^{\mu\hphantom{\nu}\lambda}_{\hphantom{\mu}\nu\hphantom{\lambda}\rho}$ is the tensor of elastic moduli, and gives the response of the stress tensor to static strains [hence the time integral in Eq.~(\ref{eq:constitutive})]. Finally, the tensor $\eta^{\mu\hphantom{\nu}\lambda}_{\hphantom{\mu}\nu\hphantom{\lambda}\rho}$ is the viscosity tensor, and will be the fundamental object of study for this work. In particular, we will mostly focus on the Hall viscosity tensor
\begin{equation}
    (\eta^\mathrm{H})^{\mu\hphantom{\nu}\lambda}_{\hphantom{\mu}\nu\hphantom{\lambda}\rho}\equiv\frac{1}{2}\left(\eta^{\mu\hphantom{\nu}\lambda}_{\hphantom{\mu}\nu\hphantom{\lambda}\rho}-\eta^{\lambda\hphantom{\rho}\mu}_{\hphantom{\lambda}\rho\hphantom{\mu}\nu}\right),
\end{equation}
which is the part of the viscosity tensor which does not contribute to power dissipation. 

\section*{Outline}

In the following sections, we will examine the constraints that point group symmetry places on the momentum density, stress, and (Hall) viscosity tensors. The structure of the paper is as follows: First, in Sec.~\ref{sec:continuumstrain}, we will show how to define the analogue of the Belinfante stress tensor in a translation-invariant, nonrelativstic anisotropic system. Then, in Section~\ref{sec:latticeformalism}, we will extend this formalism to systems with only discrete translation symmetry. With the formalism developed, we will in Sec.~\ref{sec:viscosity} investigate the constraints of point group symmetry on the viscosity tensor both phenomenologically and microscopically using the Kubo formalism. In Sec.~\ref{sec:freefermion} we will focus in particular on free fermion systems, where we can relate the Hall viscosity to band topology. Finally, in Sec.~\ref{sec:examples} we will apply these results to lattice and continuum models of interest.

\section{Continuum systems: strain and stress with anisotropy}\label{sec:continuumstrain}

To begin, let us consider a general Hamiltonian for an interacting, translation invariant system of particles in two dimensions. We will additionally assume that the particles have $N_L$ internal degrees of freedom. We can write the Hamiltonian for such a system as 
\begin{equation}
    H=\sum_{ia} \mathbf{M}_aT^a(\mathbf{p}_i)+\frac{1}{2}\sum_{i\neq j,a}\mathbf{M}_aV^a(\mathbf{x}_i-\mathbf{x}_j),\label{eq:hamiltonian}
\end{equation}
where $\mathbf{p}_i$ is the momentum operator for particle $i$, $\mathbf{x}_i$ is the position operator for particle $i$, $i,j=1,\dots,N$ index the particles, and the matrices $\mathbf{M}_a$ form a basis for $N_L\times N_L$ Hermitian matrices. We have the canonical commutation relations
\begin{equation}
    \left[x^\mu_i,p_\nu^j\right]=i\delta^\mu_\nu\delta_i^j.
\end{equation}

To compute the viscosity for the ground state of such a system, we would like to employ the Kubo formalism of Ref.~\onlinecite{bradlyn2012kubo}. To do so, we must first identify the momentum density operator $\mathbf{g}(\mathbf{r})$.

If we ignore the internal degrees of freedom of the particles, then we have that all momentum must be carried by kinetic motion. Following the logic of Ref.~\onlinecite{bradlyn2012kubo}, we would write the momentum density as the density of kinetic momentum
\begin{equation}
    \mathbf{g}_\mathrm{kin}(\mathbf{r})=\frac{1}{2}\sum_i\left\{\mathbf{p}_i,\delta(\mathbf{r}-\mathbf{x}_i)\right\},
\end{equation}
where $\{\cdot,\cdot\}$ indicates the anticommutator. We can identify the first moment of the kinetic momentum density
\begin{equation}
J^\mu_{\hphantom{\mu}\nu}=-\int d^2r r^\mu g_\nu(\mathbf{r})=-\frac{1}{2}\sum_i\{x_i^\mu,p^i_\nu\}\label{eq:kineticstrainmomentummoment}
\end{equation}
with the generators of uniform spatial strains (i.e.~position-dependent displacements). Furthermore, by taking the long wavelength limit of the Fourier transform of Eq.~(\ref{eq:continuity}), we see that the time derivative of the strain generators give the integrated (canonical) stress tensor
\begin{equation}
T^\mu_{\hphantom{\mu}\nu}=\int d^2r\tau^\mu_{\hphantom{\mu}\nu}(\mathbf{r})\equiv -i[H,J^\mu_{\hphantom{\mu}\nu}]\label{eq:kineticwardidentity}
\end{equation}
Because Eq.~(\ref{eq:kineticwardidentity}) was derived from the kinetic momentum density, we will refer to it as the kinetic stress.

However, when the internal degrees of freedom of the particles transform nontrivially under rotations, we must modify the momentum density to account for the density of \emph{internal linear momentum}. To see this most directly, let us recall that the antisymmetric part of the strain generator should generate rotations.
On the coordinates and the momenta, rotations act in the standard way as
\begin{align}
    U(\theta)\mathbf{x}_iU^\dag(\theta)&=\sum_\nu R^\mu_\nu x_i^\nu \\
    U(\theta)\boldsymbol{\pi}^iU^\dag(\theta)&=\sum_\nu R_\mu^\nu \pi^i_\nu,
\end{align}
with
\begin{equation}
    R=\left(\begin{array}{cc}
    \cos\theta & -\sin\theta \\
    \sin\theta & \cos\theta
    \end{array}\right)\label{eq:rotationmatrix}
\end{equation}
a rotation matrix. This action is generated by
\begin{equation}
    \left.i\frac{\partial U}{\partial\theta}U^\dag\right|_{\theta=0} =L_{\mathrm{orb}}=\sum_{i\mu\nu}\epsilon_\mu^{\hphantom{\mu}\nu}x_i^\mu p^i_\nu
\end{equation}
In the absence of any internal degrees of freedom, this would be the end of the story. However, when rotations act nontrivially on the internal indices in Eq.~(\ref{eq:hamiltonian}), we have that rotations are generated by the total angular momentum
\begin{equation}
\label{eq:angularmomentum}
    L\equiv L_\mathrm{orb}+L_\mathrm{int},
\end{equation}
where $L_\mathrm{int}$ is an $N_L\times N_L$ matrix acting on the internal degrees of freedom. As an example, for a spin-$1/2$ particle in two dimensions, $L_\mathrm{int}=\sigma_z/2$, the generator of rotations about the $z$-axis.

The antisymmetric part of the strain generator Eq.~(\ref{eq:kineticstrainmomentummoment}) gives only the orbital angular momentum by construction. Thus, following Ref.~\onlinecite{link2018elastic}, we seek a modified strain generator $\mathcal{J}^\mu_{\hphantom{\mu}\nu}$ satisfying
\begin{equation}
\sum_{\mu\nu}\epsilon_\mu^{\hphantom{\mu}\nu}\mathcal{J}^\mu_{\hphantom{\mu}\nu}=-L_\mathrm{orb}-L_\mathrm{int}\label{eq:antisymmodstrain}
\end{equation}
In order for this modified strain generator to be computable as a moment of the momentum density, we can define the momentum density to be
\begin{equation}
    g_\mu(\mathbf{r})\equiv g_\mathrm{kin,\mu}+\frac{1}{2}\sum_{i\nu}\epsilon_{\mu}^{\hphantom{\mu}\nu}\partial_\nu\delta(\mathbf{r}-\mathbf{x}_i)L_\mathrm{int}\label{eq:modmomentum}
\end{equation}
This is the minimal modification of the momentum density which enforces Eq.~(\ref{eq:antisymmodstrain}). Note, importantly, that this redefinition does not change the total linear momentum
\begin{equation}
    \mathbf{P}=\int d^2r \mathbf{g}(\mathbf{r})=\int d^2r \mathbf{g}_{\mathrm{kin}}(\mathbf{r})
\end{equation}

By examining the first moment of the modified momentum density, we can express the modified strain generators in terms of the kinetic strain generators as
\begin{equation}
\mathcal{J}^{\mu}_{\hphantom{\mu}\nu}=-\int{d^2r}r^\mu g_\nu(\mathbf{r})=J^\mu_{\hphantom{\mu}\nu}-\frac{1}{2}\epsilon^\mu_{\hphantom{\mu}\nu}L_\mathrm{int}.\label{eq:modstrain}
\end{equation}
Inserting the modified momentum density into the continuity equation Eq.~(\ref{eq:continuity}), Fourier transforming, and taking the long wavelength limit, we find that the modified integrated stress tensor is
\begin{equation}
T^\mu_{\mathrm{B},\nu}=-i\left[H,\mathcal{J}^{\mu}_{\hphantom{\mu}\nu}\right]=T^\mu_{\hphantom{\mu}\nu}+\frac{i}{2}\epsilon^\mu_{\hphantom{\mu}\nu}\left[H,L_\mathrm{int}\right],\label{eq:modwardidentity}
\end{equation}
where the subscript ``$\mathrm{B}$'' refers to Belinfante\cite{belinfante1940current}, for reasons we will make precise below. Because the second term in Eq.~(\ref{eq:modwardidentity}) originates from the internal angular momentum, we will refer to it as the spin stress.

So far, we have made no assumption on the rotational symmetry of the system under study. The criterion of rotational symmetry of the Hamiltonian can be expressed as
\begin{equation}
[L,H]=[L_\mathrm{orb},H]+[L_\mathrm{int},H]=0
\end{equation}
Combining this with Eq.~(\ref{eq:modwardidentity}), we see that for rotationally invariant systems, the modified stress tensor $T^\mu_{\mathrm{B},\nu}$ satisfies
\begin{equation}
\sum_{\mu\nu}T^\mu_{\mathrm{B},\nu}\epsilon_\mu^{\hphantom{\mu}\nu}=0.
\end{equation}
In other words, the stress tensor corresponding to the momentum density Eq.~(\ref{eq:modmomentum}) is symmetric in flat space (where we can raise and lower indices with impunity).
In the special case of spin-$1/2$ Dirac electrons, it was shown that the stress defined in this way is precisely the Belinfante-Rosenfeld stress tensor\cite{link2018elastic,nakahara2003geometry,belinfante1940current,rosenfeld1940tenseur}; here we have generalized this result to arbitrary representations of internal rotations. 

Going further, we can also show that if we compare this treatment with the field-theoretic formalism of Ref.~\onlinecite{bradlyn2014low}, we see that the strain generators $\mathcal{J}^{\mu}_{\hphantom{\mu}\nu}$ implement precisely the generalized Belinfante procedure described there. To see this, we can imagine defining a second-quantized version of Eq.~(\ref{eq:hamiltonian}), valid on an arbitrary curved surface. The action for this system takes the general form
\begin{equation}
    S=\int d^3x\sqrt{g} \psi^\dag\left[ie^t_0\left(\partial_t-i\omega_tL_\mathrm{int}\right)-H(\vec{e},\vec{\omega})\right]\psi,\label{eq:action}
\end{equation}
where $\psi$ is an $N_L$ component annihilation operator, $\vec{e}$ are frame fields (vielbeins), and $\vec{\omega}$ is the spin connection. We have defined $H(\vec{e},\vec{\omega})$ to be the Hamiltonian with all indices covariantly contracted with vielbeins, and all derivatives replaced by (rotationally) covariant derivatives. For the sake of this discussion, we have included in the Hamiltonian all terms which involve a spatial derivative of the fermion fields. Following Refs.~\onlinecite{Gromov20141,bradlyn2014low,Abanov2014}, the stress tensor and momentum density for this system can be derived by considering the conservation law associated with diffeomorphism invariance of the action. A priori, the spin connection and the vielbeins are treated as independent variables; doing so leads to the conservation law for the canonical stress tensor analogous to Eq.~(\ref{eq:kineticwardidentity}). However, if we demand that the torsion of space
\begin{equation}
    \vec{T}=d\vec{e}+\vec{\omega}\wedge\vec{e}
\end{equation}
remained fixed (more properly, as fixed as possible) as we vary the background geometry, then the spin connection can be expressed in terms of the vielbeins. Eliminating $\vec{\omega}$ from Eq.~(\ref{eq:action}) and applying Noether's theorem yields a stress tensor which is symmetric when the system is rotationally invariant. To check whether this field-theoretic Belinfante procedure yields the same stress tensor $T^\mu_{\mathrm{B},\nu}$ obtained from the improved strain generators, it suffices to consider the field-theoretic Belinfante momentum density. The correction to the momentum density from the vielbein dependence of the spin connection can be written
\begin{equation}
    \delta g_\mu  = -\frac{1}{\sqrt{g}}\sum_{A,\aleph=0}^{2}\int d^3 x \sqrt{g} \frac{\delta S}{\delta \omega_\aleph}\frac{\delta \omega_\aleph}{\delta e^A_0}e_\mu^A,
\end{equation}
where $A$ and $\aleph$ are internal (frame) and ambient spacetime indices, respectively. Taking the variations using the expressions in Appendix A of Ref.~\onlinecite{bradlyn2014low} for the spin connection and evaluating the result in the unstrained geometry yields
\begin{equation}
\label{eq:spinmomentumdensity}
    \delta g_\mu =
    \sum_\nu\epsilon_{\mu}^{\hphantom{\mu}\nu}\partial_\nu\left(\psi^\dag(\mathbf{r})L_{\mathrm{int}}\psi(\mathbf{r})\right),
\end{equation}
which is exactly the second-quantized form of the internal momentum in Eq.~(\ref{eq:modmomentum}). Thus, we conclude that the stress tensor Eq.~(\ref{eq:modwardidentity}) coincides with the usual Belinfante procedure, at least on the physical Hilbert space (recall that the Belinfante stress tensor as a quantum operator is only symmetric when the equations of motion are applied, i.e. when acting on physical states). 
The importance of considering the Belinfante tensor will be illustrated in Sec.~\ref{sec:examples} when we reexamine the viscosity of multicomponent lattice tight-binding systems, first considered in Ref.~\onlinecite{shapourian2015viscoelastic}. In that work the spin connection $\vec{\omega}$ is explicitly set to zero, and hence those authors use the canonical stress tensor. 

It is important to note also that there is a fundamental difference between the Belinfante procedure in relativstic and nonrelativistic systems. Due to Lorentz symmetry, in relativistic systems the Belinfante symmetrization adds a divergence free term to the entire energy-momentum tensor. As such, integrated quantities such as the tensor $T^{\mu}_{\hphantom{\mu}{\nu}}$ do not change when the stress tensor is symmetrized. By contrast, in nonrelativistic systems, there is no symmetry relating the momentum density to the stress tensor; a consequence of this is that the integrated Belinfante tensor $T^{\mu}_{\mathrm{B},{\nu}}$ is physically distinct from the integrated canonical stress tensor.

Furthermore, note that the internal angular momentum contribution to the momentum density originates solely from the time-derivative term in the action Eq.~(\ref{eq:action}). As such, even when the Hamiltonian breaks rotational symmetry explicitly, the form of the momentum density operator is unchanged. From this we conclude that the stress tensor Eq.~(\ref{eq:modwardidentity}) is the meaningful extension of the Belinfante ``symmetrization" to non-rotationally invariant situations, validating the conjecture of Ref.~\onlinecite{link2018elastic}.

However, we cannot get away with modifying the momentum density without paying some price; while the kinetic strain generators $J^\mu_{\hphantom{\mu}\nu}$ generate the algebra $\mathfrak{gl}(2,\mathbb{R})$,
\begin{equation}
i\left[J^\mu_{\hphantom{\mu}\nu},J^\lambda_{\hphantom{\lambda}\rho}\right]=\delta^\mu_\rho J^\lambda_{\hphantom{\lambda}\nu}-\delta^\lambda_\nu J^\mu_{\hphantom{\mu}\rho},
\end{equation}
of diffeomorphism of the plane, the same cannot be said of the modified strain generators. In fact, because we have chosen the internal degrees of freedom to be invariant under shears and dilatations, and to transform only under rotations, we have that
\begin{equation}
i\left[\mathcal{J}^\mu_{\hphantom{\mu}\nu},\mathcal{J}^\lambda_{\hphantom{\lambda}\rho}\right]=\delta^\mu_\rho J^\lambda_{\hphantom{\lambda}\nu}-\delta^\lambda_\nu J^\mu_{\hphantom{\mu}\rho}.\label{eq:strainalgebra}
\end{equation}
The modified strain algebra thus does not close. Furthermore, even if we were to let the internal degrees of freedom transform under shears and dilatations, we could never get the algebra of the $\mathcal{J}$s to close as long as the number of internal degrees of freedom is finite. This is because the Lie group $GL(d,\mathbb{R})$ is non-compact, and hence has no finite-dimensional unitary representations\cite{lipkin2002lie}.

We thus see that, by modifying the strain generators as per Eq.~(\ref{eq:modmomentum}), we obtain the Belinfante stress tensor $T^\mu_{\mathrm{B},\nu}$. Even in the absence of rotational symmetry, the Belinfante stress gives the conserved current associated to deformations of spacetime at fixed (reduced) torsion. In this way, we can fix the definition of the antisymmetric part of the stress tensor in rotationally non-invariant systems. By focusing on the Belinfante stress, we ensure that the torque density $\epsilon_\mu^{\hphantom{\mu}\nu}T^\mu_{\mathrm{B},\nu}$ gives only the torque due to rotational symmetry breaking in the Hamiltonian. Thus, the modified strain generators $\mathcal{J}^\mu_{\hphantom{\mu}\nu}$ and Belinfante stress tensor $T^\mu_{\mathrm{B},\nu}$ are the natural objects to consider in the study of viscosity of anisotropic systems.

Using this Belinfante formalism, we can proceed to calculate the viscosity tensor for arbitrary anisotropic continuum systems. Before moving on to examine this, however, let us recall that in condensed matter systems the most common source of rotational symmetry breaking comes from an underlying lattice. In order to consistently treat the viscosity in such systems (even in the low-energy $\mathbf{k}\cdot\mathbf{p}$ limit), we will first develop a formalism for computing momentum transport in lattice models.

\section{Can we generalize to a lattice system?}\label{sec:latticeformalism}

At first sight, the idea of quantifying momentum transport in a lattice system is fraught with difficulties. First and foremost, in the presence of translational symmetry breaking, total momentum is no longer conserved. This creates difficulties in defining a continuity equation for the density of momentum: the split between external forces $\mathbf{f}^\mathrm{ext}$ and internal forces
\begin{equation}
    f^\mathrm{int}_\nu\equiv -\sum_\mu\partial_\mu \tau^\mu_{\hphantom{\mu}\nu} \label{eq:intforcedef}
\end{equation}
becomes unnatural. Furthermore, the strain formalism of Ref.~\onlinecite{bradlyn2012kubo} and Sec.~\ref{sec:continuumstrain} is no longer available if we work with a lattice-regularized system with discrete positions, since the momentum operators are no longer well-defined.

To remedy these issues, we will in this section generalize the strain and stress formalisms to tight-binding lattice systems, and show that there exists a meaninful notion of stress tensor and momentum transport which reduce to those of Sec.~\ref{sec:continuumstrain} in an appropriate continuum limit. Our motivation for developing this formalism is twofold. First, a lattice formulation allows us to incorporate rotational symmetry breaking into the viscosity formalism in a controlled way, where the strength of the anisotropy can be quantitatively tied to the underlying crystal structure. Second, we would like to provide a framework for interpreting and understanding previous results on the viscosity of lattice systems such as Chern insulators\cite{shapourian2015viscoelastic}, and the connection to low-energy expansions of the viscosity near multiple fermi pockets such as in graphene\cite{link2018elastic}.

To begin, let us consider a general lattice Hamiltonian
\begin{align}
H&=\sum_{\mathbf{R}\mathbf{R}'nm}c^\dag_{n\mathbf{R}}f^{nm}(\mathbf{R},\mathbf{R'})c_{m\mathbf{R'}}\nonumber\\
&+\frac{1}{2}\sum_{\mathbf{RR'}plnm}[V(\mathbf{R},\mathbf{R'})]^{nmp\ell}c^\dag_{n\mathbf{R}}c^\dag_{m\mathbf{R'}}c_{p\mathbf{R}}c_{\ell\mathbf{R'}},\label{eq:latticeham}
\end{align}
where we have, for convenience, introduced a second-quantized description. Here $\mathbf{R}$ and $\mathbf{R'}$ are lattice vectors in a Bravais lattice, and the operators $c^\dag_{n\mathbf{R}}$ creates a fermion in unit cell $\mathbf{R}$ in state indexed by $n$. The fermion operators satisfy the anticommutation relations
\begin{align}
    \{c_{n\mathbf{R}},c_{m\mathbf{R'}}\}&=0 \\
     \{c^\dag_{n\mathbf{R}},c_{m\mathbf{R'}}\}&=\delta_{\mathbf{RR'}}\delta_{nm}.
\end{align}
Although we consider a fermionic system for concreteness, the generalization of all our results to bosonic systems is straightforward. Note that the states created by $c^\dag_{n\mathbf{R}}$ for the same $\mathbf{R}$ need not all be centered at the same spatial location; nevertheless, we will for simplicity refer to the degrees of freedom indexed by $n$ as ``internal'' degrees of freedom. The first term $f^{nm}(\mathbf{R},\mathbf{R}')$ is the single-particle Hamiltonian, and includes both kinetic and on-site interactions. The second term $[V(\mathbf{R},\mathbf{R'})]^{nmp\ell}$ is the (normal-ordered) two-body interaction energy. We assume that the Hamiltonian has discrete tranlation symmetry
\begin{align}
f^{nm}(\mathbf{R}+\mathbf{R}_a,\mathbf{R}'+\mathbf{R}_a)&=f^{nm}(\mathbf{R},\mathbf{R}') \\
[V(\mathbf{R}+\mathbf{R}_a,\mathbf{R}'+\mathbf{R}_a)]^{nmp\ell}&=[V(\mathbf{R},\mathbf{R'})]^{nmp\ell},
\end{align}
where 
\begin{equation}
    \mathbf{R}_a=\sum_\mu n_a^\mu\mathbf{a}_\mu,\;\;\; n^\mu_a\in\mathbb{Z}
\end{equation} 
is any Bravais lattice vector, written in terms of the primitive lattice vectors $\mathbf{a}_\mu$. Beyond this, we will make no other assumptions about symmetry. Due to the preponderance of expressions with repeated indices that will appear as we consider lattice systems, we remind the reader that \emph{we will not sum over repeated indices unless noted explicitly}. Let us also introduce a (normalized) basis $\{\mathbf{b}^\mu|\mu=1,2\}$ for the reciprocal lattice satisfying
\begin{equation}
\mathbf{a}_\nu\cdot\mathbf{b}^\mu=\delta^\mu_\nu 
\end{equation}

In order to discuss momentum transport in this system, we must first define a lattice momentum density. We would like to do this in such a way that we recover the continuum momentum density Eq.~(\ref{eq:modmomentum}) in the long wavelength limit, which is equivalently the limit that the lattice spacings $|\mathbf{a}_\mu|$ goes to zero, keeping the ratios $|\mathbf{a}_\mu|/|\mathbf{a}_\nu|$ fixed. Furthermore, we expect a hydrodynamic approach to be valid only in a ``coarse-grained'' sense, which means that we should focus primarily on the transport of momentum between unit cells (rather than within the unit cell). As we emphasized in Sec.~\ref{sec:continuumstrain}, it is also critical for us to incorporate internal angular momentum into the momentum density. In the lattice system, the internal angular momentum is determined by the representation of rotations on the internal degrees of freedom $\{n\}$; systems with the same Hamiltonian $H$ may have different internal rotation generators $L^{\mathrm{int}}_{nm}$ depending on the physical meaning (embedding) of the degrees of freedom $\{n\}$. 

Taking these considerations into account, we decompose the lattice momentum density
\begin{equation}
g^\mathrm{L}_\mu(\mathbf{R})=g_\mu^\mathrm{kin}(\mathbf{R})+g_\mu^\mathrm{int}(\mathbf{R})\label{eq:latticemomtotal}
\end{equation}
into a kinetic piece $g_\mu^\mathrm{kin}(\mathrm{R})$ and an internal angular momentum contribution $g_\mu^\mathrm{int}(\mathrm{R})$. For the kinetic momentum, we write 
\begin{widetext}

\begin{equation}
g_\mu^\mathrm{kin}(\mathbf{R})=\frac{i}{4|\mathbf{a}_\mu|}\sum_n\left(
 c^\dag_{n\mathbf{R+a_\mu}}c_{n\mathbf{R}}
-c^\dag_{n\mathbf{R-a_\mu}}c_{n\mathbf{R}}
+c^\dag_{n\mathbf{R}}c_{n\mathbf{R-a_\mu}}
-c^\dag_{n\mathbf{R}}c_{n\mathbf{R+a_\mu}}\right),\label{eq:latticekinmom}
\end{equation}
Similarly, we take for the internal momentum density 
\begin{equation}
g_\mu^\mathrm{int}(\mathbf{R})=\sum_{nm\nu}\frac{1}{4|\mathbf{a}_\nu|}\tilde{\epsilon}_\mu^{\hphantom{\mu}\nu}L_\mathrm{int}^{nm}\left(
c^\dag_{n\mathbf{R+a_\nu}}c_{m\mathbf{R}}
-c^\dag_{n\mathbf{R-a_\nu}}c_{m\mathbf{R}}
+c^\dag_{n\mathbf{R}}c_{m\mathbf{R+a_\nu}}
-c^\dag_{n\mathbf{R}}c_{m\mathbf{R-a_\nu}}
\right)\label{eq:latticeintmom},
\end{equation}
\end{widetext}
where we have defined the lattice epsilon tensor (note that for systems with threefold rotational symmetry, contractions of lattice vector indicies are not tensorial. To make contact with standard results in these cases, every index should be projected onto the vector representation of the symmetry group. We leave these insertions implicit to avoid overburdening notation) as
\begin{equation}
\tilde{\epsilon}_\mu^{\hphantom{\mu}\nu}\equiv \mathbf{a}_\mu^\mathrm{T}\cdot\boldsymbol{\epsilon}\cdot\mathbf{b}^\nu.
\end{equation} 
The sum of fermion bilinears appearing here give a symmetric discretization of the derivative. As such, in the limit of small lattice spacing $|\mathbf{a}_\mu|\rightarrow 0$, the sum of Eqs.~(\ref{eq:latticekinmom}) and (\ref{eq:latticeintmom}) coincides with the continuum momentum density Eq.~(\ref{eq:modmomentum}),
\begin{equation}
\lim_{|\mathbf{a}|\rightarrow 0} g^\mathrm{L}_\mu(\mathbf{R})=\mathbf{\hat{a}}_\mu\cdot\mathbf{g}(\mathbf{R}).
\end{equation}
Additionally, the particular discretization of the derivative that we have chosen here includes all terms that carry momentum into or out of the unit cell $\mathbf{R}$ from adjacent unit cells. This is reminiscent of the form of conserved currents in lattice gauge theories\cite{kogut1975hamiltonian}, and is necessary to ensure that we recover the proper long-wavelength properties upon coarse-graining.

Taking Eq.~(\ref{eq:latticemomtotal}) as our starting point, we can now examine its time derivative and attempt to derive an analogue of Eq.~(\ref{eq:continuity}). This will allow us to extract the integrated Belinfante stress tensor. Since we will be primarily interested in the integrated stress tensor, we procede in momentum space. We introduce the momentum space creation and annihilation operators
\begin{align}
c_{n\mathbf{k}}&=\sum_\mathbf{R} c_{n\mathbf{R}}e^{i\mathbf{k}\cdot\mathbf{R}} \\
c^\dag_{n\mathbf{k}}&=\sum_\mathbf{R} c^\dag_{n\mathbf{R}}e^{-i\mathbf{k}\cdot\mathbf{R}}. \\
\end{align}
Inserting these into Eqs.~(\ref{eq:latticekinmom}) and (\ref{eq:latticeintmom}), we find
\begin{equation}
g^\mathrm{L}_\mu(\mathbf{R})=\sum_\mathbf{q}e^{i\mathbf{q}\cdot\mathbf{R}}(g^\mathrm{kin}_\mu(\mathbf{q})+g^\mathrm{int}_\mu(\mathbf{q}))
\end{equation}
with
\begin{widetext}
\begin{equation}
    g_\mu^\mathrm{kin}(\mathbf{q})=\frac{1}{2|\mathbf{a}_\mu|}\sum_{n\mathbf{k}}\left[\sin\left(\mathbf{k}+\mathbf{q}\right)\cdot\mathbf{a}_\mu+\sin\mathbf{k}\cdot\mathbf{a}_\mu\right]c^\dag_{n\mathbf{k}}c_{n\mathbf{k}+\mathbf{q}}\label{eq:momqa}
\end{equation}
and
\begin{equation}
g_\mu^\mathrm{int}(\mathbf{q})=\sum_{\mathbf{k}\nu,m,n}\frac{i}{2|\mathbf{a}_\nu|}\tilde{\epsilon}_\mu^{\hphantom{\mu}\nu}\left[\sin\left(\mathbf{k}+\mathbf{q}\right)\cdot\mathbf{a}_\nu-\sin\mathbf{k}\cdot\mathbf{a}_\nu\right]L_{\mathrm{int}}^{nm}c^\dag_{n\mathbf{k}}c_{m\mathbf{k}+\mathbf{q}}\label{eq:momqb}
\end{equation}
Note that in this form we see clearly that the total lattice momentum, obtained by taking $\mathbf{q}\rightarrow 0$, coincides with the conventional expression for the momentum operator found in, e.g. Ref.~\onlinecite{fetter2012quantum}. In particular, the internal angular momentum does not contribute to the total momentum, as in the continuum case. 
\end{widetext}
By taking the commutator of Eqs.~(\ref{eq:momqa}) and (\ref{eq:momqb}), with the Hamiltonian (\ref{eq:latticeham}) we can attempt to extract the stress tensor and external force densities from the time derivative
\begin{equation}
    \partial_tg_\mu^\mathrm{L}(\mathbf{q})=i\left[H,g_\mu ^\mathrm{L}(\mathbf{q})\right],
\end{equation}
where $H$ is the Hamiltonian (\ref{eq:latticeham}). Our goal will be to make an analogy with the continuum continuity equation in the long-wavelength $\mathbf{q}\rightarrow 0$ limit. To do so, it is simplest to expand Eqs.~(\ref{eq:momqa}) and (\ref{eq:momqb}) to leading order in $\mathbf{q}$, which gives
\begin{widetext}
\begin{equation}
    g_\mu^\mathrm{L}(\mathbf{q})=
   P_\mu+\frac{1}{2}\sum_{nm\nu\mathbf{k}}q_\nu
    c^\dag_{n\mathbf{k}}
    \left[\left\{\frac{\sin\mathbf{k}\cdot\mathbf{a}_\mu}{|\mathbf{a}_\mu|},\partial^{\nu}\right\}
    \delta^{nm}+i\tilde{\epsilon}_\mu^{\hphantom{\mu}\nu}\cos\mathbf{k}
    \cdot\mathbf{a}_\nu L^{nm}_\mathrm{int}\right]c_{m\mathbf{k}}+\dots,
\end{equation}
where we have introduced the total momentum operator $\mathbf{P}=\mathbf{g}^\mathrm{L}(\mathbf{q}=0)$, and $\partial^\nu$ here represents a derivative with respect to $k_\nu\equiv\mathbf{k}\cdot\hat{\mathbf{a}}_\nu$ (we expect no confusion to arise between the use of $\partial$ for momentum and position derivatives--position derivatives always carry a lower index, and momentum derivatives an upper index). Examining the expression in brackets, we recognize that it is a discretized form of the strain generator Eq.~(\ref{eq:modstrain}). In particular, if we define
\begin{equation}
    \mathcal{J}^\mu_{\mathrm{L},\nu}\equiv-\frac{i}{2}\sum_{\mathbf{k}nm}c^\dag_{n\mathbf{k}}\left[\left\{\frac{\sin\mathbf{k}\cdot\mathbf{a}_\nu}{|\mathbf{a}_\nu|},\frac{\partial}{\partial k_\mu}\right\}\delta^{nm}+i\tilde{\epsilon}_\nu^{\hphantom{\nu}\mu}\cos\mathbf{k}\cdot\mathbf{a}_\mu L_\mathrm{int}^{nm}\right]c_{m\mathbf{k}}\label{eq:latticestrain}
\end{equation}
\end{widetext}
then we have
\begin{equation}
    g^\mathrm{L}_\mu(\mathbf{q})=P_\mu+i\sum_\nu q_\nu \mathcal{J}^\nu_{\mathrm{L},\mu}+\mathcal{O}(\mathbf{q}^2)
\end{equation}
Comparing with the continuity equation, this allows us to identify, in the long wavelength limit,
\begin{align}
f^\mathrm{ext}_\mu&=-i\left[H,P_\mu\right] \label{eq:latticeforce}\\
T^\mu_{\mathrm{B},\nu}&=-i\left[H,\mathcal{J}^\mu_{\mathrm{L},\nu}\right]\label{eq:latticestressdef}
\end{align}
Note that, unlike in a continuum system, the external force $f_\mu^\mathrm{ext}$ need not vanish for the Hamiltonian (\ref{eq:latticeham}). This is because the interaction term $[V(\mathrm{R},\mathrm{R'})]^{nmp\ell}$ is not invariant under infinitesimal translations. More physically, $f_\mu^\mathrm{ext}$ corresponds to the rate of momentum relaxation due to Umklapp scattering\cite{ashcroft2005solid}. Note also that all $\mathrm{O}(\mathbf{q}^2)$ contributions to the lattice momentum density can, in the long-wavelength limit, be written as
\begin{equation}
    g^\mathrm{L}_\mu(\mathbf{q})=P_\mu+i\sum_\nu q_\nu \mathcal{J}^\nu_{\mathrm{L},\mu}+\sum_\nu iq_\nu M^\nu_{\hphantom{\nu}\mu}(\mathbf{q})\label{eq:latticemomexpansion}
\end{equation}
for some $\mathbf{q}$ dependent tensor $M^\nu_{\hphantom{\nu}\mu}(\mathbf{q})$ which vanishes at least linearly as $\mathbf{q}\rightarrow 0$. This tensor is in principle computable by Taylor expanding Eqs.~(\ref{eq:momqa}) and (\ref{eq:momqb}). By taking the time-derivative of $M^{\mu}_{\hphantom{\mu}\nu}$, we can compute the long-wavelength stress tensor $\tau^\mu_{\mathrm{B},\nu}(\mathbf{q})$ to arbitrary order in $\mathbf{q}$. 

For interacting systems, defining the stress tensor and internal forces in this way is tantamount to making a particular choice for how to split the force density between internal and external contributions. Although there is no ambiguity in attributing uniform forces to $\mathbf{f}^\mathrm{ext}$, solely to the interaction term via Eq.~(\ref{eq:latticeforce}), at higher order in $\mathbf{q}$ both viscous and external forces may appear as divergences of functions. Our formalism following from Eq.~(\ref{eq:latticemomexpansion}) chooses to attribute \emph{all} nonuniform forces to internal forces Eq.~(\ref{eq:intforcedef}). This nonuniqueness is at the heart of the claim that the viscosity requires translational invariance to be well-defined\cite{read2011hall,bradlyn2012kubo}. Nevertheless, we see that in the long-wavelength limit there exists a natural choice for the integrated stress tensor, and hence a natural definition for the viscosity and elastic moduli tensors.

Finally, we point out that, due to the factor of $\cos\mathbf{k}\cdot\mathbf{a}_\mu$ multiplying the internal angular momentum in (\ref{eq:latticestrain}), the internal angular mommentum contribution to the lattice stress tensor is not purely antisymmetric, in contrast to the continuum case. Mathematically, this factor originated from our choice of symmetric discretization of the derivative in Eq.~(\ref{eq:latticeintmom}). Physically, it reflects the fact that $\mathcal{J}^\mu_{\mathrm{L},\nu}$ is not a generator of simple uniform deformations of the plane beyond leading order in $\mathbf{k}\cdot\mathbf{a}$, which can be seen by expanding the kinetic strain.

By carrying out the commutator in Eq.~(\ref{eq:latticestressdef}), we can derive a general expression for the lattice stress tensor. It is beneficial to carry this computation out in two parts, first for the non-interacting contribtuion to the stress, and second for the interaction contribution. Introducing the Fourier transform
\begin{equation}
    f^{nm}(\mathbf{k})=\sum_\mathbf{R}e^{-i\mathbf{k}\cdot\mathbf{R}}f^{nm}(\mathbf{R},0)
\end{equation}
of the single particle Hamiltonian in Eq.(\ref{eq:latticeham}), we find rather directly that the single-particle contribution $T^\mu_{\mathrm{B}0,\nu}$ to the stress tensor is given by
\begin{widetext}
\begin{equation}
    T^\mu_{\mathrm{B}0,\nu}=\sum_{nm\mathbf{k}}c^\dag_{n\mathbf{k}}\left(\partial^\mu f_{nm}(\mathbf{k})\frac{\sin\mathbf{k}\cdot \mathbf{a}_\nu}{|\mathbf{a}_\nu|}+\frac{i}{2}\tilde{\epsilon}^\mu_{\hphantom{\mu}\nu}\cos\mathbf{k}\cdot\mathbf{a}_\mu[f(\mathbf{k}),L_\mathrm{int}]_{nm}\right)c_{m\mathbf{k}}\label{eq:fflatticestress}
\end{equation}
Combined with Eq.~(\ref{eq:latticeforce}), we see from this result that in a noninteracting system, lattice effects enter only via anisotropy of the stress tensor.

Next, we can compute the interaction contributions to the lattice stress tensor. The algebra is a bit more involved, and it is most convenient to use the representation
\begin{equation}
    \mathcal{J}^{\mu}_{\mathrm{L},\nu}=-\sum_R R^\mu g_\nu(\mathbf{R})
\end{equation}
for the lattice strain generator. Defining the shorthand 
\begin{align}
    V_{\mathbf{R}}^{\mathbf{k}\mathbf{q}}&\equiv \sum_{\mathbf{R'}nmp\ell}[V(\mathbf{R}',\mathbf{R})]^{nmp\ell}c^\dag_{n\mathbf{R'}}c^\dag_{m\mathbf{k}}c_{p\mathbf{R'}}c_{\ell\mathbf{k+q}} \\
    (VL)_\mathbf{R}^{\mathbf{kq}}&=\sum_{\mathbf{R'}nmp\ell n'}[V(\mathbf{R}',\mathbf{R})]^{nmpn'}L_{\mathrm{int}}^{n'\ell}c^\dag_{n\mathbf{R'}}c^\dag_{m\mathbf{k}}c_{p\mathbf{R'}}c_{\ell\mathbf{k+q}}\\
    (LV)_\mathbf{R}^{\mathbf{kq}}&=\sum_{\mathbf{R'}nmp\ell n'}L_{\mathrm{int}}^{mn'}[V(\mathbf{R}',\mathbf{R})]^{nn'p\ell}c^\dag_{n\mathbf{R'}}c^\dag_{m\mathbf{k}}c_{p\mathbf{R'}}c_{\ell\mathbf{k+q}}
\end{align}
we find that $T^{\mu}_{\mathrm{BIK},\nu}$, the lattice form of the ``Belinfante-Irving-Kirkwood\cite{irving1950statistical}'' is
\begin{align}
    T^{\mu}_{\mathrm{BIK},\nu}&=\sum_{\mathbf{Rkq}}\frac{1}{4|\mathbf{a}_\nu|}R^\mu e^{i\mathbf{q}\cdot\mathbf{R}}\left[\left(V^{\mathbf{kq}}_{\mathbf{R+a_\nu}}-V^{\mathbf{kq}}_{\mathbf{R}}\right)\left(e^{i\mathbf{(k+q)\cdot a_\nu}}+e^{-i\mathbf{k\cdot a_\nu}}\right)-\left(V^{\mathbf{kq}}_{\mathbf{R-a_\nu}}-V^{\mathbf{kq}}_{\mathbf{R}}\right)\left(e^{-i\mathbf{(k+q)\cdot a_\nu}}+e^{i\mathbf{k\cdot a_\nu}}\right)\right] \nonumber \\
    &+\sum_{\mathbf{Rkq}\lambda}\frac{i}{4|\mathbf{a}_\lambda|}R^\mu\tilde{\epsilon}^\lambda_{\hphantom{\lambda}{\nu}}e^{i\mathbf{q\cdot R}}\left[ e^{i\mathbf{k}\cdot\mathbf{a}_\lambda}\left((LV)^{\mathbf{kq}}_\mathbf{R}-(VL)^{\mathbf{kq}}_{\mathbf{R-a_\lambda}}\right)+e^{-i\mathbf{k}\cdot\mathbf{a}_\lambda}\left((VL)^{\mathbf{kq}}_\mathbf{R+a_\lambda}-(LV)^{\mathbf{kq}}_{\mathbf{R}}\right)\right.\nonumber\\
    &\hphantom{+\sum_{\mathbf{Rkq}\lambda}\frac{i}{4|\mathbf{a}_\lambda|}R^\mu\tilde{\epsilon}^\lambda_{\hphantom{\lambda}{\nu}}e^{i\mathbf{q\cdot R}}}\left.+e^{i(\mathbf{k+q})\cdot\mathbf{a}_\lambda}\left((VL)^{\mathbf{kq}}_\mathbf{R}-(LV)^{\mathbf{kq}}_{\mathbf{R+a_\lambda}}\right)+e^{-i(\mathbf{k+q})\cdot\mathbf{a}_\lambda}\left((LV)^{\mathbf{kq}}_\mathbf{R-a_\lambda}-(VL)^{\mathbf{kq}}_{\mathbf{R}}\right)
    \right].
    \label{eq:latticeIKstress}
\end{align}
\end{widetext}
In deriving this expression we have exploited the translational and exchange symmetries of the interaction $[V(\mathbf{R,R'})]^{nmp\ell}$ in order to obtain a somewhat compact expression. Nevertheless, we will spare the reader any explicit use of Eq.~(\ref{eq:latticeIKstress}) going forward.

\subsection{Addition of a rational magnetic flux
}

We remark here on the addition of a magnetic field $B$ of rational flux $\phi = p/q$ (in units of the flux quantum $\phi_0$) to our formalism for lattice elasticity. After a Peierls substitution, our formalism as it stands with $L_{int} = 0$ (the one-band case) reproduces the results of Refs.~\onlinecite{shapourian2015viscoelastic,tuegel2015hall} for the Hofstadter model on a square lattice\footnote{for a model with a symmetric stress tensor and no spin, our stress tensor matches Ref~.\onlinecite{shapourian2015viscoelastic}.}. Within our formalism, it is the lattice vectors (enlarged) magnetic unit cell that go into defining the lattice momentum Eq.~(\ref{eq:latticemomtotal}.

The only subtelty in the more general case is to ensure that the generator of internal angular momentum $\hat{L}_{int}$ is not artificially modified by the addition of the magnetic field. As the creation operators, annihilation operators, and hamiltonian increase in dimension proportional to $q$ when the Brillouin zone is replaced by the contracted magnetic Brilloiun zone, we must ensure that $L_{int} = L_{int}(B\rightarrow 0) \otimes \mathbb{I}_{\, p \times p}$  -- the internal angular momentum is determined in the zero-field limit (corresponding to spin or pseudospin degrees of freedom) and is trivial (block-diagonal) in the space of the magnetic subbands. With these caveats in mind, the application of our formalism to a multiband lattice system in the presence of a magnetic field is straightforward. Further, all the general considerations of this and the following sections will continue to hold in the presence of a magnetic field.

\subsection{Connecting the lattice to the continuum}

With this method of defining the momentum density and long-wavelength stress tensor, we see that the lattice forces enter into hydrodynamics only through Umklapp scattering in interacting systems. In a noninteracting system the total momentum $\mathbf{P}$ defined here is a conserved quantity. We thus recover in our formalism that hydrodynamics applies when the impurity scattering rate (here set to zero from the outset) and the Umklapp scattering rate go to zero. This observation, along with the similarities between Eqs.~(\ref{eq:modstrain}) and (\ref{eq:latticestrain}) suggest a concrete connection between the lattice and continuum formulations.

We must be careful, however, to distinguish two different points of view on the continuum limit. In the first, which we will call the ``lattice gauge theory'' point of view, the lattice Hamiltonian Eq.~(\ref{eq:latticeham}) represents a regularized UV completion of some underlying continuum theory. This is the perspective that was taken in Ref.~\onlinecite{shapourian2015viscoelastic}, where the lattice model for a Chern insulator was interpreted as the UV completion of a theory of free Dirac fermions. In this point of view, the continuum model is recovered by taking the appropriate $|\mathbf{a}_\mu|\rightarrow 0$ limit. This perspective is also appropriate for metallic systems (such as graphene\cite{link2018elastic}), where we can view the $|\mathbf{a}|\rightarrow 0$ limit of Eq.~(\ref{eq:latticeham}) as reproducing low-energy Fermi pockets with anisotropic dispersion. In the lattice gauge theory point of view, we see that as $|\mathbf{a}_\mu|\rightarrow 0$, the free-fermion integrated stress tensor becomes
\begin{align}
    T^\mu_{\mathrm{B}0,\nu}&\rightarrow\sum_{nm\mathbf{k}}c^\dag_{\mathbf{k}n}\left(k_\nu\partial^\mu f^0_{nm}(\mathbf{k})\vphantom{\frac{1}{2}}\right.\nonumber \\
    &\left.+\frac{i}{2}\tilde{\epsilon}^{\mu}_{\hphantom{\mu}\nu}\left[f^0(\mathbf{k}),L_\mathrm{int}\right]_{nm}\right)c_{m\mathbf{k}}, \\
    &=\hat{\mathbf{b}}^\mu\cdot\mathbf{T}_\mathrm{B},\cdot\hat{\mathbf{a}}_\nu,\label{eq:ffstresscontlimit}
\end{align}
where $f^0$ is the continuum limit of the single-particle Hamiltonian. Eq.~(\ref{eq:ffstresscontlimit}) is nothing other than the continuum free fermion stress tensor projected along the direct and reciprocal lattice directions. Similarly, the interaction contribution to the stress tensor reproduces the (Belinfante symmetrized) Irving-Kirkwood stress tensor\cite{bradlyn2012kubo}. This corresponds to a coarse-graining where we ignore physics at the lattice scale and focus instead only on inter-unit cell dynamics.

There is, however, a different notion of continuum limit in condensed matter, where we view the Hamiltonian (\ref{eq:latticeham}) as a tight-binding or Wannier approximation to an underlying continuous Schr\"{o}dinger equation with a periodic potential. In this case, taking the continuum limit corresponds to leaving the lattice constant fixed, but letting the number of internal degrees of freedom go to infinity, such that the operators $c_{n\mathbf{R}}\rightarrow c_{\sigma\mathbf{r}\mathbf{R}}$, with $\mathbf{r}$ a vector within the unit cell, and $\sigma$ is a spin index. The operator $c_{\sigma\mathbf{r}\mathbf{R}}$ annihilates an electron in a delta-function localized orbital with spin $\sigma$ at position $\mathbf{r}+\mathbf{R}$. We refer to this method of taking the continuum limit as the ``tight-binding'' continuum limit. Examining Eqs.~(\ref{eq:fflatticestress}) and (\ref{eq:latticeIKstress}), we see that in the tight-binding continuum limit the sums over internal degrees of freedom are replaced with real-space integrals. The lattice momentum density Eq.~(\ref{eq:latticemomtotal}) is, in this limit, a unit-cell averaged momentum; it follows that other observables such as the stress correspond to unit-cell averages in this limit. We thus see in the tight-binding continuum limit that our lattice formalism is able to produce a well-defined stress tensor even in the presence of a continuous periodic potential.

When we are interested in the viscosity of a particular model, we should ask ourselves which of these continuum limits is the most appropriate for interpreting our results. The lattice gauge theory continuum limit produces a translationally-invariant but potentially anisotropic system, where UV divergences from an unbounded negative spectrum have been regularized by the lattice. The tight-binding continuum limit, on the other hand, produces a system with only discrete translational symmetry, and no momentum conservation. Conversely, as long as we start with a basis of Wannier functions that fully captures the low-energy behavior of our system, taking the tight-binding continuum limit should not quantitatively change any observables. This is not true of the lattice gauge theory continuum limit, where the continuum model has a very different set of symmetries and degrees of freedom than the  lattice approximation.

As we will explore in Sec.~\ref{sec:chern}, we expect both continuum limits to give similar results for metallic systems with small Fermi pockets, where we can Taylor expand observables in $\mathbf{k}$. For band insulators, special care must be taken with the lattice gauge theory continuum limit. This is because a band insulator is characterized by an integer filling
\begin{equation}
    \nu=N_e/N_c
\end{equation}
of $N_e$ electrons in $N_c$ unit cells. Rewriting the number of unit cells in terms of the total volume and the volume $V_c$ of a unit cell, we find that
\begin{equation}
    \nu=V_c\bar{n},
\end{equation}
where $\bar{n}$ is the particle density. Now, in the lattice-gauge continuum limit, $V_c\rightarrow 0$. This presents a conundrum for an insulator, where $\nu\in\mathbb{Z}$ implies that we must also take the density $\bar{n}\rightarrow\infty$. Intuitively, this corresponds to treating the filled bands as forming, in the limit, a uniform Dirac sea. While this perspective is useful for some purposes, it is at odds with the reality that most continuum fluids of interest have fixed, finite density. This difficulty does not arise in the tight-binding continuum limit, since the unit-cell volume stays fixed. In that case, however, taking the continuum limit requires introducing additional degrees of freedom which are not contained in the original model, and there is no guarantee or expectation that such a procedure is unique. For insulators whose low-energy physics is dominated by a set of discrete band inversions, however, we can circumvent these difficulties by Taylor expanding the Hamiltonian and stress tensors around the band inversion (analogous to our discussion of metals). We will see an example of this in Sec.~\ref{sec:examples} when we examine the Chern insulator.

\section{Viscosity tensor in anisotropic systems}\label{sec:viscosity}

Now that we have established a formalism for long-wavelength hydrodynamics both in the continuum and on the lattice, we can move on to examine the viscosity tensor. Recall from Eq.~(\ref{eq:constitutive}) that the viscosity $\eta$ and elastic moduli $\kappa$ govern the change in the average stress tensor,

\begin{equation}
    \langle\stress{\mu}{\nu}\rangle=\langle\stress{\mu}{\nu}\rangle_0-\sum_{\lambda\rho}\left[\visc{\kappa}{\mu}{\nu}{\lambda}{\rho}\partial_\lambda u^\rho+\visc{\eta}{\mu}{\nu}{\lambda}{\rho}\partial_\lambda v^\rho+\dots\right]\label{eq:viscdef},
\end{equation}
where $u^\rho$ is a displacement field, and $v^\rho$ is its time derivative, which gives a velocity field. We now have all the tools necessary to compute the viscosity for both lattice and continuum systems. To do this, we will in subsection~\ref{subsec:kubo} make use of the Kubo formalism of Ref.~\onlinecite{bradlyn2012kubo}. Next, in subsection~\ref{subsec:decomp} we will discuss the decomposition of the viscosity tensor into physically significant components. Finally, in subsection~\ref{subsec:symmetry} we will discuss symmetry constraints on these components. All of what follows will apply equally well in the continuum or on the lattice, provided the appropriate definitions of strain and stress are used.

\subsection{Kubo Formalism}\label{subsec:kubo}

 We will add to our Hamiltonian [either Eq.~(\ref{eq:hamiltonian}) or (\ref{eq:latticeham})] a strain perturbation
\begin{equation}
    H_1=\sum_{\mu\nu}\frac{\partial\lambda_\mu^{\hphantom{\mu}\nu}}{\partial t}\mathcal{J}^\mu_{\hphantom{\mu}\nu},
\end{equation}
where $\mathcal{J}$ is the strain generator, either in the continuum or the lattice as appropriate. The field $\partial\lambda_\mu^{\hphantom{\mu}\nu}/\partial{t}$ represents the uniform part of a velocity gradient in the electron system. Up to a time-dependent gauge transformation\cite{bradlyn2012kubo}, this perturbations gives rise to the coupling between the spatial metric (or, more generally, vielbeins) and the Belinfante tensor $T^\mu_{\mathrm{B},\nu}$ which arises naturally from the field theory discussed in Sec.~\ref{sec:continuumstrain}.  The field $\partial\lambda_\mu^{\hphantom{\mu}\nu}/\partial{t}$ depends on time, and we write
\begin{equation}
    \frac{\partial\lambda_\mu^{\hphantom{\mu}\nu}}{\partial t}(t)=\int dt \frac{\partial\lambda_\mu^{\hphantom{\mu}\nu}}{\partial t}(\omega)e^{-i\omega t}
\end{equation}
We then compute the linear response of the Belinfante stress $T^\mu_{\mathrm{B},\nu}$. Making use of the definitions Eqs.~(\ref{eq:modwardidentity}) and (\ref{eq:latticestressdef}), we write
\begin{align}
    \delta\langle T^\mu_{\mathrm{B},\nu}\rangle(\omega) &= \sum_{\lambda\rho}X^{\mu\hphantom{\nu}\lambda}_{\hphantom{\mu}\nu\hphantom{\lambda}\rho}(\omega)\frac{\partial\lambda_\lambda^{\hphantom{\lambda}\rho}}{\partial t}(\omega) \\
    X^{\mu\hphantom{\nu}\lambda}_{\hphantom{\mu}\nu\hphantom{\lambda}\rho}(\omega)&=\frac{1}{V\omega^+}\left(\langle \left[T^\mu_{\mathrm{B},\nu},\mathcal{J}^\lambda_{\hphantom{\lambda}\rho}\right]\rangle_0\right.\nonumber \\
    &+\left.\int_0^\infty dt e^{i\omega^+t}\langle\left[T^\mu_{\mathrm{B},\nu}(t),\mathcal{T}^\lambda_{\mathrm{B},\rho}(0)\right]\rangle_0\right)\label{eq:kubo}
\end{align}
where time evolution and averages $\langle\cdot\rangle_0$ are evaluated with respect to the unperturbed Hamiltonian, and $V$ is the volume of the system. Furthermore $\omega^+=\omega+i\epsilon,$ and it is understood that $\epsilon$ is taken to zero at the end of the calculation.

The response function $X^{\mu\hphantom{\nu}\lambda}_{\hphantom{\mu}\nu\hphantom{\lambda}\rho}(\omega)$ incorporates both elastic moduli
\begin{equation}
    \kappa^{\mu\hphantom{\nu}\lambda}_{\hphantom{\mu}\nu\hphantom{\lambda}\rho}=\frac{1}{V}\langle T^\mu_{\mathrm{B},\nu}\rangle_0\delta^\lambda_\rho + \lim_{\omega\rightarrow 0}i\omega X^{\mu\hphantom{\nu}\lambda}_{\hphantom{\mu}\nu\hphantom{\lambda}\rho}(\omega)
\end{equation}
and the zero-frequency viscosity tensor
\begin{equation}
 \eta^{\mu\hphantom{\nu}\lambda}_{\hphantom{\mu}\nu\hphantom{\lambda}\rho}=\lim_{\omega\rightarrow 0}\left[
 X^{\mu\hphantom{\nu}\lambda}_{\hphantom{\mu}\nu\hphantom{\lambda}\rho}
 +\frac{i}{\omega^+}\kappa^{\mu\hphantom{\nu}\lambda}_{\hphantom{\mu}\nu\hphantom{\lambda}\rho}\right],\label{eq:hallresponsedef}
\end{equation}

While this formula is valid for both dissipative and nondissipative responses, we will focus here on the nondissipative Hall response 
\begin{equation}
(X^\mathrm{H})^{\mu\hphantom{\nu}\lambda}_{\hphantom{\mu}\nu\hphantom{\lambda}\rho}\equiv\frac{1}{2}\left(X^{\mu\hphantom{\nu}\lambda}_{\hphantom{\mu}\nu\hphantom{\lambda}\rho}-X^{\lambda\hphantom{\rho}\mu}_{\hphantom{\lambda}\rho\hphantom{\mu}\nu}\right)    
\end{equation}

For the Hall response, we can use the Jacobi identity along with the Ward identity Eq.~(\ref{eq:modwardidentity}) to simplify the contact term
\begin{align}
    C^{\mu\hphantom{\nu}\lambda}_{\hphantom{\mu}\nu\hphantom{\lambda}\rho}&\equiv\frac{1}{\omega}\lim_{\omega\rightarrow 0} i\omega(X^\mathrm{H})^{\mu\hphantom{\nu}\lambda}_{\hphantom{\mu}\nu\hphantom{\lambda}\rho}\nonumber \\&=\frac{i}{2\omega}\left(\langle \left[T^\mu_{\mathrm{B},\nu},\mathcal{J}^\lambda_{\hphantom{\lambda}\rho}\right]\rangle_0-\left[T^\lambda_{\mathrm{B},\rho},\mathcal{J}^\mu_{\hphantom{\mu}\nu}\right]\rangle_0\right)
\end{align}

\subsection{Decomposition of the Hall Viscosity Tensor}\label{subsec:decomp}

Before applying this Kubo formula to some model systems, let us first analyze the properties of the nondissipative response tensor. We will be primarily interested in fluids, where we expect the Hall elastic moduli to vanish in the long-wavelength limit (we will revisit the elastic moduli in the next section). Focusing then on the Hall viscosity, without any symmetries we have the decomposition
\begin{align}
    \visc{(\eta^\mathrm{H})}{\mu}{\nu}{\lambda}{\rho}
    &\equiv\frac{1}{2}\left(\visc{\eta}{\mu}{\nu}{\lambda}{\rho}-\visc{\eta}{\lambda}{\rho}{\mu}{\nu}\right)\label{eq:hallviscdef}\\
    &=\eta^\mathrm{H}\visc{(\sigma^z\wedge\sigma^x)}{\mu}{\nu}{\lambda}{\rho}+\gamma\visc{(\sigma^z\wedge\epsilon)}{\mu}{\nu}{\lambda}{\rho}  \nonumber \\
    & +\Theta\visc{(\sigma^x\wedge\epsilon)}{\mu}{\nu}{\lambda}{\rho}\nonumber +\bar{\eta}^\mathrm{H}\visc{(\delta\wedge\epsilon)}{\mu}{\nu}{\lambda}{\rho}+\bar{\gamma}\visc{(\delta\wedge\sigma^x)}{\mu}{\nu}{\lambda}{\rho}\nonumber  \\ &+\bar{\Theta}\visc{(\sigma^z \wedge \delta)}{\mu}{\nu}{\lambda}{\rho}, \label{eq:hallviscgeneral}
\end{align}
where $\sigma^x$ and $\sigma^z$ are Pauli matrices, $\epsilon$ is the two-dimensional Levi-Civita symbol, $\delta$ is the Kronecker delta, and $\wedge$ represents the antisymmetrized tensor product. We see that there are six independent Hall viscosity coefficients. First, the three coefficients $\eta^\mathrm{H},\gamma$ and $\Theta$ are traceless on the last pair of indices, and so do not involve compression of the fluid. $\eta^\mathrm{H}$ is the ordinary isotropic Hall viscosity, while $\gamma$ and $\Theta$ explicitly break three and fourfold rotational symmetry, and involve the antisymmetric stress. The three ``barred" coefficients, $\bar{\eta}^\mathrm{H}, \bar{\gamma}$ and $\bar{\Theta}$ have nonvanishing trace. As with the unbarred coefficients, $\bar{\eta}^\mathrm{H}$ is rotationally invariant, while $\bar{\gamma}$ and $\bar{\Theta}$ explicitly break three and fourfold rotational symmetry. However, although the tensor structure of $\bar{\eta}^\mathrm{H}$ is isotropic, it explicitly generates antisymmetric stress. Thus, in a rotationally invariant system we must have $\bar{\eta}^\mathrm{H}=0$ when the symmetric (Belinfante) stress tensor is used, even for a compressible fluid. This is our first indication that a proper examination of anisotropic Hall viscosity invovles a careful extension of the Belinfante stress tensor to non-rotationally-invariant systems. To our knowledge the ``barred" Hall viscosities have not been emphasized previously in the quantum mechanics literature (though see Ref.~\onlinecite{souslov2019anisotropic}). Consequently we believe the anomalous antisymmetric stress in  Ref.~\onlinecite{yu2017emergent} can  now be attributed to $\bar{\eta}^\mathrm{H}$.

To understand the meaning of these viscosity coefficients, let us return to the continuity equation (\ref{eq:continuity}). In the presence of a fluid velocity gradient $\partial_\mu v^\nu$, the viscosity creates an additional stress
\begin{equation}
    \delta\tau^\mu_{\mathrm{B},\nu}=-\sum_{\lambda\rho}\eta^{\mu\hphantom{\nu}\lambda}_{\hphantom{\mu}\nu\hphantom{\lambda}\rho}\partial_\lambda v^\rho.
\end{equation}
Inserting this into the continuity equation gives a viscous contribution to the force density
\begin{equation}
    f^\eta_\nu=\sum_{\lambda\rho\mu}\eta^{\mu\hphantom{\nu}\lambda}_{\hphantom{\mu}\nu\hphantom{\lambda}\rho}\partial_\lambda\partial_\mu v^\rho.\label{eq:viscforce}
\end{equation}
We see that only the components of the viscosity tensor that are symmetric under the exchange $\mu\rightarrow\lambda$ contribute to the viscous forces. If we focus explicitly on the non-dissipate forces
\begin{equation}
    f^{\mathrm{H},\eta}_\nu= \sum_{\lambda\rho\mu}(\eta^\mathrm{H})^{\mu\hphantom{\nu}\lambda}_{\hphantom{\mu}\nu\hphantom{\lambda}\rho}\partial_\lambda\partial_\mu v^\rho.
\end{equation}
we find, from our decomposition Eq.~(\ref{eq:hallviscgeneral}) that
\begin{align}
    f^{\mathrm{H},\eta}_\mathrm{\nu}&=\sum_{\substack{\mu\nu'\rho'\\ \rho\lambda}}\frac{1}{2}\left(\epsilon^{\nu'\rho'}\visc{(\eta^\mathrm{H})}{\mu}{\nu'}{\lambda}{\rho'}\right)\partial_\mu\partial_\lambda(\epsilon_{\nu\rho}v^\rho) \\
    &\equiv\sum_{\mu\lambda\rho}\eta_\mathrm{H}^{\mu\lambda}\partial_\mu\partial_\lambda(\epsilon_{\nu\rho}v^\rho).
\end{align}
In going from the first to the second line we have used the antisymmetry of the Hall viscosity tensor, as well as the explicit symmetry of Eq.~(\ref{eq:viscforce}) under the interchange of $\mu$ and $\lambda$. This defines the \emph{Hall tensor}
\begin{equation}
    \eta_\mathrm{H}^{\mu\nu}=\frac{1}{4}\sum_{\lambda\rho}\epsilon^{\lambda\rho}\left(\visc{\eta}{\mu}{\lambda}{\nu}{\rho}+\visc{\eta}{\nu}{\lambda}{\mu}{\rho}\right),
\end{equation}
which coincides with the contracted Hall viscosity introduced in the study of the quantum Hall effect in Refs.~\onlinecite{haldane2009hall,haldane2011geometrical,haldane2015geometry}. In those works, the Hall tensor was introduced to parameterize the quadrupole ("internal" or ``second" metric) degrees of freedom in anisotropic quantum Hall states; here we see that this same tensor governs the relationship between velocity gradients and internal forces in the fluid. 

As a rank-two symmetric tensor, we see that the Hall tensor has at most three independent components, even in the absence of any symmetries. This means that in anisotropic systems, not all odd viscosities result in independent forces in the bulk. As an example, let us consider a (possibly compressible) twofold symmetric fluid. Using Eq.~(\ref{eq:hallviscgeneral}) we find for the Hall tensor
\begin{equation}
    \label{eq:hallvisc2tensor}
    \eta_\mathrm{H}^{\mu\nu}=(\eta^\mathrm{H}+\bar{\eta}^\mathrm{H})\delta^{\mu\nu}+(\gamma+\bar{\gamma})\sigma_z^{\mu\nu}+(\Theta+\bar{\Theta})\sigma_x^{\mu\nu}.
\end{equation}
We see that, at the level of forces, the compressive ``barred" Hall viscosities are indistinguishable from the traceless ``unbarred" viscosities. This means that we should treat both viscosities on equal footing.

Going further, we can see that we can shift contributions between the barred and unbarred viscosity coefficients by adding divergenceless terms to the stress tensor. Explicitly, consider a contact term which redefines the stress tensor by
\begin{equation}
    \delta\tau^\mu_{\hphantom{\mu}\nu}=\sum_{\lambda\rho}\epsilon^{\mu\lambda}C_{\nu\rho}\partial_\lambda v^\rho,\label{eq:divfreecontact}
\end{equation}
where $C_{\nu\rho}$ is a generic symmetric tensor. Because of the epsilon tensor, $\sum_\mu\partial_\mu\delta\tau^\mu_{\hphantom{\mu}\nu}=0$ by construction, and so this redefinition of the stress tensor does not change the viscous force density Eq.~(\ref{eq:viscforce}). At the level of viscosities, however, the contact term Eq.~(\ref{eq:divfreecontact}) contributes to the \emph{difference} between the barred and unbarred viscosity coefficient. Concretely, if we write 
\begin{equation}
    C_{\nu\rho}=C_0\delta_{\nu\rho}+C_x \sigma^x_{\nu\rho}+C_z\sigma^z_{\nu\rho},
\end{equation}
then by contracting the tensor $\epsilon^{\mu\lambda}C_{\nu\rho}$
with the antisymmetric products of matrices in the decomposition Eq.~(\ref{eq:hallviscgeneral}), we find that the contact term shifts the viscosities by

\begin{eqnarray}
\eta^\mathrm{H}&\rightarrow \eta^\mathrm{H}+C_0    \;\;\;\;\;\;\bar{\eta}^\mathrm{H}&\rightarrow \bar{\eta}^\mathrm{H}-C_0 \\
\gamma&\rightarrow \gamma+C_z   \;\;\;\;\;\;\bar{\gamma}&\rightarrow \bar{\gamma}-C_z \\
\Theta&\rightarrow \Theta+C_x   \;\;\;\;\;\;\bar{\Theta}&\rightarrow \bar{\Theta}-C_x
\end{eqnarray}

Because we define the stress tensor directly through the continuity equation (\ref{eq:continuity}), the value of the tensor $C_{\nu\rho}$ is a priori undetermined. While in field theories there may be underlying principles for coupling particles to geometry that fix the divergence-free contact terms, condensed matter models--especially those derived from lattice models--lack such guiding principles. As such, the sum of the barred and unbarred viscosities appearing in Eq.~(\ref{eq:hallvisc2tensor}) give precisely the responses that can be calculated and physically probed in a model-independent fashion. It is only in the presence of dynamical gravity sourced by the energy-momentum tensor that the effect of $C_{\mu\nu}$ can be observed in the bulk. In a fluid with boundaries, however, these redundant terms will result in different contributions to the tangent and normal forces on the surface. In this case, a general principle regarding the choice of boundary conditions is still needed to disambiguate barred and unbarred contributions to the viscosity tensor. 

It is also important to note that the contact term (\ref{eq:divfreecontact}) can be nonzero even for incompressible flows satisfying $\nabla\cdot\mathbf{v}=0$. This means that we should expect both the barred and unbarred viscosities to be nonzero even in incompressible anisotropic systems.

Finally, while we will not focus on the dissipative viscosity
\begin{equation}
(\eta^\mathrm{D})^{\mu\hphantom{\nu}\lambda}_{\hphantom{\mu}\nu\hphantom{\lambda}\rho}\equiv\frac{1}{2}\left(\eta^{\mu\hphantom{\nu}\lambda}_{\hphantom{\mu}\nu\hphantom{\lambda}\rho}+\eta^{\lambda\hphantom{\rho}\mu}_{\hphantom{\lambda}\rho\hphantom{\mu}\nu}\right)
\end{equation}
in this work, it is curious to note that there exists a similar ambiguity there as well. In particular, choosing the tensor $C_{\mu\nu}$ to be \emph{antisymmetric} gives a dissipative viscous stress that does not appear as a viscous force. This means that of the ten possible independent dissipative viscosities without symmetry, only nine appear in the continuity equation. The physical implications of this for hydrodynamics and entropy production will be a focus of future work.

\subsection{Symmetry Constraints on the Viscosity Tensor}\label{subsec:symmetry}

Finally, let us analyze the symmetries of the viscosity coefficients in Eq.~(\ref{eq:hallviscgeneral}) in more detail. We would like to analyze how point group symmetries constrain the decomposition of the Hall viscosity (\ref{eq:hallviscgeneral}). A useful point of view for this is to regard the two-tensors $\sigma^x,\sigma^z,\epsilon$ and $\delta$ as Clebsch-Gordan coefficients: each tensor is a map which takes two vectors as an input, and returns an element in a representation of the point group. For instance, the Kronecker delta $\delta^\mu_\nu$ is a map which takes a vector $v^\nu$ and a dual vector $u_\mu$ as inputs, and returns an object $v^\mu u_\mu$ which is a scalar under all point group symmetries. In order for the viscosity tensor to respect the point group symmetries of the system, we demand that the fully contracted object
\begin{equation}
\sum_{\mu\nu\lambda\rho}(\eta^\mathrm{H})^{\mu\hphantom{\nu}\lambda}_{\hphantom{\mu}\nu\hphantom{\lambda}\rho}a_\mu b^\nu c_\lambda d^\rho
\end{equation}
transform in the trivial (scalar) representation of the point group. Applying this logic, we then must ask which of the terms in Eq.~(\ref{eq:hallviscgeneral}) project onto the trivial representation. Note that for lattice systems with threefold rotational symmetry, there is an additional subtlety in identifying the proper Clebsch-Gordan coefficients because the indices on the stress tensor and momentum density do not transform tensorially under threefold rotational symmetry. As this will not be relevant for the examples treated here, we will defer the systematic treatment of that case to future work.

Let us focus first on rotational symmetry. Under the action of the continuous rotational symmetry group, the tensors $\delta^\mu_\nu$ and $\epsilon^\mu_{\hphantom{\mu}\nu}$ are both maps to the scalar representation. On the other hand, the two Pauli matrices $(\sigma^x,\sigma^z)$ transform as a two-dimensional quadrupole representation
\begin{equation}
    \left(\begin{array}{c}
    \sigma_x\\
    \sigma_z
    \end{array}\right)\rightarrow\left(\begin{array}{cc}
    \cos 2\theta & \sin2\theta \\
    -\sin 2\theta & \cos2\theta\end{array}\right)\left(\begin{array}{c}
    \sigma_x\\
    \sigma_z
    \end{array}\right)
\end{equation}
under a rotation by angle $\theta$.
Since both the product of two scalars and the antisymmetric product of two vectors are again scalars, we deduce that with continuous $SO(2)$ rotational symmetry only $\eta^\mathrm{H}$ and $\bar{\eta}^\mathrm{H}$ can be nonzero. Our Belinfante construction assures additionally that $\bar{\eta}^\mathrm{H}=0$ in this case, as we can deduce from the symmetry of $T^\mu_{\mathrm{B},\nu}$ entering the Kubo formula Eq.~(\ref{eq:kubo}).

Furthermore, for any discrete rotational symmetry group including a rotation of order $3$ or greater (except the group $C_4$ generated by a fourfold rotation), the quadrupole representation remains an irreducible representation. For the group $C_4$, the quadrupole representation reduces to two copies of the sign representation $\Delta$ with $\Delta(C_4)=-1$\cite{Bilbao1,Bilbao2,Bilbao3}. In both cases, we find that the only scalar viscosities are $\eta^\mathrm{H}$ and $\bar{\eta}^\mathrm{H}$, with all other coefficients vanishing identically. Unlike with continuous rotational symmetry, with discrete symmetry we must allow $\bar{\eta}^\mathrm{H}\neq 0$.

In addition to the discrete rotational symmetries, we can also think about the action of parity $P$ and time reversal $T$ on the viscosity tensor, as well as the composite symmetry $PT$. We will see that parity and $PT$ in particular trim down the number of nonzero components of the Hall viscosity tensor \cite{golan2019boundary}. By Onsager reciprocity\cite{forster1975hydrodynamic}, we know that the Hall viscosity is odd under time-reversal symmetry,

\beq
T: (\eta^\mathrm{H})^{\mu\hphantom{\nu}\lambda}_{\hphantom{\mu}\nu\hphantom{\lambda}\rho} \rightarrow -(\eta^\mathrm{H})^{\mu\hphantom{\nu}\lambda}_{\hphantom{\mu}\nu\hphantom{\lambda}\rho}
\eeq

Let us now now consider the effects of parity symmetry. In two spatial dimensions, parity coincides with a crystallographic mirror symmetry and hence requires a choice of axis. For concreteness, let us for now take $P$ to be the mirror symmetry $M_x$, 

\beq
P\equiv M_x : \,\, x \rightarrow -x \;\; \& \;\;  y \rightarrow y 
\eeq

Under the action of this transformation,

\beq
\bal
\delta &\rightarrow P^T \delta P = \delta \\
\epsilon &\rightarrow P^T\epsilon P= - \epsilon \\
\sigma_x &\rightarrow P^T\sigma_xP=-\sigma_x\\
\sigma_z &\rightarrow P^T\sigma_zP=\sigma_z
\eal
\eeq
The diagonal Clebsch-Gordan coefficients $\delta, \sigma^z$ are parity even while the off-diagonal ones $\epsilon, \sigma^x$ are parity odd. When combined with the decomposition (\ref{eq:hallviscgeneral}), this means that under parity symmetry, $\eta^\mathrm{H},\bar\eta^\mathrm{H},\gamma,$ and $\bar{\gamma}$ are odd, while $\Theta$ and $\bar{\Theta}$ are even. This means that in a $P$-symmetric system, only $\Theta$ and $\bar{\Theta}$ can ever be nonzero.

Furthermore, since all Hall viscosity coefficients are $T$-odd, we can easily deduce the effect of the composite symmetry $PT$: it flips the sign of $\Theta$ and $\bar{\Theta}$, while leaving the other Hall viscosities invariant. This means that in systems with $PT$ symmetry but neither $P$ or $T$ individually, there are only four possible nonzero viscosity coefficients, $\eta^\mathrm{H},\bar\eta^\mathrm{H},\gamma,$ and $\bar{\gamma}$. This is consistent with the symmetry analysis of Refs.~\onlinecite{haldane2015geometry,golan2019boundary}. We will often work with $PT$-invariant systems, where we can say in general that $\Theta = \bar{\Theta} = 0$. 

The analysis so far was done for parity symmetry defined to be the mirror reflection $M_x$. For a more general reflection axis, it is still true that four viscosity coefficients are parity odd, and two viscosity coefficients are parity-even. More concretely, let us consider the reflection $M_{Rx}$ defined as
\begin{equation}
    M_{Rx}= R(\theta)M_xR^T(\theta),
\end{equation}
in terms of the rotation matrix $R(\theta)$ in Eq.~(\ref{eq:rotationmatrix}). Under this reflection, the Clebsch-Gordan tensor $\delta$ is invariant, $\epsilon$ is odd, and the coefficients $(\sigma^x,\sigma^z)$ transform as
\begin{equation}
    \left(\begin{array}{c}\sigma^x \\ \sigma^z\end{array}\right)\rightarrow\left(\begin{array}{cc}-\cos2\theta & -\sin2\theta \\-\sin2\theta & \cos2\theta\end{array}\right)\left(\begin{array}{c}\sigma^x \\ \sigma^z\end{array}\right),
\end{equation}
meaning that the linear combination $\cos\theta\sigma^z-\sin\theta\sigma^x$ is invariant under $M_{Rx}$, while the linear combination $\cos\theta\sigma^x+\sin\theta\sigma^z$ is odd under $M_{Rx}$. It follows then that $\eta^\mathrm{H},\bar{\eta}^\mathrm{H}$ as well as
\begin{align}
    \gamma_R&\equiv\gamma\cos\theta-\Theta\sin\theta,\label{eq:trans1} \\
    \bar{\gamma}_R&\equiv\bar{\gamma}\cos\theta+\bar{\Theta}\sin\theta
\end{align}
 are odd under $M_{Rx}$, while the orthogonal linear combinations
 \begin{align}
 \Theta_R&\equiv\Theta\cos\theta+\gamma\sin\theta,\\
 \bar{\Theta}_R&\equiv\bar{\Theta}\cos\theta-\bar{\gamma}\sin\theta\label{eq:trans2}
 \end{align}
 are even under $M_{Rx}$. Our previous results on PT symmetry thus readily generalize by replacing $\gamma,\bar{\gamma},\Theta$ and $\bar{\Theta}$ by their transformed counterparts (\ref{eq:trans1}--\ref{eq:trans2}) respectively.

We are now able to begin a discussion on the viscosity of lattice and continuum systems, keeping the constraints provided by rotational and $PT$ symmetries in mind. When dealing with systems with internal degrees of freedom, such as spin or pseudospin, it will be important to establish the representation of rotations and reflections in order to ultimately verify the symmetries of the viscosity tensor.

\subsection{Viscosity-conductivity relation, and other tensors}
We note that the viscosity-conductivity relation (Eq. (4.10) in Ref.~\onlinecite{bradlyn2012kubo}) links the finite-wavevector conductivity to the stress response $X$, which is related to the viscosity, as (in the case of zero external magnetic field)
\beq
\frac{\partial^2\sigma_{\nu\rho}}{\partial q_\mu\partial q_\lambda} \propto X_{\mu\nu\lambda\rho}+X_{\lambda\nu\mu\rho},\label{eq:visccond}
\eeq
Therefore the ($q=0$) Hall viscosity has a direct connection to the $q^2$ part of the conductivity tensor. Note that the symmetry of Eq.~(\ref{eq:visccond}) under the exchange of $\mu\leftrightarrow\lambda$ implies that only the Hall tensor (i.e. only the combinations of viscosity coeffiencts that contribute to forces) enter into the conductivity. This holds generally for Galilean-invariant systems with or without rotational symmetry. Hence, this relation holds for the anisotropic systems we consider, provided the viscosity response is expanded to include the novel coefficients in Eq.~\eqref{eq:hallviscgeneral}. Going to higher orders in momentum, one can consider the link between the full finite-wavector viscosity and conductivity, noting that at non-zero momentum, the transformation of the viscosity tensor under $PT$-symmetry changes accordingly\cite{golan2019boundary}. 

Regarding rotational symmetry, we note that through the viscosity-conductivity relation, the viscosity is connected to another fourth rank tensor which gives the stress response to electric field gradients. The same component-by-component symmetry analysis we consider can be applied here (and in fact, to any response function), with the caveat that properties under time-reversal are not the same as that of the viscosity tensor (electric field gradients and strain rates do not have the same behavior under time reversal). It would be interesting to connect irreducible representations of the viscosity tensor to those of the electric field gradient response tensor using the viscosity-conductivity relation as a mapping. 

As a last point, we mention that the fourth rank tensor that shares the most properties with the viscosity is the tensor of elastic moduli $\kappa$. For example, for the Hall elastic modulus\cite{scheibner2019odd,offertaler2019viscoelastic} $(\kappa^\mathrm{H})^{\mu\hphantom{\nu}\lambda}_{\hphantom{\mu}\nu\hphantom{\lambda}\rho}$, the force redundancy statement in Eq.~\eqref{eq:viscforce} also holds. Similar to the Hall viscosity, the Hall elastic modulus is antisymmetric under $(\mu \nu) \leftrightarrow (\lambda \rho)$ (except unlike the viscosity, it is only nonvanishing in driven systems.) 

\section{Free Fermions: Hall viscosity and Berry curvature}\label{sec:freefermion}

Using this general formalism, we can now examine the Hall viscosity in anisotropic free-fermion systems, both in the continuum and on the lattice. We take as our unperturbed Hamiltonian
\begin{equation}
    H_0=\sum_{nm\mathbf{k}}c^\dag_{n\mathbf{k}}f^{nm}(\mathbf{k})c_{m\mathbf{k}}.\label{eq:freeham}
\end{equation}
Below we will first consider the stress response in anisotropic continuum systems, where $f^{nm}(\mathbf{k})$ can be any smooth function of $\mathbf{k}$. Second, we will consider lattice systems, where $f^{nm}(\mathbf{k})$ satisfies the additional reciprocal lattice periodicity constraints. In both formalisms, we will see how the Hall viscosities simultaneously capture the effects of Berry curvature and internal angular momentum.

\subsection{Continuum Viscosity}\label{subsec:continuumviscff}

For a continuum free-fermion system with Hamiltonian Eq.~(\ref{eq:freeham}) we can compute the stress tensor explicity from the form of the strain generator Eq.~(\ref{eq:modstrain}) and the definition (\ref{eq:modwardidentity}) to find
\begin{align}
    T^\mu_{\mathrm{B},\nu}&=-i\left[H_0,\mathcal{J}^\mu_{\hphantom{\mu}\nu}\right]\nonumber \\
    &=\sum_{nm\mathbf{k}}c^\dag_{n\mathbf{k}}\left(k_\nu\partial^\mu f^{nm}(\mathbf{k})+\frac{i}{2}\epsilon^{\mu}_{\phantom{\mu}\nu}[f(\mathbf{k}),L_\mathrm{int}]_{nm}\right)c_{m\mathbf{k}}\label{eq:ffcontstress}
\end{align}

\subsubsection{Equilibrium Stress}

Before inserting this expression into the Kubo formula Eq.~(\ref{eq:kubo}), it is worth looking at some general properties of the free fermion stress. First, let us reiterate that for a rotationally invariant system,
\begin{equation}
    i[f(\mathbf{k}),L_\mathrm{int}]_{nm}\rightarrow k_x\partial^yf_{nm}(\mathbf{k})-k_y\partial^xf_{nm}(\mathbf{k}),
\end{equation}
guaranteeing that the stress tensor will be symmetric in this case. Next, let us look at the equilibrium stress $\langle T^\mu_{\mathrm{B},\nu}\rangle_0$ in a zero temperature ground state with chemical potential $E_F$. We introduce the diagonalizing unitary transformation
\begin{equation}
    f^{nm}(\mathbf{k})=\sum_\alpha U^\dag_{n\alpha}(\mathbf{k})U_{\alpha m}(\mathbf{k})\epsilon_\alpha(\mathbf{k}),\label{diagonalization}
\end{equation}
where $\epsilon_{\alpha}(\mathbf{k})$ are the single-particle eigenvalues of $H_0$. This transformation allows us to write the single-particle Green's function as
\begin{equation}
    G_{nm}(\mathbf{k})=\langle c^\dag_{n\mathbf{k}} c_{m\mathbf{k}}\rangle_0=\sum_\alpha n_F(\epsilon_\alpha(\mathbf{k}))U^\dag_{m\alpha}(\mathbf{k})U_{\alpha n}(\mathbf{k})\label{eq:spgf},
\end{equation}
where $n_F$ is the Fermi distribution function. Taking the ground state average of Eq.~(\ref{eq:ffcontstress}) and using Eq.~(\ref{eq:spgf}) yields, after some simplification
\begin{equation}
    \langle T^\mu_{\mathrm{B},\nu}\rangle_0=\sum_{\alpha \mathbf{k}} n_F(\epsilon_\alpha(\mathbf{k}))k_\nu\partial^\mu\epsilon_\alpha(\mathbf{k})\label{eq:ffavgstress}
\end{equation}
Importantly, the spin stress does not contribute to the ground state average, since the average of a commutator of (finite rank) operators vanishes. Using this expression, we can prove an important theorem about the ground state stress in a continuum free-fermion system,
\begin{equation}
    \langle T^\mu_{\mathrm{B},\nu}\rangle_0=\frac{1}{2}\delta^\mu_\nu\langle \mathrm{tr} (T_\mathrm{B})\rangle_0.
\end{equation}
In words, this means that the average stress of a free fermion system is isotropic. To prove this, let us first rewrite Eq.~(\ref{eq:ffavgstress}) as an integral over momentum space,
\begin{equation}
    \langle T^\mu_{\mathrm{B},\nu}\rangle_0 = \frac{V}{4\pi^2}\int d^2k n_F(\epsilon_\alpha)k_\nu v^{\mu}_{\alpha}\label{eq:ffstressintegral1},
\end{equation}
where $v^{\mu}_{\alpha}\equiv\partial^\mu\epsilon_\alpha$ is the group velocity of band $\alpha$. To simplify this further, let us do a coordinate transformation to a basis of constant-energy contours. Let us write
\begin{equation}
\mathbf{k}_{a\alpha}=k_{a\alpha}(\theta,\epsilon)(\cos\theta,\sin\theta),\label{eq:coordtranspolar}
\end{equation}
where the new index $a$ to run over possible disjoint branches of the inverse of $\epsilon_{\alpha}(\mathbf{k})$. Note that care must be taken due to the inverse function, and we cannot assume that partial derivatives with respect to $\theta$ and $\epsilon$ commute. Since the number of values of the index $a$ may depend on $\epsilon$, we must also take care to integrate over $\theta$ and sum over $a$ before integrating over $\epsilon$. With these caveats in mind, Eq.~(\ref{eq:ffstressintegral1}) becomes
\begin{equation}
     \langle T^\mu_{\mathrm{B},\nu}\rangle_0=\frac{V}{4\pi^2}\int_0^\infty d\epsilon n_F(\epsilon)\sum_{\alpha a}\int_0^{2\pi} d\theta |J_{a\alpha}|k_{a\alpha\nu} v_{a\alpha}^{\mu},
\end{equation}
where $|J_{a\alpha}|$ is the Jacobian determinant for the transformation, given by
\begin{equation}
    |J_{a\alpha}|=k_{a\alpha}\partial_\epsilon k_{a\alpha}.
\end{equation}
Additionally, the inverse of the Jacobian matrix yields the velocity
\begin{equation}
    v_{a\alpha}^{\mu}=\sum_\nu\epsilon^{\mu\nu}\frac{\partial_\theta k_{a\alpha\nu}}{|J_{a\alpha}|}
\end{equation}
Putting these pieces together gives
\begin{equation}
    \langle T^{\mu}_{\mathrm{B},\nu}\rangle0 = \frac{V}{4\pi^2}\int_0^\infty d\epsilon n_F(\epsilon)\sum_{a \alpha\lambda}\int_0^{2\pi}d\theta \epsilon^{\mu\lambda}k_{a\alpha\nu}\partial_\theta k_{a\alpha\lambda}\label{eq:ffstressintegral2}
\end{equation}
When contracted with $\epsilon_\mu^{\hphantom{\mu}\nu},\sigma_\mu^{x,\nu}$ and $\sigma_\mu^{z,\nu}$, the integrand in (\ref{eq:ffstressintegral2}) can be rewritten as a total $\theta$ derivative, and hence the integral vanishes. On the other hand, when contracted with $\delta^\mu_\nu$, the integrand in (\ref{eq:ffstressintegral2}) becomes the differential area element $\mathbf{k}_{a\alpha}\times\partial_\theta\mathbf{k}_{a\alpha}d\theta$ enclosed by the constant energy surfaces in Eq.~(\ref{eq:coordtranspolar}). We thus conclude that
\begin{equation}
   \langle T^{\mu}_{\mathrm{B},\nu}\rangle_0=\frac{V}{8\pi^2}\delta^\mu_\nu\int_0^\infty d\epsilon n_F(\epsilon) A(\epsilon),
\end{equation}
where $A(\epsilon)$ is the momentum-space area enclosed by the constant-$\epsilon$ contour. Note that this result holds at any temperature, and even in the absence of rotational symmetry. We thus see that equilibrium stress in an anisotropic Fermi gas is isotropic. We will now make use of this result to simplify the contact term in the nondissipative stress response.

\subsubsection{Stress response} 

Now we will examine the Hall stress response (\ref{eq:hallresponsedef}). First, let us focus on the contact term in the Kubo formula, which gives the ``Hall elastic modulus.\cite{offertaler2019viscoelastic,scheibner2019odd}'' Starting from Eq.~(\ref{subsec:kubo}) and focusing on the antisymmetric contribution to the contact term,
\begin{equation}
    C^{\mu\hphantom{\nu}\lambda}_{\hphantom{\mu}\nu\hphantom{\lambda}\rho}=\frac{i}{2\omega}\left(\langle \left[T^\mu_{\mathrm{B},\nu},\mathcal{J}^\lambda_{\hphantom{\lambda}\rho}\right]\rangle_0- \langle\left[T^\lambda_{\mathrm{B},\rho},\mathcal{J}^\mu_{\hphantom{\mu}\nu}\right]\rangle_0\right)
\end{equation}
we can use the Jacobi identity and the Ward identity Eq.~(\ref{eq:modwardidentity}) to find
\begin{align}
C^{\mu\hphantom{\nu}\lambda}_{\hphantom{\mu}\nu\hphantom{\lambda}\rho}&=\frac{1}{2\omega}\left(
\langle\left[\left[H_0,\mathcal{J}^\mu_{\hphantom{\mu}\nu}\right],\mathcal{J}^\lambda_{\hphantom{\lambda}\rho}\right]\rangle_0-\langle\left[\left[H_0,\mathcal{J}^\lambda_{\hphantom{\lambda}\rho}\right],\mathcal{J}^\mu_{\hphantom{\mu}\nu}\right]\rangle_0
\right) \nonumber \\
&=\frac{1}{2\omega}\left(\langle\left[\left[H_0,\mathcal{J}^\mu_{\hphantom{\mu}\nu}\right],\mathcal{J}^\lambda_{\hphantom{\lambda}\rho}\right]\rangle_0+\langle\left[\left[\mathcal{J}^\lambda_{\hphantom{\lambda}\rho},H_0\right],\mathcal{J}^\mu_{\hphantom{\mu}\nu}\right]\rangle_0
\right) \nonumber \\
&=-\frac{1}{2\omega}\langle\left[\left[\mathcal{J}^\mu_{\hphantom{\mu}\nu},\mathcal{J}^\lambda_{\hphantom{\lambda}\rho}\right],H_0\right]\rangle_0
\end{align}
The commutator of strain generators can then be simplified using the algebra Eq.~(\ref{eq:strainalgebra}). We find using the kinetic Ward identity Eq.~(\ref{eq:kineticwardidentity})
\begin{align}
C^{\mu\hphantom{\nu}\lambda}_{\hphantom{\mu}\nu\hphantom{\lambda}\rho}&=\frac{i}{2\omega}\left(
\langle\left[J^\lambda_{\hphantom{\lambda}\nu},H_0\right]\rangle_0\delta^\mu_\rho-\langle\left[J^\mu_{\hphantom{\mu}\rho},H_0\right]\rangle_0\delta^\lambda_\mu
\right) \nonumber \\
&= \frac{1}{2\omega}\left(\langle T^{\lambda}_{\hphantom{\lambda}\nu}\rangle_0\delta^\mu_\rho-\langle T^{\mu}_{\hphantom{\mu}\rho}\rangle_0\delta^\lambda_\nu
\right)
\end{align}
Finally, using the results of Eq.~(\ref{eq:ffavgstress}) that
\begin{equation}
\langle T^{\mu}_{\hphantom{\mu}\nu}\rangle_0=\langle T^{\mu}_{\mathrm{B},\nu}\rangle_0\propto\delta^\mu_\nu,
\end{equation}
we see that
\begin{equation}
C^{\mu\hphantom{\nu}\lambda}_{\hphantom{\mu}\nu\hphantom{\lambda}\rho}=0.
\end{equation}
Thus, the anisotropic free fermi gas does not have a Hall elastic modulus. This is consistent with the interpretation of the continuum free fermi gas as a fluid, and should be contrasted with the results of Ref.~\onlinecite{offertaler2019viscoelastic} for a quasi-two dimensional system.

Finally, we can compute the Hall viscosity of the continuum free Fermi gas,
\begin{align}
 (\eta^\mathrm{H})^{\mu\hphantom{\nu}\lambda}_{\hphantom{\mu}\nu\hphantom{\lambda}\rho}=\frac{1}{2\omega^+V}\int_0^\infty dt& e^{i\omega^+t}\left(\langle [T^\mu_{\mathrm{B},\nu}(t),T^{\lambda}_{\mathrm{B},\rho}(0)]\right.\nonumber \\
 &\left.-[T^\lambda_{\mathrm{B},\rho}(t),T^{\mu}_{\mathrm{B},\nu}(0)]\rangle\right).\label{eq:ffhallvisckubo}
\end{align}
Since all the operators in the Kubo formula (\ref{eq:kubo}) are single-particle operators, we can evaluate the viscosity using the single-particle Green's function Eq.~(\ref{eq:spgf}).It is convenient to evaluate separately the two contributions
\begin{equation}
     (\eta^\mathrm{H})^{\mu\hphantom{\nu}\lambda}_{\hphantom{\mu}\nu\hphantom{\lambda}\rho}=(\eta^\mathrm{H}_{\mathrm{kin}})^{\mu\hphantom{\nu}\lambda}_{\hphantom{\mu}\nu\hphantom{\lambda}\rho}+(\eta^\mathrm{H}_{\mathrm{int}})^{\mu\hphantom{\nu}\lambda}_{\hphantom{\mu}\nu\hphantom{\lambda}\rho}
\end{equation}

where we define the kinetic viscosity $\eta^\mathrm{H}_\mathrm{kin}$ to be the time integral evaluated with the kinetic (or strain) stress tensor $T^\mu_{\hphantom{\mu}\nu}$ only, and $\eta^\mathrm{H}_\mathrm{int}$ to be the contributions to the viscosity from the internal angular momentum. Because the spin stress in Eq.~(\ref{eq:ffcontstress}) is purely antisymmetric, it does not contribute to the symmetric viscosity coefficients $\eta^\mathrm{H},\bar{\gamma},$ or $\bar{\Theta}$, which cam be extracted entirely from the kinetic contributions. Evaluating the kinetic contribution first, we can insert a complete set of states and use the eigenbasis Eq.~(\ref{diagonalization}) to carry out the time integral. We find that
\begin{widetext}
\begin{equation}
   (\eta^\mathrm{H}_{\mathrm{kin}})^{\mu\hphantom{\nu}\lambda}_{\hphantom{\mu}\nu\hphantom{\lambda}\rho} =\frac{1}{4\pi^2}\int d^2k \sum_{nm\alpha\beta}\frac{(\epsilon_\alpha-\epsilon_\beta)^2(n(\epsilon_\beta)-n(\epsilon_\alpha))}{\omega^2+(\epsilon_\alpha-\epsilon_\beta)^2} k_{\nu}(\partial^{\mu}U_{\beta n})U_{n\alpha}^\dag  k_{\rho}(\partial^\lambda U_{\alpha m})U_{m\beta}^\dag,\label{eq:ffviscunsimplified1}
\end{equation}
\end{widetext}
In the limit of zero frequency and zero temperature, this expression simplifies considerably. Recalling that the Berry connection $\mathbf{A}_{\alpha\beta}$ is given by
\begin{equation}
    (\partial_{\nu}U_{\beta n})U_{n\alpha}^\dag=\langle \partial^\nu u_{\beta\mathbf{k}}| u_{\alpha\mathbf{k}}\rangle\equiv iA^\nu_{\beta\alpha},
\end{equation}
in terms of the single-particle eigenstates $|u_{\alpha\mathbf{k}}\rangle$, we find that at zero temperature
\begin{equation}
     (\eta^\mathrm{H}_{\mathrm{kin}})^{\mu\hphantom{\nu}\lambda}_{\hphantom{\mu}\nu\hphantom{\lambda}\rho}=\frac{1}{4\pi^2}\int_{\mathrm{occ}} d^2k  k_\nu k_\rho \epsilon^{\mu\lambda}\mathrm{tr}(\Omega),\label{eq:ffcontstrainvisc}
\end{equation}
where $\Omega$ is the non-Abelian Berry curvature viewed as a matrix in the space of occupied states at $\mathbf{k}$\cite{avron1987adiabatic}, and $\int_{\mathrm{occ}}$ denotes an integral over occupied states. Recalling Eq.~(\ref{eq:hallviscgeneral}) and noting again that $(\eta^\mathrm{H}_{\mathrm{kin}})^{\mu\hphantom{\nu}\lambda}_{\hphantom{\mu}\nu\hphantom{\lambda}\rho}$ is the only contribution to the symmetric Hall viscosity coefficients, we find that
\begin{align}
    \eta^\mathrm{H}&=\frac{1}{16\pi^2}\int_{\mathrm{occ}} d^2{k}( k_x^2+ k_y^2)\mathrm{tr}(\Omega)\label{eq:ffeta} \\
    \bar{\gamma}&=\frac{1}{16\pi^2}\int_{\mathrm{occ}} d^2{k}( k_x^2- k_y^2)\mathrm{tr}(\Omega)\label{eq:ffgammabar} \\
    \bar{\Theta}&=\frac{1}{8\pi^2}\int_{\mathrm{occ}} d^2{k} k_x k_y\mathrm{tr}(\Omega) \label{eq:ffthetabar}\\
\end{align}
We see that the coefficients $\eta^\mathrm{H}$, $\bar{\gamma}$, and $\bar{\Theta}$ are given by the quadrupole moments of the Berry curvature. This is one of the main results of this work. 

Note that since the Berry curvature $\mathrm{tr}(\Omega)$ is $PT$-even, we see here explicitly that $\bar{\Theta}$ is $PT$-odd while $\eta^\mathrm{H}$ and $\bar{\gamma}$ are $PT$-even, consistent with the results of Section~\ref{subsec:symmetry}.

However, note that our expression Eq.~(\ref{eq:ffcontstrainvisc}) is antisymmetric on the indices $\mu$ and $\lambda$. From our discussion in Eq.~(\ref{eq:divfreecontact}), we see that this is precisely the tensor structure for a term that \emph{does not contribute to the equations of motion}. That is, we see from (\ref{eq:viscforce}) for the viscous forces that the antisymmetric viscosities $\bar{\eta}^\mathrm{H},\gamma,$ and $\Theta$ computed from Eq.~(\ref{eq:ffcontstrainvisc}) exactly cancel the viscous forces due to the symmetric viscosities (\ref{eq:ffeta}--\ref{eq:ffthetabar}). This leads us to our second main result, namely that \emph{the continuum Hall tensor is entirely determined from the internal angular momentum contribution to the stress}. This holds true even at finite frequency and temperature, due to the antisymmetry of Eq.~(\ref{eq:ffviscunsimplified1}) under the exchange of $\mu\leftrightarrow\lambda$. 

At zero frequency and temperature, we can evaluate the remainder of the Kubo formula in a simplified form. Using the same spectral decomposition leading to (\ref{eq:ffviscunsimplified1}) we find in the limit of zero frequency and zero temperature that

\begin{align}
    (\eta^\mathrm{H}_{\mathrm{int}})^{\mu\hphantom{\nu}\lambda}_{\hphantom{\mu}\nu\hphantom{\lambda}\rho}&=\sum_{\alpha\beta}\frac{1}{4\pi^2}\int d^2k (n(\epsilon_\alpha)-n(\epsilon_\beta)) L_\mathrm{int}^{\alpha\beta}\times \nonumber \\
    &
    \times\left(\epsilon^{\lambda}_{\hphantom{\lambda}\rho}k_\nu A_{\beta\alpha}^\mu-\epsilon^{\mu}_{\hphantom{\mu}\nu}k_\rho A_{\beta\alpha}^\lambda\right),
\end{align}
where we have introduced
\begin{equation}
    L_\mathrm{int}^{\alpha\beta}=\langle u_{\alpha\mathbf{k}}|L_\mathrm{int}|u_{\beta\mathbf{k}}\rangle=U_{\alpha n}L_\mathrm{int}^{nm}U^\dag_{m\beta}
\end{equation}
as the matrix elements of the internal angular momentum in the basis of single-particle energy eigenstates $|u_{\alpha\mathbf{k}}\rangle$. To see that this expression is gauge-invariant, we can rewrite it in terms of projectors 
\begin{equation}
    P(\mathbf{k})=\sum_{\mathbf{k}\alpha}n_F(\epsilon_\alpha)|u_{\alpha\mathbf{k}}\rangle\langle u_{\alpha\mathbf{k}}|
\end{equation} 
onto the set of occupied bands to find
\begin{widetext}
\begin{align}
     (\eta^\mathrm{H}_{\mathrm{int}})^{\mu\hphantom{\nu}\lambda}_{\hphantom{\mu}\nu\hphantom{\lambda}\rho}=\frac{1}{2\pi^2}\mathrm{Re}\sum_{\alpha}\int d^2k\; n_F(\epsilon_\alpha)\left(k_\nu\epsilon^\lambda_{\hphantom{\lambda}\rho}\langle u_{\alpha\mathbf{k}}|L(1-P(\mathbf{k}))|\partial^\mu u_{\alpha\mathbf{k}}\rangle_0-k_\rho\epsilon^\mu_{\hphantom{\mu}\nu}\langle u_{\alpha\mathbf{k}}|L(1-P(\mathbf{k}))|\partial^\lambda u_{\alpha\mathbf{k}}\rangle_0\right)
\end{align}

In this form, invarinace under unitary transformations in the space of occupied bands is manifest due to the factor of $1-P(\mathbf{k})$ which annihilates the inhomogeneous part of the transformation of $|\partial^\mu u_{\alpha\mathbf{k}}\rangle$.

By applying the contractions introduced in Eq.~(\ref{eq:hallviscdef}) and recombining the kinetic and internal angular momentum contributions to the viscosity we arrive at the following relations between the barred and unbarred viscosity coefficients:
\begin{align}
    \bar{\eta}^H&=\sum_{\alpha\beta}\frac{1}{16\pi^2}\int d^2k (n(\epsilon_\alpha)-n(\epsilon_\beta)) L_\mathrm{int}^{\alpha\beta}\mathbf{k}\cdot\mathbf{A}_{\beta\alpha}-\eta^\mathrm{H} \label{eq:ffetabar}\\
    \gamma &=\sum_{\alpha\beta}\frac{1}{16\pi^2}\int d^2k (n(\epsilon_\alpha)-n(\epsilon_\beta)) L_\mathrm{int}^{\alpha\beta}\mathbf{k}\cdot\sigma_z\cdot\mathbf{A}_{\beta\alpha}-\bar{\gamma} \label{eq:ffgamma}\\
    \Theta &=\sum_{\alpha\beta}\frac{1}{16\pi^2}\int d^2k (n(\epsilon_\alpha)-n(\epsilon_\beta)) L_\mathrm{int}^{\alpha\beta}\mathbf{k}\cdot\sigma_x\cdot\mathbf{A}_{\beta\alpha}-\bar{\Theta} \label{eq:fftheta}\\
\end{align}
We see then that the effect of the internal angular momentum contribution to the stress tensor is, for free fermions, to shift the values of the barred viscosity coefficients relative to their unbarred counterparts. In a rotationally invariant system, the fact that $\bar{\eta}^\mathrm{H}=\gamma=\bar{\gamma}=\Theta=\bar{\Theta}=0$ implies 
\begin{align}
&\sum_{\alpha\beta}\int d^2k (n(\epsilon_\beta)-n(\epsilon_\alpha)) L_\mathrm{int}^{\alpha\beta}\mathbf{k}\cdot\mathbf{A}_{\beta\alpha}\rightarrow\int_{\mathrm{occ}} d^2\mathbf{k}( k_x^2+ k_y^2)\mathrm{tr}(\Omega)\\
&\sum_{\alpha\beta}\int d^2k (n(\epsilon_\beta)-n(\epsilon_\alpha)) L_\mathrm{int}^{\alpha\beta}\mathbf{k}\cdot\sigma_z\cdot\mathbf{A}_{\beta\alpha} \rightarrow 0 \\
&\sum_{\alpha\beta}\int d^2k (n(\epsilon_\beta)-n(\epsilon_\alpha)) L_\mathrm{int}^{\alpha\beta}\mathbf{k}\cdot\sigma_x\cdot\mathbf{A}_{\beta\alpha} \rightarrow 0
\end{align}

Before moving on to discuss the analogous expressions in the lattice formalism, let us attempt to recast Eqs.~(\ref{eq:ffeta}--\ref{eq:ffthetabar}) and (\ref{eq:ffetabar}--\ref{eq:fftheta}) in terms of Fermi surface integrals, in the spirit of Haldane's approach to the anomalous Hall conductance\cite{haldaneahe}. Let us assume for simplicity of illustration that we have a rotationally invariant system with a single occupied band. In this case the Berry curvature is Abelian, only the isotropic Hall viscosity $\eta^\mathrm{H}$ is nonzero. Inserting 
\begin{equation}
\mathrm{tr}(\Omega)=\sum_{\mu\nu}\epsilon_{\mu\nu}\partial^\mu A^\nu
\end{equation}
into the expression (\ref{eq:ffeta}) and integrating by parts yields
\begin{align}
\eta^\mathrm{H}&=\sum_{\mu\nu}\frac{1}{16\pi}\epsilon_{\mu\nu}\int d^2k n_F(\mathbf{k})|\mathbf{k}|^2\partial^\mu A^\nu\\
&=-\sum_{\mu\nu}\frac{1}{16\pi}\epsilon_{\mu\nu}\int d^2k \partial^\mu(n_F(\mathbf{k})|\mathbf{k}|^2)A^\nu \\
&=\sum_{\mu\nu}\frac{1}{16\pi}\epsilon_{\mu\nu}\int d^2k\delta(\epsilon(\mathbf{k})-\epsilon_F)k_F^2v_{\mathrm{F}}^\mu A^\nu\nonumber\\&-\sum_{\mu\nu}\frac{1}{8\pi}\int_{\mathrm{occ}} d^2k \epsilon^{\mu}_{\hphantom{\mu}\nu}k_\mu A^\nu \\
&=\frac{k_F^2}{16\pi}\int_{\mathrm{FS}}{d\mathbf{k}_F}\cdot\mathbf{A} -\frac{1}{8\pi}\int_\mathrm{occ}d^2k \mathbf{k}\times\mathbf{A} \\
&=\frac{k_F^2}{8}\sigma^{AHE}-\frac{1}{8\pi}\int_\mathrm{occ}d^2k\mathbf{k}\times\mathbf{A},\label{eq:etahaldane}
\end{align}
\end{widetext}

where $\sigma^{AHE}$ is Haldane's expression for the Fermi-surface contribution to the anomalous Hall conductvitiy. Note that rotational symmetry was essential in deriving this expression, since only for rotationally invariant systems is the Fermi momentum independent of position along the Fermi surface. In addition to the Fermi surface contribution, however, we are left with an additional bulk contribution given by the second term. For rotationally-invariant systems each of these is separately gauge-invariant (mod $2\pi$), and neither is in general zero.

We can see here that the obstacle to obtaining a purely Fermi-surface expression for the Hall viscosity arises from the explicit factors of $\mathbf{k}$ in the integrands (\ref{eq:ffeta}--\ref{eq:ffthetabar}). Once we consider a fully anisotropic system, the situation becomes even worse: the Fermi surface and bulk terms will no longer be separately gauge invariant, and only the difference in Eq.~(\ref{eq:etahaldane}) remains physically meaningful. 

\subsection{Lattice Viscosity}
 Next, let us turn to the lattice version of the Kubo formula for viscosity Eq.~(\ref{subsec:kubo}). For free fermions, our starting point is the lattice stress tensor given in Eq.~(\ref{eq:fflatticestress}), which we repeat here:

\begin{align}
    T^\mu_{\mathrm{B}0,\nu}&=\sum_{nm\mathbf{k}}c^\dag_{n\mathbf{k}}\left(\partial_\mu f^{nm}(\mathbf{k})\frac{\sin\mathbf{k}\cdot \mathbf{a}_\nu}{|\mathbf{a}_\nu|}\right.\nonumber\\
    &\left.+\frac{i}{2}\tilde{\epsilon}^{\mu}_{\hphantom{\mu}\nu}\cos\mathbf{k}\cdot\mathbf{a}_\mu[f(\mathbf{k}),L_\mathrm{int}]^{nm}\right)c_{m\mathbf{k}}\label{eq:fflatticestressredux}
\end{align}

The fundamental differences between this and the continuum expression Eq.~(\ref{eq:ffcontstress}) are the replacement of $k_\mu$ by $\sin (\mathbf{k}\cdot\mathbf{a}_\mu)/|\mathbf{a}_\mu|$, and the factor of $\cos\mathbf{k}\cdot\mathbf{a}_\mu$ in the spin-stress. These similarities in structure allow us to repeat much of the same logic as in the continuum case for evaluating expectation values of the stress tensor  and commutators. In particular, the single particle lattice Green's function $G_{nm}(\mathbf{k})$ has a form identical to Eq.~(\ref{eq:spgf}), with $\epsilon_\alpha$ and $U_{\alpha n}$ periodic functions of $\mathbf{k}$. As in Sec.~\ref{subsec:continuumviscff}, we will evaluate the equilibrium stress, Hall elastic modulus, and Hall viscosity for a generic free-fermion lattice system.

\subsubsection{Equilibrium Stress}
Let us begin by evaluating the ground state average $\langle T^\mu_{\mathrm{B}0,\nu}\rangle_0$. As in the continuum case, we find using Eq.~(\ref{eq:spgf}) that the average of the spin-stress term vanishes, and
\begin{align}
    \langle T^\mu_{\mathrm{B}0,\nu}\rangle_0&=\sum_{\alpha\mathbf{k}}n_F(\epsilon_\alpha(\mathbf{k}))\frac{\sin\mathbf{k}\cdot\mathbf{a}_\nu}{|\mathbf{a}_\nu|}\partial^\mu\epsilon_\alpha(\mathbf{k}) \\
    &=\frac{V}{4\pi^2}\sum_{\alpha}\oint d^2k n_F(\epsilon_\alpha(\mathbf{k}))\frac{\sin\mathbf{k}\cdot\mathbf{a}_\nu}{|\mathbf{a}_\nu|}\partial^\mu\epsilon_\alpha(\mathbf{k})\label{eq:avglatticestress}
\end{align}
Unlike in the continuum, a coordinate transformation such as Eq.~(\ref{eq:coordtranspolar}) does not simplify our lives for two reasons. First, in the Brillouin zone constant energy countours may form non-contractible closed curves, rendering the coordinate transformation ill-defined. Second, even for contractible Fermi pockets, we can no longer exploit symmetry to explain the vanishing of the integral due to the $\sin$ function. As such, we cannot conclude that the average stress is isotropic in a lattice system. This is reasonable, since in a fundamental sense a lattice system is generically similar to a solid. Note, however, that in the ``lattice gauge theory'' continuum limit with $|\mathbf{a}_\mu|\rightarrow 0$, the stress tensor acquires a continuum form, and so we recover an isotropic expectation value. This supports the notion that at wavelengths long compared to the lattice spacing (i.e. for metallic systems with small Fermi surfaces) we recover a hydrodynamic picture of electron transport.

However, if we assume that our Hamiltonian has $3$-fold or higher rotational symmetry, we can greatly constrain the average stress. To see this, we follow the logic of Sec.~\ref{subsec:symmetry}. Writing in general
\begin{equation}
    \langle T^\mu_{\mathrm{B}0,\nu}\rangle_0=T_\delta \delta^\mu_\nu + T_\epsilon\tilde{\epsilon}^\mu_{\hphantom{\mu}\nu}+T_x\sigma^\mu_{x,\nu}+T_z\sigma^\mu_{z,\nu},
\end{equation}
We see that the coefficients $(T_x,T_z)$ must transform in a nontrivial representation of the group of n-fold rotations whenever $n\geq 3$. Since these quantities must be scalars by definition, we conclude that $T_x=T_z=0$ in the presence of more than threefold rotational symmetry. Furthermore, mirror symmetry (along any direction) forces $T_\epsilon=0$ by the same logic.

\subsubsection{Stress Response}
Let us now move on to examine the Hall response of lattice stress response the Kubo formula (\ref{eq:kubo}). As in the continuum, we will separately analyze the contact term and the time-integral contribution to the response function. Starting with the contact term, we will be rather brief. Unlike in the continuum, the lattice strain generators $\mathcal{J}^\mu_{\mathrm{L},\nu}$ do not form a closed algebra. As such, while we can still use the Jacobi identity to write the Hall contribution of the contact term as
\begin{align}
C^{\mu\hphantom{\nu}\lambda}_{\hphantom{\mu}\nu\hphantom{\lambda}\rho}&=\frac{1}{2\omega}\left(
\langle\left[\left[H_0,\mathcal{J}^\mu_{\mathrm{L},\nu}\right],\mathcal{J}^\lambda_{\mathrm{L},\rho}\right]\rangle_0-\langle\left[\left[H_0,\mathcal{J}^\lambda_{\mathrm{L},\rho}\right],\mathcal{J}^\mu_{\mathrm{L},\nu}\right]\rangle_0
\right) \nonumber \\
&=-\frac{1}{2\omega}\langle\left[\left[\mathcal{J}^\mu_{\mathrm{L},\nu},\mathcal{J}^\lambda_{\mathrm{L},\rho}\right],H_0\right]\rangle_0,\label{eq:latticecontact}
\end{align}
we cannot reduce this to an expression involving only the ground state average of the lattice stress tensor. Instead, we have
\begin{widetext}
\begin{align}
    \left[\mathcal{J}^\mu_{\mathrm{L},\nu},\mathcal{J}^\lambda_{\mathrm{L},\rho}\right]=-&\frac{1}{4}\sum_{\mathbf{k}n}c^\dag_{n\mathbf{k}}\left(\delta^\mu_\rho\left\{\left\{\partial^\lambda,\cos\mathbf{k}\cdot\mathbf{a}_\mu\right\},\sin\mathbf{k}\cdot\mathbf{a}_\nu\right\}-\delta^\lambda_\nu\left\{\left\{\partial^\mu,\cos\mathbf{k}\cdot\mathbf{a}_\nu\right\},\sin\mathbf{k}\cdot\mathbf{a}_\rho\right\}\right)c_{n\mathbf{k}} \nonumber \\
    &+\frac{i}{2}\sum_{\mathbf{k}nm}\delta^{\mu\lambda}c^\dag_{n\mathbf{k}}L_\mathrm{int}^{nm}c_{m\mathbf{k}}\sin\mathbf{k}\cdot\mathbf{a}_\mu\left(\hat{\epsilon}^{\mu}_{\hphantom{\mu}\nu}\sin\mathbf{k}\cdot\mathbf{a}_\rho-\hat{\epsilon}^{\mu}_{\hphantom{\mu}\rho}\sin\mathbf{k}\cdot\mathbf{a}_\nu\right)
\end{align}
We can then insert this expression into Eq.~(\ref{eq:latticecontact}). Fortunately, like in the continuum the internal angular momentum does not contribute to the contact term. We find
\begin{align}
    C^{\mu\hphantom{\nu}\lambda}_{\hphantom{\mu}\nu\hphantom{\lambda}\rho}&=\frac{1}{2\omega}\sum_{\mathbf{k}nm}G_{nm}(\mathbf{k})\left(\delta^\mu_\rho\cos\mathbf{k}\cdot\mathbf{a}_\rho\sin\mathbf{k}\cdot\mathbf{a}_\nu\partial^\lambda f_{nm}(\mathbf{k})-\delta^\lambda_\nu\cos\mathbf{k}\cdot\mathbf{a}_\nu\sin\mathbf{k}\cdot\mathbf{a}_\rho\partial^\mu f_{nm}(\mathbf{k})\right)\label{eq:latticehallmodulus}
\end{align}
which we cannot simplify further in general.
\end{widetext}

For the Hall viscosity Eq.~(\ref{eq:ffhallvisckubo}), we are considerably more lucky. To evaluate the Kubo formula, it is convenient to treat the kinetic and internal angular momentum contributions to the stress tensor separately. We can write
\begin{equation}
    (\eta^\mathrm{H})^{\mu\hphantom{\nu}\lambda}_{\hphantom{\mu}\nu\hphantom{\lambda}\rho}=(\eta^\mathrm{H}_{\mathrm{kin}})^{\mu\hphantom{\nu}\lambda}_{\hphantom{\mu}\nu\hphantom{\lambda}\rho}+(\eta^\mathrm{H}_{\mathrm{int}})^{\mu\hphantom{\nu}\lambda}_{\hphantom{\mu}\nu\hphantom{\lambda}\rho},
\end{equation}
where $(\eta^\mathrm{H}_{\mathrm{kin}})^{\mu\hphantom{\nu}\lambda}_{\hphantom{\mu}\nu\hphantom{\lambda}\rho}$ is defined by Eq.~(\ref{eq:ffhallvisckubo}) with only the kinetic contribution to the stress tensor included. For the kinetic contribution, we can follow the same logic as in Sec.~(\ref{subsec:continuumviscff}). We find in complete analogy with that section
\begin{equation}
(\eta^\mathrm{H}_{\mathrm{kin}})^{\mu\hphantom{\nu}\lambda}_{\hphantom{\mu}\nu\hphantom{\lambda}\rho}=\frac{1}{4\pi^2}\int_{\mathrm{occ}}d^2k \sin\mathbf{k}\cdot\mathbf{a}_\nu\sin\mathbf{k}\cdot\mathbf{a}_\rho \hat{\epsilon}^{\mu\lambda}\mathrm{tr}(\Omega)\label{eq:lattickineticviscosity}
\end{equation}
Thus, like in the continuum, the strain viscosity is related to (periodic) moments of the trace of the Berry curvature. Next, we can compute the internal angular momentum contributions to the viscosity. Unlike in the continuum, these will contribute to both the ``barred'' and ``unbarred'' viscosity coeffiicents (since, as noted in Sec.~\ref{sec:latticeformalism}, the internal angular momentum contribution to the stress tensor is not purely antisymmetric on the lattice). Following the computation in Sec.~\ref{subsec:continuumviscff}, we find that at zero temperature
\begin{widetext}
\begin{equation}
    (\eta^\mathrm{H}_{\mathrm{int}})^{\mu\hphantom{\nu}\lambda}_{\hphantom{\mu}\nu\hphantom{\lambda}\rho}=\sum_{\alpha\beta}\frac{1}{4\pi^2}\int d^2 k (n(\epsilon_\alpha)-n(\epsilon_\beta))L_\mathrm{int}^{\alpha\beta}\left(\hat{\epsilon}^\lambda_{\hphantom{\lambda}\rho}\cos\mathbf{k}\cdot\mathbf{a}_\lambda\sin\mathbf{k}\cdot\mathbf{a}_\nu A_{\alpha\beta}^\mu-\hat{\epsilon}^\mu_{\hphantom{\mu}\nu}\cos\mathbf{k}\cdot\mathbf{a}_\mu\sin\mathbf{k}\cdot\mathbf{a}_\rho A_{\alpha\beta}^\lambda\right).\label{eq:fflatticeinteta}
\end{equation}
\end{widetext}
As in the continuum, we have that the kinetic viscosity Eq.~(\ref{eq:lattickineticviscosity}) does not contribute to the Hall tensor Eq.~(\ref{eq:hallvisc2tensor}), and hence does not contribute to viscous forces; the internal angular momentum is responsible for all nondissipative viscous forces that enter the equations of motion. One important difference on the lattice, However, is that the internal viscosity $ (\eta^\mathrm{H}_{\mathrm{int}})^{\mu\hphantom{\nu}\lambda}_{\hphantom{\mu}\nu\hphantom{\lambda}\rho}$ on the lattice contributes to \emph{both} the barred and unbarred viscosity coefficients, due to the extra cosine factors in Eq.~(\ref{eq:fflatticeinteta})

\section{Example systems}\label{sec:examples}
\subsection{Continuum massive Dirac fermion}

As a first example, let us consider a massive Dirac fermion in two dimensions, with Hamiltonian
\begin{equation}
    H=\sum_{k\sigma\sigma'}c^\dag_{\sigma\mathbf{k}}\left(\vec{d}(\mathbf{k})\cdot\vec{\sigma}^{\sigma\sigma'}\right)c_{\sigma'\mathbf{k}}\label{eq:dirachamgen}
\end{equation}
where $\vec{d}(\mathbf{k})$ is a three-vector of functions $\vec{d}=(d_x,d_y,d_z)\equiv (\mathbf{d},d_z)$, $\vec{\sigma}=(\sigma_x,\sigma_y,\sigma_z)\equiv(\boldsymbol{\sigma},\sigma_z)$, and we view this internal degree of freedom as a real spin. Note that we use $\mathbf{v}$ for two-dimensional vectors, and $\vec{v}$ for three-dimensional vectors. The single-particle eigenvalues and eigenfunctions of this Hamiltonian have the well-known form
\begin{align}
    |\pm\mathbf{k}\rangle&=\sum_\sigma u_{\sigma\mathbf{k}}c^\dag_{\sigma\mathbf{k}}|0\rangle\label{eq:diracprops1} \\
    H|\pm\mathbf{k}\rangle&= \pm E_{\mathbf{k}}|\pm\mathbf{k}\rangle
\end{align}
with 
\begin{align}
    u_{\uparrow\mathbf{k}}&=\left(\cos\frac{\theta}{2}, e^{i\phi} \sin\frac{\theta}{2}\right) \\
    u_{\downarrow\mathbf{k}}&=\left(\sin\frac{\theta}{2}, -e^{i\phi} \cos\frac{\theta}{2}\right) \\
    E_k&=\left|\vec{d}(\mathbf{k})\right| \\
    \tan\theta&=\frac{|\boldsymbol{d}|}{m} \\
    \tan\phi&=\frac{d_y}{d_x}\label{eq:diracprops2}
\end{align}

Since we view $\sigma,\sigma'$ as real spin, rotations in the plane are generated by both the kinetic angular momentum and the internal angular momentum operator given by
\begin{equation}
L_\mathrm{int}=\frac{1}{2}\sigma^z.
\end{equation}

Using Eq.~(\ref{eq:ffcontstress}), we can compute the Belinfante stress tensor for this model. We find
\begin{align}
    T^\mu_{\mathrm{B},\nu}&=\sum_{\mathbf{k}\sigma\sigma'}c^\dag_{\sigma\mathbf{k}}\left(k_\nu\partial^\mu\vec{d}\cdot\vec{\sigma}^{\sigma\sigma'}+\frac{1}{2}\epsilon^\mu_{\hphantom{\mu}\nu}\left(\mathbf{d}\times\boldsymbol{\sigma}^{\sigma\sigma'}\right)\right)c_{\sigma'\mathbf{k}} \\
    &\equiv \sum_{\mathbf{k}\sigma\sigma'}c^\dag_{\sigma\mathbf{k}}[T^{\sigma\sigma'}]^\mu_{\mathrm{B},\nu}(\mathbf{k})c_{\sigma'\mathbf{k}}\label{eq:dirackstress}
\end{align}
We will use this to evaluate the viscosity for both isotropic and anisotropic systems
\subsubsection{Isotropic Dirac Point}\label{sec:isodirac}
Let us first focus on the isotropic case
\begin{equation}
    \vec{d}=(v_\mathrm{F}\mathbf{k},m)
\end{equation}
with $m$ a constant mass. This choice of $\vec{d}$ corresponds to an isotropic massive Dirac fermion in two dimensions. The stress tensor in our formalism for this case coincides with the usual Belinfante tensor for a Dirac theory\cite{nakahara2003geometry}
\begin{equation}
    T^\mu_{\mathrm{B},\nu}(\mathbf{k})=\frac{v_\mathrm{F}}{2}\left(k_\nu \sigma^\mu + k_\mu \sigma^\nu\right)
\end{equation}
where, since this is a flat-space expression, the distinction between upper and lower indices is immaterial. We have also suppressed the $\sigma,\sigma'$ indices on the Pauli matrices for notational convenience. Since this system is invariant under continuous rotational symmetries, and the stress tensor is symmetric, we can make use of Eqs.~(\ref{eq:ffeta}--\ref{eq:ffthetabar}) and (\ref{eq:ffetabar}--\ref{eq:fftheta}) to evaluate the viscosity. In fact, the symmetry of the problem immediately implies that only $\eta^\mathrm{H}$ can be nonzero. Let us consider the Fermi level $E_\mathrm{F}$ in the lower band of the model. We can write the Berry curvature for this band as
\begin{align}
    \Omega_-&=\frac{1}{2}\sum_{\mu\nu}\epsilon_{\mu\nu}\sin\theta\partial^\mu\theta\partial^\nu\phi\label{eq:Berrygeneral} \\
    &=\frac{mv_\mathrm{F}^2}{2E_\mathbf{k}^3}
\end{align}
Substituting this into Eq.~(\ref{eq:ffeta}), we find
\begin{align}
    \eta^\mathrm{H}&=\frac{1}{16\pi^2}\int_\mathrm{occ} d^2k |\mathbf{k}|^2\Omega_- \\
    &=\frac{1}{4\pi^2}\int_\mathrm{occ} d^2k \frac{mv_\mathrm{F}^2|\mathbf{k}|^2}{8E_\mathbf{k}^3}\label{eq:isodiraccont}
\end{align}
consistent with the known result\cite{shapourian2015viscoelastic,hughesleighfradkin}. Because the energy spectrum of a Dirac fermion is unbounded from below, we need to introduce a momentum cutoff to evaluate this expression. Perhaps more useful, however, is to examine the differential viscosity
\begin{equation}
    \frac{d\eta^\mathrm{H}}{d E_\mathrm{F}}=\frac{m}{16\pi E_\mathrm{F}}(m^2-E_\mathrm{F}^2)
\end{equation}
the derivative of the Hall viscosity as a function of chemical potential. Since we assumed that only the lower band is occupied, this expression is valid only when $E_\mathrm{F}\leq -|m|$. For $-|m|\leq E_\mathrm{F}\leq |m|$, we have that
\begin{equation}
    \frac{d\eta^\mathrm{H}}{dE_\mathrm{F}}=0.
\end{equation}
Finally, when the Fermi surface is in the upper band ($E_\mathrm{F}\geq|m|$)
\begin{equation}
    \frac{d\eta^\mathrm{H}}{dE_\mathrm{F}}=\frac{m}{16\pi E_\mathrm{F}}(E_\mathrm{F}^2-m^2),
\end{equation}
where we have used the fact that $\Omega_+=-\Omega_-$.

\subsubsection{$C_4$ Warping}
Next, let us consider an anisotropic Dirac Hamiltonian with only fourfold rotational symmetry. The simplest such model is given by
\begin{align}
    d_x&=v_\mathrm{F}k_x+uk_x^3 \\
    d_y&=v_\mathrm{F}k_y+uk_y^3 \\
    d_z&=m.
\end{align}

The stress tensor determined from Eq.~(\ref{eq:dirackstress}) is
\begin{align}
    T^\mu_{\mathrm{B},\nu}(\mathbf{k})&=\frac{v_\mathrm{F}}{2}\left(k_\mu\sigma^\nu+k_\nu\sigma^\mu\right)+\frac{3u}{2}\left(k_\nu k_\mu^2\sigma^\mu + k_\mu k_\nu^2\sigma^\nu\right)\nonumber\\
    &+\frac{u}{2}\epsilon^\mu_{\hphantom{\mu}\nu}\left(k_y\sigma^x(3k_x^2-k_y^2)-k_x\sigma^y(3k_y^2-k_x^2)\right)
\end{align}
Since this model does not have continuous rotational symmetry, even the Belinfante stress tensor has an explicit antisymmetric contribution. Since we still have fourfold rotational symmetry, however, we expect the only nonzero viscosities to be $\eta^\mathrm{H}$ and $\bar{\eta}^\mathrm{H}$. To compute these, we note that in addition to Eqs.~(\ref{eq:diracprops1}--\ref{eq:diracprops2}), the expression Eq.~(\ref{eq:Berrygeneral}) also holds true even for this anisotropic model. Using this along with either the expressions Eq.~(\ref{eq:ffeta}) and (\ref{eq:ffetabar}) or directly from the Kubo formula (\ref{eq:kubo}) we find that when the chemical potential is in the valence band
\begin{align}
    \eta^\mathrm{H}&=\frac{1}{4\pi^2}\int_{\mathrm{occ}}d^2k\frac{m|\mathbf{k}|^2(3uk_x^2+v_{\mathrm{F}})(3uk_y^2+v_\mathrm{F})}{8E_\mathbf{k}^3} \\
    \bar{\eta}^\mathrm{H}&=\frac{1}{4\pi^2}\int_{\mathrm{occ}}d^2k\frac{mu}{8E_\mathbf{k}^3}\left(\vphantom{\frac{m}{E_k^3}}3|\mathbf{k}|^2((k_x^2-k_y^2)^2-2k_x^2k_y^2)\right.\nonumber\\
    &\left.\vphantom{\frac{m}{E_k^3}}+v_F \left((k_x^2-k_y^2 )^2-4k_x^2k_y^2 \right)\right).
\end{align}
As in the isotropic case, these integrals require a momentum cutoff to be evaluated, although in the interest of brevity we will not pursue this further in this section. The physically relevant combination $\eta^\mathrm{H}_\mathrm{total}=\eta^\mathrm{H}+\bar{\eta}^\mathrm{H}$ that determines the Hall tensor (\ref{eq:hallvisc2tensor}) and viscous forces (\ref{eq:viscforce}) is then
\begin{equation}
    \eta^\mathrm{H}_\mathrm{total}=\frac{m}{4\pi^2}\int_{\mathrm{occ}}d^2k\frac{3u^2(k_x^6+k_y^6)+4uv_\mathrm{F}(k_x^4+k_y^4)+v_\mathrm{F}^2|\mathbf{k}|^2}{8E_\mathbf{k}^3}.
\end{equation}
We see that in addition to the isotropic contribution proportional to $|\mathbf{k}|^2$, the integrand also contains the rotational-symmetry breaking function $k_x^4+k_y^4$ and $k_x^6+k_y^6$, which make the fourfold rotational symmetry manifest.

\subsubsection{$C_2$ Symmetry}

Finally, we will consider a simple model with only twofold rotational symmetry, by taking
\begin{align}
    d_x&=v_\mathrm{F}^xk_x\\
    d_y&=v_\mathrm{F}^yk_y\\
    d_z&=m.
\end{align}
Becuase this system differs by only a momentum rescaling from an isotropic Dirac point, the calculation of the viscosity coefficients proceeds in much the same way as in Sec~\ref{sec:isodirac}, with the exception being that now the $\bar{\eta}^H$ viscosity is non-zero. Although the stress tensor is now
\begin{equation}
    T^\mu_{\mathbf{B},\nu}(\mathbf{k})=k_\nu\sigma^\mu\left(v_\mathrm{F}^\mu-\frac{1}{2}v_{\mathrm{F}}^{\nu}\right)+\frac{1}{2}v_{\mathrm{F}}^{\nu}k^\mu\sigma^\nu,
\end{equation}
which contains an explicit antisymmetric contribution, we find for the \textit{total} Hall viscosity an analogous expression to the unbarred Hall viscosity for the isotropic case in Sec~\ref{sec:isodirac}. It is given by
\begin{align}
    \eta_{\mathrm{total}}^\mathrm{H}\equiv \eta^\mathrm{H}+\bar{\eta}^\mathrm{H}&=\frac{1}{16\pi^2}\int_\mathrm{occ} d^2k |\mathbf{k}|^2\Omega_- \\
    &=\frac{1}{4 \langle v_\mathrm{F}\rangle_\mathrm{g}^2 \pi^2}\int_\mathrm{occ} d^2u \frac{m|\mathbf{u}|^2}{8E_\mathbf{u}^3}\label{eq:c2diraccont}
\end{align}
We have introduced transformed coordinates $u_x = v_F^x k_x$ and $u_y = v_F^y k_y$, and we have also introduced the geometric mean
\begin{equation}
    \label{eq:vfgeommean}
    \langle v_\mathrm{F}\rangle_g\equiv\sqrt{v_\mathrm{F}^xv_\mathrm{F}^y}.
\end{equation}
Thus we see that for this simple $C_2$ symmetric model, the \textit{total} Hall viscosity matches the isotropic result upon replacing the isotropic Fermi velocity with the geometric mean of the two anisotropic Fermi velocities.

In the next section, we will examine a lattice-regularization for the Dirac Hamiltonian Eq.~(\ref{eq:dirachamgen}), which is of considerable interest in its own right as a model for a Chern insulator. We will highlight what differs between the lattice and continuum models, and show how to take the continuum limit of the lattice viscosities in the limit of large bandwidth.

\subsection{Lattice models for a Chern insulator}\label{sec:chern}
We now move to our first example without full translational invariance -- a square lattice Chern insulator. We take for our Bravais lattice vectors
\begin{equation}
    \mathbf{a}_1=a\hat{\mathbf{x}},\;\;\;\mathbf{a}_2=a\hat{\mathbf{y}},
\end{equation}
and thus we can avoid the complications associated with non-Cartesian bases. This model represents the lattice regularization of the massive Dirac fermion in the previous section -- heuristically, we map the momenta in to crystal momenta $k_\mu  \rightarrow\sin(\mathbf{k\cdot a_\mu})= \sin(k_\mu a)$ and add a momentum-dependent mass to the model. We continue to use the tight-binding Hamiltonian form of Eq. \eqref{eq:dirachamgen}, 

\beq
\bal
\label{eq:cilatham}
d_x &= t \sin(k_x a)\\
d_y &= t' \sin(k_y a)\\
d_z &= m - r\left\lbrace{}\cos(k_x a) + \cos(k_y a)\right\rbrace
\eal
\eeq
This model still has internal degrees of freedom that need to be taken into account when considering rotations of the system, meaning that the generator of rotations is the full angular momentum operator Eq. \eqref{eq:angularmomentum}, and the proper rotation operator is thus 

\beq
\hat{R}(\phi) = R_{\mathrm{orb}}(\phi) \otimes e^{i\phi \sigma^z/2}
\eeq
The orbital rotation matrix is of the usual form $R_{\mathrm{orb}}(\phi)\in SO(2)$ as in Eq. \eqref{eq:rotationmatrix}. The square lattice Chern insulator model we have described is always invariant twofold rotations. The point group that leaves the Hamiltonian invariant is therefore $C_2$. When $t=t'$, the model acquires a fourfold rotational symmetry as well, expanding the point group to $C_4$. The model also always has $PT$ symmetry, which can be represented as
\begin{equation}
    \Delta(PT)=\sigma_z\mathcal{K}
\end{equation}
where $K$ is complex conjugation, and $PT$ takes $k_x\rightarrow-k_x$ and leaves $k_y$ invariant.
For any $m\neq 0,\pm 2r$, this model is fully gapped. When $|m|\geq 2r$ the valence and conduction band have Chern number zero, while when $|m|\le 2r$ the valence and conduction band each have Chern number of magnitude 1.

The stress tensor, computed with Eq.~ \eqref{eq:fflatticestressredux} in components is then given by
\begin{equation}
    T^\lambda_{\mathrm{B},\rho}=\sum_{\mathbf{k}\sigma\sigma'}c^\dag_{\sigma\mathbf{k}}[T^{\sigma\sigma'}]^\mu_{\mathrm{B},\nu}(\mathbf{k})c_{\sigma'\mathbf{k}}
\end{equation}
with components (repeated indices are {\it not} summed over)
\beq
\bal
\label{eq:latticestressci}
\hat{T}^{\lambda}_{\mathrm{B},{\rho}}({\bf k}) &= t^\lambda \cos(k_\lambda a) \sin(k_\rho a) \sigma^\lambda + r \sin(k_\lambda a) \sin(k_\rho a)\, \sigma^z \\ &+ \frac{1}{2} \cos(k_\lambda a) \left(t^\lambda\sin(k_\lambda a) \sigma^\rho   - t^\rho\sin(k_\rho a) \sigma^\lambda \right)\\
&\equiv (\hat{T}^{\mathrm{kin}}({\bf k}))^{\lambda}_{\hphantom{\lambda}{\rho}} + (\hat{T}^{\mathrm{\, spin}}({\bf k}))^{\lambda}_{\hphantom{\lambda}{\rho}}
\eal
\eeq
The first two terms above are the stress tensor calculated from the purely kinetic part of the strain generator and are therefore labelled $
(\hat{T}^{\mathrm{kin}})^{\lambda}_{\hphantom{\lambda}{\rho}}$ whereas the spin contribution is given by $(\hat{T}^{\mathrm{\, spin}})^{\lambda}_{\hphantom{\lambda}{\rho}}$. We make this distinction to stress the importance of including the spin contribution for viscosity computations -- showing the difference between what we call the kinetic viscosity ($\eta$ calculated from solely the strain part of the stress tensor) and the overall viscosity. We can tailor the Kubo formula Eqn.~\eqref{eq:kubo} to our system by writing 
\begin{align}
\label{eq:cikubo}
(\eta^\mathrm{H}({\bf k}))^{\lambda\hphantom{\rho}\mu}_{\hphantom{\lambda}\rho\hphantom{\mu}\nu} & = -\frac{\Im\left[\left(T^{-+}({\bf k})\right)^{\lambda}_{\mathrm{B},\rho} \left(T^{+-}({\bf k})\right)^{\mu}_{\mathrm{B},\nu} \right]}{2\epsilon_{\bf k}^2} \\
(\eta^\mathrm{H})^{\lambda\hphantom{\rho}\mu}_{\hphantom{\lambda}\rho\hphantom{\mu}\nu} & = \frac{1}{4\pi^2}\int_{\mathrm{occ}}d^2k (\eta^\mathrm{H}({\bf k}))^{\lambda\hphantom{\rho}\mu}_{\hphantom{\lambda}\rho\hphantom{\mu}\nu}
\end{align}
Above, and henceforth, we use the notation 
\beq
\left(T^{-+}({\bf k})\right)^{\lambda}_{\mathrm{B},\rho} = \mel{-,{\bf k}}{\left(T^{\lambda}_{\mathrm{B},\rho}({\bf k}) \right)}{+,{\bf k}}
\eeq    
We focus on the linear combinations of the viscosity tensor which enter into the Hall tensor (\ref{eq:hallvisc2tensor}). For this gapped, time-reversal odd model the isotropic contribution to the Hall tensor is given by the sum of the two isotropic Hall viscosities:

\beq
\label{eq:etatotdefn}
\eta^{H}_{\, \mathrm{total}} = \eta^\mathrm{H} + \bar{\eta}^H = \frac{1}{2} \left(\eta^{1\hphantom{1}1}_{\hphantom{1}1\hphantom{1}2}-
\eta^{2\hphantom{2}2}_{\hphantom{2}2\hphantom{2}1}\right)
\eeq
In general, the anisotropic components $\gamma$ and $\bar{\gamma}$ may also be non-zero, and they are given by the opposite combination of components
\beq
\gamma_{\, \mathrm{total}} = \gamma +\bar{\gamma} = \frac{1}{2}\left(\eta^{1\hphantom{1}1}_{\hphantom{1}1\hphantom{1}2} + \eta^{2\hphantom{2}2}_{\hphantom{2}2\hphantom{2}1}\right)
\eeq
We can view the above as a measure for the anisotropy of a system -- in particular, it represents how far away a system is from having fourfold rotation symmetry.

\subsubsection{$C_4$ (t=t') system}

\begin{figure}[t]
    \centering
    \includegraphics[width=0.4\textwidth]{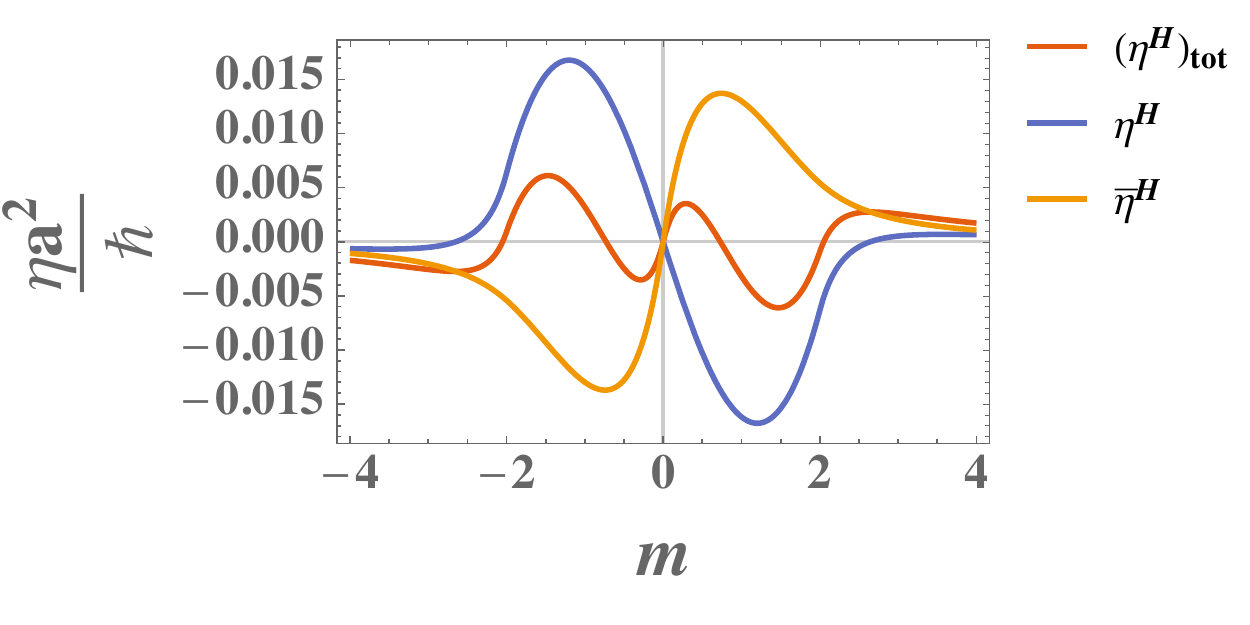}
    \centering
    \caption{A plot of the total Hall viscosity $\eta^\mathrm{H}_{\, \mathrm{total}}$, $\eta^\mathrm{H}$, and $\bar{\eta}^H$ as a function of mass $m \in [-4,4]$ for the $C_4$ symmetric Chern insulator. We fix the hopping parameters $t = r = 1$ and set the lattice constant $a=1$. The viscosity is measured in units of $\hbar/a^2$. We see that each viscosity is odd in the mass parameter $m$, and all go to zero as $|m|\rightarrow\infty$.} 
    \label{fig:etatotc4}
\end{figure}

We start by considering the $C_4$ symmetric system with hoppings $t=t'$. First, we note that the Hall modulus Eq.~(\ref{eq:latticehallmodulus}) vanishes, since the integrand is odd in momentum. The total lattice Hall viscosity computed from Eqs.~\eqref{eq:cikubo} \& \eqref{eq:etatotdefn} is a non-trivial expression, which we plot as a function of $m$ in Figure~\ref{fig:etatotc4}. As in Ref.~\onlinecite{shapourian2015viscoelastic}, we see that $\eta^\mathrm{H}, \bar{\eta}^\mathrm{H},$ and $\eta^\mathrm{H}_\mathrm{total}$ are all smooth functions of $m$ across all phase boundaries $m=0,\pm2 r$. 

As we saw in Eq.~(\ref{eq:lattickineticviscosity}), if instead we focus on the contribution to the viscosity from the kinetic stress we find that the contribution $\eta^\mathrm{H}_{\mathrm{kin}}+\bar{\eta}^\mathrm{H}_\mathrm{kin}\equiv(\eta^\mathrm{H}_{\mathrm{total}})_{\mathrm{kin}}$ to the Hall tensor has the property that 
\beq
(\eta^\mathrm{H}_{\mathrm{total}})_{\mathrm{kin}} = 0 \leftrightarrow \eta^{H}_{\mathrm{kin}} = -\bar{ \eta}^{H}_{\mathrm{kin}}.
\eeq
This is what was computed in Ref.~\onlinecite{shapourian2015viscoelastic}.

\begin{figure}[t]
\includegraphics[width=0.35\textwidth]{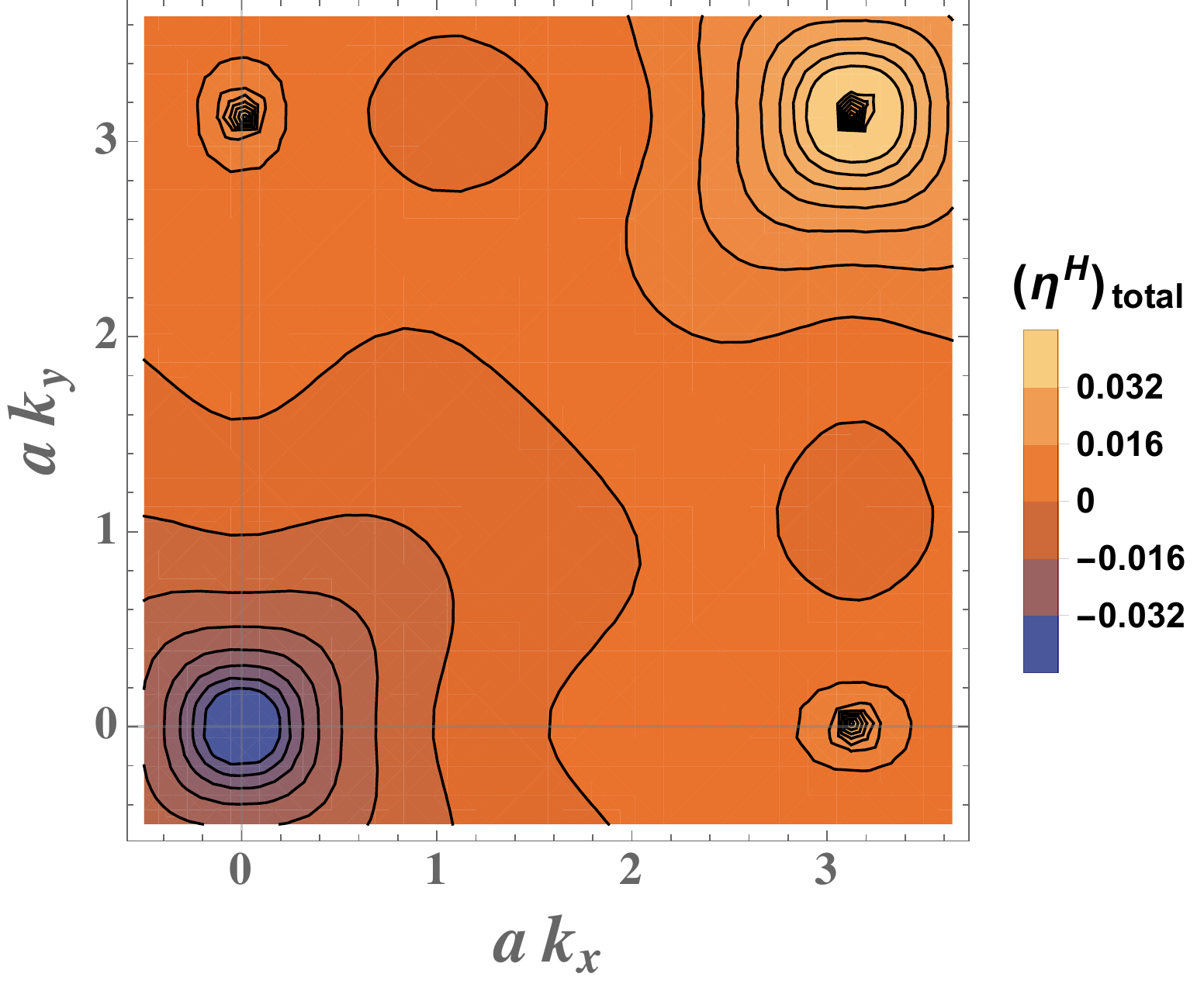}
\caption{Conour plot of the density $\eta^\mathrm{H}_\mathrm{total}(\mathbf{k})$ for the $C_4$ symmetric Chern insulator. The plot is shown with the hopping parameters $t=t'=20$, $m=0.25$, and $r=1$.}\label{fig:c4contours}
\end{figure}

\textit{Continuum limit}: To extract some of the underlying physics of the problem, we consider the continuum limit of this model by approximating the lattice  fermion operators by slowly-varying field operators $\psi_{ni}$ defined around the Dirac points in momentum space $\mathbf{K}_i = (k_x, k_y) = \lbrace (0,0), (\pi,0),(0,\pi),(\pi,\pi)\rbrace$:

\beq
\label{eq:contops}
c_{n\bf R} \approx \sum\limits_i e^{i {\bf K_i \cdot R}} \psi_{ni}({\bf R})
\eeq
We can proceed to compute the viscosity in this language, but must be careful as to how we do so, since the lattice strain generator Eq.~\eqref{eq:latticestrain} has momentum dependence which much be considered to find the proper stress tensor Eq.~(\ref{eq:latticestressci}). To proceed, we independently expand the lattice Hamiltonian and stress tensor, compute the viscosity around each Dirac point, and then finally sum up the resulting coefficients to find the continuum Hall viscosity. To first order, the Hamiltonian for each Dirac point is given by
\beq
H^{(i)} = v_F \left(\alpha^i_x k_x \sigma^x + \alpha^i_y k_y \sigma^y \right) + M^i \sigma^z \label{eq:expandeddiracham}
\eeq
where $v_F \equiv ta$ and the sign coefficients about each Dirac point are given by $\alpha_x^i = \lbrace 1, -1 , 1, -1 \rbrace$ and $\alpha_y^i = \lbrace 1, 1, -1, -1 \rbrace$, and the Dirac mass changes across the points as $M^i = \lbrace m-2r, m, m, m+2r \rbrace$. The dispersion for each of these points is of the same form, 

\beq
\epsilon^{(i)}_{\bf k} = \sqrt{v_F^2 \mathbf{|k|}^2 + (M^i)^2}.
\eeq  
The expansion of this tensor around the Dirac points can be written as

\begin{widetext}
\beq
\bal
(T^{(i)})^{\lambda}_{\mathrm{B},\rho}(\mathbf{k}) &= v_F \left[ \alpha_\lambda^i \alpha_\rho^i  \sigma^\lambda k_\rho + \frac{1}{2} \alpha_\lambda^i \left(\alpha_\lambda^i k_\lambda \sigma^\rho - \alpha_\rho^i k_\rho \sigma^\lambda \right) \right]
\\ &=  \frac{v_F}{2} \alpha^i_\lambda \left[  \alpha_\rho^i \sigma^\alpha k_\rho + \alpha_\lambda^i k_\lambda \sigma^\rho\right]
\eal
\eeq

The resulting viscosity can be easily computed, noting that the Pauli matrices satisfy $\Im{
(\sigma^a)^{-+} (\sigma^b)^{+-}} = M \epsilon_{ba} / \epsilon_{\bf k}$

\beq
\bal
\label{eq:viscgeneraldp}
(\eta^{(i)})^{\lambda\hphantom{\rho}\mu}_{\hphantom{\lambda}\rho \hphantom{\mu}\nu} = \frac{v_F^2 M^i}{4 \epsilon_{\bf k}^3} \left[\alpha^i_\lambda \alpha_\rho^i \alpha_\mu^i \alpha_\nu^i k_\rho k_\nu \epsilon^{\lambda}_{\hphantom{\lambda}\nu} + \alpha_\lambda^i \alpha_\rho^i k_\rho k_\mu \epsilon^{\lambda}_{\hphantom{\lambda}\nu} + \alpha_\mu^i \alpha_\nu^i k_\lambda k_\nu \epsilon_{\rho}^{\hphantom{\rho}\mu} + k_\lambda k_\mu \epsilon_{\rho}^{\hphantom{\rho}\mu} \right]
\eal
\eeq

\end{widetext}
Still, the only possible non-zero viscosities are $\eta^\mathrm{H}$ and $\bar{\eta}^H$, and we need to consider their sum Eq.~\eqref{eq:etatotdefn}. For the expansion, we need to sum this over all the Dirac points. The general viscosity-integrand across the Dirac points is then given by

\beq
\bal
\label{eq:viscdirac}
(\eta^{H}_{\, \mathrm{total}}({\bf k}))^{(i)} &=  \left(\eta^\mathrm{H}({\bf k}) + \bar{\eta}^H({\bf k})\right)^{(i)} \\&= \frac{v_F^2 M^i}{8 \epsilon_{\bf k}^3} \mathbf{|k|}^2
\eal
\eeq
Eq \eqref{eq:viscdirac} shows that each Dirac point contributes to $\eta^\mathrm{H}_\mathrm{total}$ identically to Eq.~(\ref{eq:isodiraccont}) for an isolated isotropic Dirac point in the continuum. This result is distinct from that in Ref.~\onlinecite{shapourian2015viscoelastic} for the Hall viscosity obtained using only the kinetic stress tensor and focusing only on the isotropic viscosity. The distinction arises due to our careful treatment of the spin stress. We note that, as shown in Appendix~\ref{app:othersymmetrization} that there exists an alternative definition for the spin stress which reproduces the results of Ref.~\onlinecite{shapourian2015viscoelastic} for the viscosity of the Chern insulator model when both $\eta^\mathrm{H}$ and $\bar{\eta}^\mathrm{H}$ are computed. However, this comes at the expense of the simplicity of Eq.~(\ref{eq:latticeintmom}) in real space. 

We expect this continuum approximation to be accurate when the mass gap at each Dirac point is small compared to the bandwidth, meaning $t\gg m,r$. In this limit, the low-energy behavior of the model is approximately that of four independent massive Dirac points. To verify this, we show in Fig.~\ref{fig:c4contours} a plot of the density $\eta^\mathrm{H}_\mathrm{total}(\mathbf{k})$ for $t=t'=20, m=r=1$. We see that near the Dirac points, the viscosity is isotropic and matches the form of the integrand in Eq.~(\ref{eq:isodiraccont}).

Returning to Eq.~(\ref{eq:viscdirac}), we can introduce a momentum cutoff $\Lambda$, integrate over momentum, and sum over the Dirac points to obtain.

\beq
\label{eq:etatotaldiracptintegrated}
\eta^{H}_{\, \mathrm{total}} = \frac{1}{2} \sum\limits_i \left[\frac{M^i \Lambda}{8\pi v_F} - \frac{M^i|M^i|}{4\pi v_F^2}\right]
\eeq
The momentum cut-off $\Lambda$ is introduced to perform the integrals about each Dirac point, and the overall viscosity is given by a cut-off dependent piece and a universal part
\beq
\bal
\label{eq:etatotalfinalsum}
\eta^{H}_{\, \mathrm{total}} &= \left[\frac{m \Lambda}{4\pi v_F} - \sum_i \frac{M^i|M^i|}{8\pi v_F^2}\right]
\eal
\eeq
 Unlike in Ref.~\onlinecite{shapourian2015viscoelastic}, however, we see that the contributions to the viscosity above the momentum cutoff cannot be neglected. If we focus on the barred and unbarred Hall viscosities separately, we see that the end points $(0,0)$ and $(\pi,\pi)$ are described fully isotropic theories, so the unbarred Hall viscosity is the only non-zero contribution $(\eta^\mathrm{H}_{\mathrm{total}})^{(0,3)} = \eta^\mathrm{H}$. At the corner points, however, $(\pi,0)$ and $(0,\pi)$, $(\eta^\mathrm{H}_{\mathrm{total}})^{(1,2)} = \bar{\eta}^H$. Thus the total viscosity from each Dirac point is the same, but the isotropic points express their viscosity through $\eta^\mathrm{H}$ and the anisotropic ones express their viscosity through the rotational-symmetry breaking coefficient $\bar{\eta}^H$. Examining the contour plot Fig.~\ref{fig:c4contours}, we can interpret the roll of the momentum cutoff $\Lambda$ as limiting the domain of integration to the regions of the Brillouin zone where the viscous density $\eta^\mathrm{H}_\mathrm{total}(\mathbf{k})$ is concentrated near the (massive) Dirac points.

Therefore we see that the barred viscosity is non-zero is crucial for achieving the result Eq. \eqref{eq:etatotalfinalsum}. If we were to consider only the unbarred viscosity $\eta^\mathrm{H}$, we would be neglecting half of the physical effect of the viscosity. Our results here should be contrasted with those of Ref.~\onlinecite{shapourian2015viscoelastic}, which computed the kinetic viscosity $\eta^\mathrm{H}_\mathrm{kin}$ only. This viscosity does not contribute to viscous forces, since as mentioned previously, $\bar{\eta}^\mathrm{H}_\mathrm{kin}=-\eta^\mathrm{H}_\mathrm{kin}$. Nevertheless, those authors found that the continuum limit expression for $\eta^\mathrm{H}_\mathrm{kin}$ is cutoff independent for the Chern insulator. Here, on the other hand, we find the equally appealing result that $\eta^\mathrm{H}_\mathrm{total}$ for the Chern insulator reduces in the continuum limit to the total viscosity of four uncoupled isotropic massive Dirac points. This is consistent with the expansion Eq.~(\ref{eq:contops}) about the four massive Dirac points in the model.

\subsubsection{$C_2$ anisotropy and comments on physical responses}
\label{sec:c2lat}
With the intuition from the fourfold rotation-invariant case, we now consider the general model in Eq.~(\ref{eq:cilatham}). We first mention that this model realizes the anisotropic viscosity components $\gamma$ and $\bar{\gamma}$ in addition to $\eta^\mathrm{H}_{\, \mathrm{total}}$ (all others are forbidden by $PT$ symmetry). We plot the total anisotropic viscosity $\gamma_{\mathrm{total}}$ as a function of $m$ in Fig.~\ref{fig:gammatotc2} below.

\begin{figure}[t]
    \centering
    \includegraphics[width=0.45\textwidth]{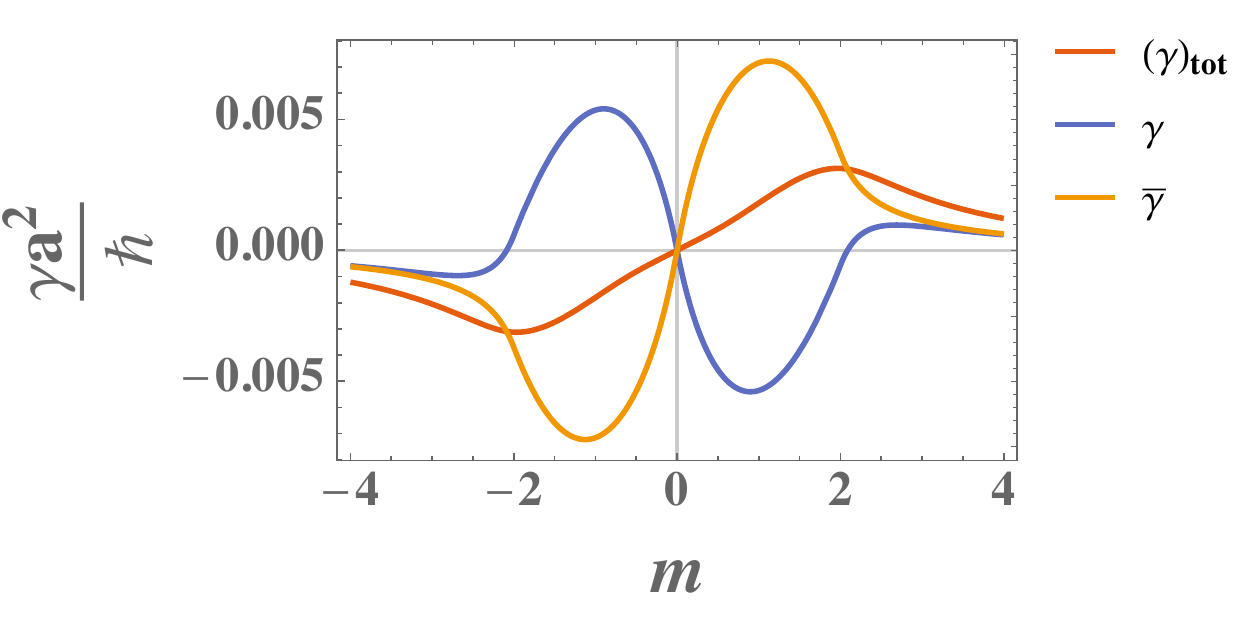}
    \centering
    \caption{The anisotropic viscosities $\gamma_{\mathrm{total}}, \gamma$ and $\bar{\gamma}$ as a function of the mass $m \in [-4,4]$ for the $C_2$ symmetric Chern insulator. The hoppings are fixed to be $t = 1.5$ and $t = r =1$, while the lattice constant is again $a=1$. Again we see that all viscosities are odd in $m$.}
    \label{fig:gammatotc2}
\end{figure}

\begin{figure}[t]
\includegraphics[width=0.35\textwidth]{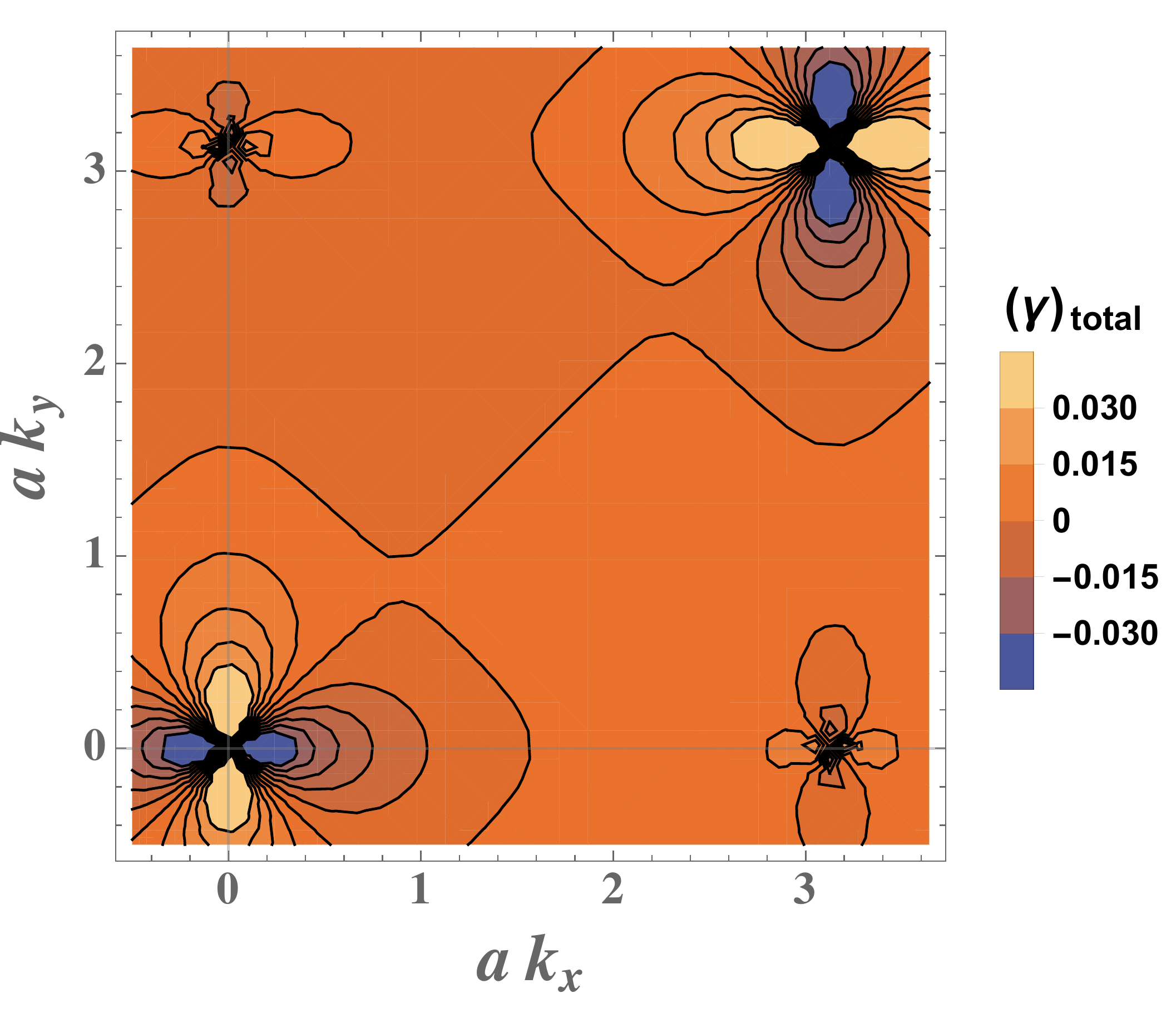}
\caption{Conour plot of the density $\gamma     _\mathrm{total}(\mathbf{k})$ for the $C_2$ symmetric Chern insulator. The plot is shown with the hopping parameters $t=20$ and $t'=21$, $m=0.25$, $a=1$, and $r=1$.}\label{fig:c2contours}
\end{figure}

\textit{Continuum limit}: Again, we consider the Dirac point expansion of the model, now noting that we can absorb the anisotropy into a non-uniform Fermi velocity ${\bf v_F} = (ta, t'a)$. Here, we find an analogous result for the physical viscosity:

\beq
\label{eq:viscdiracc2}
 \left(\eta_{\mathrm{total}}({\bf k}) \right)^{(i)} \\= \frac{M^i }{8 \epsilon_{\bf k}^3} \left((v_F^x k_x)^2 + (v_F^y k_y)^2\right)
\eeq
If we make the transformations $u_x = v_F^x k_x$ and $u_y= v_F^y k_y$, then the viscosity takes the same form as the isotropic case,

\beq
\bal
  \left(\eta_{\mathrm{total}}({\bf k}) \right)^{(i)} &= \frac{M^i}{8}\frac{(u_x^2 + u_y^2)}{(u_x^2 + u_y^2 + (M^i)^2)^{3/2}} \\
  &= \frac{M^i}{8}\frac{|{\bf u}|^2}{\left(|{\bf u}|^2 + (M^i)^2\right)^{3/2}}
\eal
\eeq
In the last line, we have moved to polar coordinates in $u_x, u_y$. Performing the integration yields an analogous result to Eq. \eqref{eq:etatotaldiracptintegrated}, 

\beq
\bal
(\eta^\mathrm{H}_{\mathrm{total}})^{(i)} = \frac{1}{2 } \sum\limits_i\left[\frac{m \tilde{\Lambda}}{8\pi \ev{v_F}_g} - \sum\limits_i \frac{M^i|M^i|}{4\pi \ev{v_F}_g^2}\right]
\eal
\eeq
As in the case of the isolated $C_2$ Dirac fermion, the viscosity in this case is proportional to the inverse geometric mean of Fermi velocities Eq.~\eqref{eq:vfgeommean} when the Fermi velocity is non-uniform. The cutoff in this case is of the form 
\beq \tilde{\Lambda} = a \sqrt{t^2 \Lambda_x + t'^2 \Lambda_y^2} \eeq
where $\Lambda_{x,y}$ are momentum cut-offs. We are left with an analogous expression to the $C_4$ result, simply replacing the cut-off and fermi velocity with their anisotropic counterparts in $C_2$.

We saw in Fig.~\ref{fig:gammatotc2} that the lattice version of this model gives rise to anomalous viscosities $\gamma$ and $\bar{\gamma}$ -- this changes when moving to the long-wavelength continuum expansion. When summed over the Dirac points and transformed to variables $u_x$ and $u_y$, we find strikingly for both coefficients that
\beq
\bal
\gamma({\bf k}) &= - \bar{\gamma}({\bf k}) =  \sum\limits_i \gamma^{(i)}({\bf k}) = \frac{m}{16\epsilon_{\bf u}
^3} \left(u_x^2 - u_y^2 \right)\\
&= \frac{m |{\bf u}|^2}{16\epsilon_{\bf u}^3} \cos(2 \theta_{{\bf u}})
\eal
\eeq
As these expressions transform non-trivially under roations of the new coordinates, they vanishes upon integration over any isotropic surface. We thus have that $\bar{\gamma} = \gamma = 0$ in the continuum limit.  

As in the case of the isotropic viscosity, we expect the continuum expansion to capture the behavior of the viscosity when the bandwidth is large compared to the gap. We plot the density $\gamma_\mathrm{total}$ for hopping parameters $t=t'+1=20,r=1,m=1$ in Fig.~\ref{fig:c2contours}. We see that the majority of the variation in the density is captured by the quadrupolar lobes near each of the Dirac points, whose contributions integrate to zero. Beyond the momentum cutoff, this cancellation is imperfect and we recover the nonvanishing of $\gamma_\mathrm{total}$ shown in Fig.~\ref{fig:gammatotc2}.

To summarize, we have found that in the continuum limit, the anisotropic viscosities of the Chern insulator vanish just as they do for a $C_2$-anisotropic Dirac theory in the continuum.

\subsection{Anisotropic superfluids}

Paired superfluids, which sponetaneously break parity and time reversal symmetry in the superfluid phase, also naturally realize a Hall viscosity. If we include anisotropy in the system, these superfluids can have nonzero values for the viscosity coefficients in Eq.~(\ref{eq:hallviscgeneral}). The general mean-field Hamiltonian we work with is given by\cite{read2000paired}
\beq
\bal
H = \int d^2x \; \psi^\dagger({\bf x}) \left(-\frac{1}{2m} \mathcal{E}^{ij} \p_i \p_j - \mu \right) \psi({\bf x}) \\+ \frac{1}{2} \int d^2 x' \int d^2 x \left[ \Delta({\bf x - x'})\psi^\dagger({\bf x}) \psi^\dagger({\bf x'}) + h.c. \right]\label{eq:bdgham}
\eal
\eeq
The rotational symmetry group of this Hamiltonian depends on the symmetric mass tensor $\mathcal{E}^{ij}$. We can write (defining parameters $\alpha$ and $\beta$ for the parity even and odd anisotropic perturbations, respectively)

\beq
\label{eq:disp}
\mathcal{E}_{ij} = \delta_{ij} + \kappa \left \lbrace \beta \, \sigma^x_{ij} + \alpha \, 
\sigma^z_{ij}\right\rbrace
\eeq
To break time reversal symmetry, we require a complex gap function, and we consider a paired $\ell$-wave complex superfluid, with gap function:
\beq
\Delta_{\bf k} = |\Delta_{\bf k}| e^{i \ell \phi}
\eeq
This breaks both parity and time-reversal, but is even under the composite symmetry $PT$, whereas the kinetic term in the Hamiltonian is only $PT$-even when $\beta = 0$. We can utilize a Bogoliubov transformation to find the quasiparticle excitation spectrum. In this basis, the Hamiltonian takes the form,

\beq
H = \sum_{\bf k} E_{\bf k} d^\dagger_{\bf k} d_{\bf k} + E_0
\eeq
The transformation made was

\beq
\bal
c_{\bf k}  &= u^*_{\bf k} d_{\bf k} - v_{-{\bf k}} d^\dagger_{-\bf k} \\
c^\dagger_{\bf k}  &= u_{\bf k} d^\dagger_{\bf k} - v^*_{-{\bf k}} d_{-\bf k}
\eal
\eeq
There is a redundancy  for the functions $u_{\bf k}, v_{\bf k}$, and we can choose a gauge in which $u_{\bf k}$ is real. In this gauge, we have that
\begin{equation}
v_{\bf k}=|v_{\bf k}|e^{i\ell\phi}
\end{equation}
transforms as an $\ell$-pole under rotations. Its magnitude is given by\cite{read2000paired}

\beq
|v_{\bf k}|^2  = \frac{1}{2} \left(1 - \frac{\epsilon_{\bf k}}{E_{\bf k}}\right)
\eeq 
In this gauge $v_{\bf k}$ transforms in the same way under rotations as the gap $\Delta_{\bf k}$. The dispersion relation for the Hamiltonian (\ref{eq:bdgham}) is given by

\beq
\bal
E_{\bf k} = \sqrt{\left(\frac{1}{2m}(\mathcal{E}^{ij} k_i k_j) - \mu \right)^2 + |\Delta_{\bf k}|^2}\\
\equiv \sqrt{\epsilon^2_{\bf k} +| \Delta_{\bf k}|^2}
\eal
\eeq

To compute the viscosity for this model, we will use the fact that there are no internal degrees of freedom, so that we can make use of the formalism of Ref.~\onlinecite{bradlyn2012kubo} directly. Integrating the Kubo formula (\ref{eq:kubo}) by parts,  we can use the strain-strain form of the viscosity, 

\beq
\eta^{\mu \hphantom{\nu} \lambda}_{\hphantom{\mu} \nu \hphantom{\lambda}\rho}(\omega\rightarrow 0) = -i \ev{\comm{J^{\mu}_{\hphantom{\mu}\nu}}{J^{\lambda}_{\hphantom{\lambda}\rho}}}_0 
\eeq
This is odd under time reversal, so the full viscosity is a Hall viscosity at zero frequency. In terms of quasiparticle operators, the ground state expectation value of the strain operators is, from Refs.~\onlinecite{read2009non,bradlyn2012kubo},

\beq
\ev{J^\mu_{\hphantom{\mu}\nu}}_0 =\frac{-iL^2}{4} \int \frac{d^2k}{(2\pi)^2} \, k_\nu \left(v_{\bf k} \p_{k_\mu} v^*_{\bf k} - v^*_{\bf k} \p_{k_\mu} v_{\bf k} \right).
\eeq
Using the commutation relation between strain generators,

\beq
\comm{J^{\mu}_{\hphantom{\mu}\nu}}{J^{\lambda}_{\hphantom{\lambda} \rho}} = -i\left(\delta^{\mu}_{ \rho} J^{\lambda}_{\hphantom{\lambda}\nu} - \delta_{\nu}^{ \lambda}J^{\mu}_{\hphantom{\mu}\rho}\right)
\eeq
We can write the zero-frequency viscosity for the superfluid as

\beq
(\eta^\mathrm{H})^{\mu \hphantom{\nu} \lambda}_{\hphantom{\mu} \nu \hphantom{\lambda}\rho} = \delta_{\nu}^\lambda \ev{J^{\mu}_{\hphantom{\mu}\rho}}_0 - \delta^{\mu}_{\rho} \ev{J^{\lambda}_{\hphantom{\lambda}\nu}}_0
\eeq
We find that the independent components of the ground state strain give exactly the three traceless components of the Hall viscosity, hence we can decompose the strain expectation into irreducible representations as 
\beq
\ev{J^{\mu}_{\hphantom{\mu}\nu}}_0 = \eta^\mathrm{H} \epsilon^{\mu}_{\hphantom{\mu}\nu} + \gamma \, (\sigma^x)^{\mu}_{\hphantom{\mu}\nu} - \Theta \, (\sigma^z)^{\mu}_{\hphantom{\mu}\nu} 
\eeq
The viscosity coefficients are then given by

\beq
\bal
\label{eq:sfviscex}
\eta^\mathrm{H} / L^2 &= \frac{l}{4} \int \frac{d^2 k}{(2\pi)^2}  |v_{\bf k}|^2  \\
\gamma / L^2 &= -\frac{l}{4} \int \frac{d^2k}{(2\pi)^2} \cos(2\phi) \,  |v_{\bf k}|^2 \\
\Theta / L^2 &= -\frac{l}{4}\int \frac{d^2k}{(2\pi)^2} \sin(2\phi) \,  |v_{\bf k}|^2 \\
\eal
\eeq
We can see anisotropy of the anomalous coefficients explicitly, and it is clear that they vanish if the dispersion $\epsilon_{\bf k}$ is isotropic. To lowest order in the small parameter $\kappa$,

\beq
\bal
\label{eq:sfvisc}
\eta^\mathrm{H} &= \frac{1}{2} \bar{n}\bar{s}  \\
\gamma &= \frac{l}{32} \kappa \alpha \mathcal{I} \\
\Theta &= \frac{l}{32}\kappa \beta \mathcal{I} \\
\eal
\eeq
The isotropic Hall viscosity is quantized as in\cite{bradlyn2012kubo} by the mean particle number density and angular momentum per particle, while the anomalous coefficients enter at first order in the anisotropy and both depend on the integral 

\beq
\mathcal{I} = \frac{L^2}{(2\pi)^2} \int dk \frac{\pi k^3 |\Delta_{\bf k}|^2}{m \, E^3_{{\bf k}0}},
\eeq
where $E_{{\bf k}0}=\lim_{\kappa\rightarrow 0} E_{{\bf k}}$. This is consistent with the results of Ref.~\onlinecite{golan2019boundary}.

\section{Experimental Implications}

A large motivation for our study is experiment, especially as the Hall viscosity has seen a bevy of theoretical proposals for measurement and an initial positive measurement in graphene. We see our work as an opportunity to extend the opportunities for this developing field of the experimental study of quantum fluids without time-reversal symmetry. Our lattice formalism gives a starting point for researchers looking at hydrodynamic and ballistic regimes in quantum materials, and importantly can shed light on whether anisotropic corrections to $\mathbf{k} \cdot \mathbf{p}$ theories, often invoked in materials like graphene, are important. In particular, in materials where 'warping terms' -- non-rotationally invariant contributions to continuum hamiltonians -- are important, considering anomalous Hall viscosity components is necessary, as we have done in this work. In this section, we will comment on the experimental implications of our formalism.

First, in Subsection~\ref{subsec:pipe}, we present an anisotropic generalization of the width dependent corrections to the resistivity in the hydrodynamic regime~\cite{scaffidi2017hydrodynamic,delacretaz2017transport}, which are experimentally accessible. Second, following the framework of Ref.~\cite{holder2019unified}, we remark on implications of our anistropic viscosity for non-hydrodynamic transport in Subsection~\ref{subsec:moments}. Third, in Subsection~\ref{subsec:torque} we will argue that Hall viscosity may be detectable in future torque magnetometry experiments in clean two-dimensional systems. Finally, in Subsection~\ref{subsec:materials} we remark on new material platforms in which our formalism for anisotropic and lattice viscosity may yield new probes topological physics beyond clean, noninteracting electron systems.

\subsection{Anisotropic pipe flow: the (viscous) Hall bar}\label{subsec:pipe}

In relaxing the assumption of isotropy, we found six independent viscosities that provide three viscous forces. Generalizing the approach of Ref.~\onlinecite{scaffidi2017hydrodynamic}, we can consider the classical viscous Hall effect in a rectangular Hall bar sample  with width $W$ along the $y$ direction and obtain viscous corrections to the classical Hall resistivity. For our set of viscous forces dictated by the tensor Eq.~\eqref{eq:viscforce}, we have three distinct of viscous corrections to the resistivity, two of them being anisotropic. If we orient our Cartesian axes at an angle $\alpha$ with respect to the $C_2$ axes of our sample, we find that the change in resistivity from the 
classical value $\Delta \rho_{xy} = \rho_{xy} - \rho_{xy}^{\mathrm{bulk}}$, 
with $\rho^{\mathrm{bulk}}_{xy} = -\frac{m\omega_c}{e^2 n}$, is given by

\begin{widetext}
\beq
\bal
\label{eq:resistivity}
\Delta\rho_{xy} = - \frac{\rho^{\mathrm{bulk}}_{xy}}{A\omega_c} \frac{12 }{W^2 - 12 l_s^2}\biggr[ (\eta^H)_{\mathrm{total}} - \gamma_{\mathrm{total}} \cos(2\alpha) - \Theta_{\mathrm{total}} \sin(2\alpha)\biggr]
\eal
\eeq
\end{widetext}
Above we used partial slip boundary conditions with slip length $l_s$\cite{delacretaz2017transport}. Our result differs by a factor of area $A=LW$ since Ref.~\onlinecite{scaffidi2017hydrodynamic} works with kinematic rather than dynamic viscosities. We can thus pick out the different anisotropic contributions to the viscosity by modifying the sample angle $\alpha$ with respect to the $C_2$ axes. The effect of the total anisotropic viscosities $\gamma_\mathrm{total}$ and $\Theta_\mathrm{total}$ can thus be experimentally probed by varying the orientation of the pipe relative to the sample. This could be used to probe both lattice anisotropy and nematicity, for instance in anisotropic quantum Hall states at $\nu=7/3$\cite{xia2011evidence}.

For lattice systems, the formalism of Sec.~\ref{sec:latticeformalism} allows us to explore momentum transport beyond the hydrodynamic regime as well. In particular, by quantifying the effect of Umklapp scattering on momentum relaxation in Eq.~(\ref{eq:latticeforce}), our formalism allows for a quantitative exploration of the crossover between diffusive and hydrodynamic transport. This can be accessed experimentally in systems with tunable disorder such as incommensurate TMD heterostructures. 

Finally, implicit in all pipe-flow predictions is a choice of boundary conditions at the edge of the pipe, such as our partial-slip boundary conditions above. Different boundary conditions on the tangential and normal stresses can be reinterpreted as bulk divergenceless contributions to the stress tensor, and hence correspond to different choices for the splitting between ``barred'' and ``unbarred'' viscosity coefficients. This has recently been explored (albeit not with this perspective) in the propagation of surface waves in isotropic chiral fluids\cite{sriramhydro,abanov2018odd,abanov2019free,abanov2019hydrodynamics}. Using our formalism, the redundancy in nondissipative (and dissipative) viscosities can be explored quantitatively through both wave propagation at the surface of classical chiral active fluids\cite{vitelli2017odd}, as well as through the boundary behavior of currents in pipe flow experiments and fluid flow around obstacles.

\subsection{Beyond hydrodynamic flow}\label{subsec:moments}

Following Ref.~\onlinecite{holder2019unified}, we can study the Hall viscosity in anisotropic systems in the semiclassical regime. Ref.~\onlinecite{holder2019unified} establishes a relationship between the experimentally measurable angular moments of the semiclassical distribution function and the Hall viscosity in the isotropic case, where those moments can be expanded in terms angular harmonics. An extension of this treatment to the anistropic case promises to yield a relationship between the now more complicated moments of the distribution function and our anisotropic Hall viscosities, effectively generalizing the relationship between transverse electric field response and Hall viscosity. Such relationships are valid even beyond the hydrodynamic regime, and are therefore of great experimental interest.  

\subsection{Torque magnetometry}\label{subsec:torque}
Another property of quantum fluids without time-reversal symmetry that we can probe is magnetization. In the presence of a non-uniform electric field, there is a viscous response that gives rise to a momentum-dependent contribution to the Hall conductivity\cite{bradlyn2012kubo}. The momentum-dependent current density that arises as a result will give rise to a manifestly viscous contribution to the bulk magnetization of the sample, one which we can directly couple to the ``total" non-dissipative Hall viscosities in Eq.~\eqref{eq:hallvisc2tensor}, and measure via torque magnetometry. 

In particular, the magnetization density 
\beq
\delta M = \frac{1}{2} \int d^2 r \left({\bf r \times j}\right)_z,
\eeq
receives contributions from the wavevector dependent conducitivity, and hence from the Hall viscosity. In a $C_2$ invariant system, one can analyze the different types of viscous magnetization responses, noting that the anisotropic viscosities will produce magnetizations that transform non-trivially under rotations, i.e. in-plane multipole moments. Torque magnetometry would be able to probe these higher-moment responses\cite{harris2000magnetization,mumford2019cantilever,PhysRevLett.79.3238}.

\subsection{Viscosity as a Probe of New Material Systems}\label{subsec:materials}

There are a host of materials that could provide an domain for realizing the predicitions in this work. Black phosphoprus is one example of an anisotropic material that supports hydrodynamic flow\cite{wang2015highly,li2016quantum}. Further, PdCoO$_2$ is another material where electron viscosity plays a role in transport and has a hexagonal Fermi surface\cite{moll2016evidence}. In both of these systems, the anisotropic Hall viscosity can be used, as a probe of the interplay between Berry curvature and intra-unit cell angular momentum, following our results in Section~\ref{sec:freefermion}. 

Beyond free fermion systems, our formalism has more general implications for topological physics in strongly correlated systems in a magnetic field, such as fractional quantum Hall states in twisted bilayer graphene and graphene-boron nitride moir\'{e} heterostructures\cite{spanton2018observation,zibrov2017tunable,serlin2019intrinsic}. At the fundamental level, our formalism allows for the identification of the Belinfante stress tensor in models of these systems both with and without a magnetic field. Going further, the rich emergent symmetry structure in fractional quantum Hall and fractional Chern insulator states in these materials could allow for the evaluation of the anisotropic Hall viscosity, similar to Ref.~\onlinecite{read2009non,read2011hall}. As graphene-based Chern insulators are both anisotropic and amenable to hydrodynamic measurements\cite{berdyugin2019measuring,bandurin2016negative}, our formalism could thus enable the experimental investigation of the interplay between geometry and topology in these systems. Finally, cold-atom systems with artificial gauge fields can be used to realize the Hofstadter model on a variety of lattices, both with and without interactions\cite{Ane1602685,RevModPhys.91.015005}. These systems are a promising platform to investigate the experimental implications of our lattice formulation of momentum transport.

\section{Outlook}

In this work, we have extended the Kubo formalism for viscosity to systems with rotational-symmetry breaking anisotropy, discrete translational symmetry, and internal rotational degrees of freedom. At the microscopic level, we have shown that the symmetrization procedure for the stress tensor given in Ref.~\onlinecite{link2018elastic} is equivalent to the generalized Belinfante procedure in nonrelativistic field theory, which involves restricting to strain perturbations which do not change the torsion of space. Furthermore, we have argued that this choice for the stress tensor applies to systems without rotational symmetry as well. We have shown that in systems with discrete translational symmetry, it is possible to define a course-grained momentum density and stress tensor, and hence compute viscoelastic response functions. We also argued that our coarse-grained lattice formalism reduces to the continuum formalism in the hydrodynamic limit.

More generally, we have also analyzed the symmetry of the viscosity tensor, highlighting the importance of the six independent Hall viscosity coefficients. We have shown that only three linear combinations of these coefficients enter into the hydrodynamic equations of motion in the bulk. The other three linear combinations can be changed by adding divergenceless terms to the viscous stress tensor, and hence they have no measurable effects in the absence of a physical principle (such as gravity or boundary conditions) fixing the divergenceless part of the stress tensor. We show that the three linear combinations that determine the forces can be assembled into a rank two ``Hall tensor'' which coincides with the tensor first highlighted by Haldane in Ref.~\onlinecite{haldane2011geometrical}. Finally, we applied our formalism to a variety of free-fermion systems, both in the continuum and on the lattice. We showed that free fermions in the continuum always have vanishing Hall elastic moduli, consistent with expectations for a fluid. We also showed that the six viscosity coefficients can be expressed in terms of quadrupole moments of the Berry curvature and matrix elements of the internal angular momentum operator. When focusing on the Hall tensor (i.e. the viscous forces), we found that the Berry curvature contribution to the viscous forces cancelled, and only the internal angular momentum contributed. We showed how this works in practice by analyzing an anisotropic Dirac fermion. We performed a similar analysis for free fermions on the lattice, and compared our results for the lattice regularized Dirac fermion (Chern insulator) with those of Ref.~\onlinecite{shapourian2015viscoelastic}. Crucially, we find within our formalism that the low-energy expansion of the Hall tensor in the Chern insulator coincides with the Hall tensor for four isolated Dirac fermions in the continuum, representing a consistency check on the formalism. Lastly we showed how our formalism applies to anisotropic superfluids, obtaining results consistent with the symmetry analysis of Ref.~\onlinecite{golan2019boundary}.

Going forward, our results raise several important questions for the Hall viscosity of time-reversal symmetry broken anisotropic systems. First, our emphasis on viscous forces and the Hall tensor has direct relevance for experiments geared to extract the viscosity from hydrodynamic flow. As we argued, in a threefold or higher symmetric system, an analysis of flow can only ever extract the combination $\eta^\mathrm{H}+\bar{\eta}^\mathrm{H}=\eta^\mathrm{H}_\mathrm{total}$ that enters the Hall tensor. This fact has been overlooked in the current theoretical analysis of flow in graphene and other materials, which assume rotational symmetry and fixed boundary conditions. We sketched the experimental implications of this in Subsection~\ref{subsec:pipe}. Additionally, we have shown in Subsection~\ref{subsec:torque} that our results have implications for torque magnetometry experiments on anisotropic quantum Hall systems. Furthermore, our formalism can be used to explain current and future experiments on viscous transport in anisotropic lattice systems such as twisted bilayer graphene, black phosphorus, and graphene-boron nitride based fractional Chern insulators.

 For free fermion systems, we have seen that working with only the kinetic stress leads to $\eta^\mathrm{H}_\mathrm{total}$. This implies that for free fermions the ``torsional viscosity'' computed with the spin connection set to zero (i.e.~using only the kinetic stress) does not enter into the equations of motion for momentum transport. Furthermore, our lattice formalism suggests that the derivation of the stress tensor in low-energy theories of Bloch electrons should be revisited. The momentum dependence of the lattice strain generators Eq.~(\ref{eq:latticestrain}) suggests that the displacement of Fermi surface pockets away from zero momentum (as in graphene) may have a significant effect on viscous response functions. Additionally, the lattice formulation of momentum transport will allow for the exploration of the crossover between ballistic and diffusive transport in lattice models of quantum systems. Another interesting direction for future work is to extend our formalism to three dimensional systems such as topological semimetals\cite{landsteiner-weyl-visc,vozmediano-rotational-strain,copetti2019anomalous}.
Furthermore, the redundancy Eq.~(\ref{eq:divfreecontact}) in the dissipative and nondissipative viscous force densities could have interesting implications for hydrodynamic waves and entropy production in anisotropic fluids.
 Finally, there appears to be some connection to explore between our formulation of lattice momentum transport, and the K\"ahler-Dirac formalism of lattice gauge theory\cite{becher1982dirac}.

\begin{acknowledgments}
BB would like to thank F.~D.~M.~Haldane and N.~Read for conversations which inspired this work. The authors also thank J. Schmalian for pointing out the crucial Ref.~\onlinecite{link2018elastic}. Additionally, the authors acknowledge fruitful discussions with A.~Abanov, E.~Fradkin, T.~L.~Hughes, D.~Huse, and M. Stone. This material is based upon work supported by the National Science Foundation Graduate Research Fellowship Program under Grant No. DGE – 1746047.  
\end{acknowledgments}

\appendix
\section{Alternative form of the Lattice Spin Stress}\label{app:othersymmetrization}
Re-examining Eq.~(\ref{eq:latticestrain}) for the strain generators on the lattice, we note that another convenient choice of spin-strain would be
\begin{widetext}
\begin{equation}
\mathcal{J}^\mu_{\mathrm{L},\nu}\equiv-\frac{i}{2}\sum_{\mathbf{k}nm\nu}c^\dag_{n\mathbf{k}}\left[\left\{\frac{\sin\mathbf{k}\cdot\mathbf{a}_\nu}{|\mathbf{a}_\nu|},\frac{\partial}{\partial k_\mu}\right\}\delta^{nm}+i\tilde{\epsilon}_\nu^{\hphantom{\nu}\mu}\cos\mathbf{k}\cdot\mathbf{a}_\nu L_\mathrm{int}^{nm}\right]c_{m\mathbf{k}},\label{eq:latticealternativestrain}
\end{equation}
which differs from Eq.~(\ref{eq:latticestrain}) in the index of the cosine term in the spin stress. The convenience of this term becomes manifest when we compute the free fermion stress tensor
\begin{equation}
     T^\mu_{\mathrm{B}0,\nu}=\sum_{nm\mathbf{k}}c^\dag_{n\mathbf{k}}\left(\partial^\mu f^{nm}(\mathbf{k})\frac{\sin\mathbf{k}\cdot \mathbf{a}_\nu}{|\mathbf{a}_\nu|}+\frac{i}{2}\tilde{\epsilon}^\mu_{\hphantom{\mu}\nu}\cos\mathbf{k}\cdot\mathbf{a}_\nu[f(\mathbf{k}),L_\mathrm{int}]_{nm}\right)c_{m\mathbf{k}}=\sum_{nm\mathbf{k}}c^\dag_{n\mathbf{k}}[T^{\, \lambda}_{\mathrm{B} \rho}({\bf k})]^{nm}c_{m\mathbf{k}}
\end{equation}
which follows from this definition. We now see that the index of $\mathbf{k}$ in the cosine prefactor of the spin stress matches the second, lower index on the stress tensor. If we were to expand this new lattice stress about the time-reversal invariant momenta in the Brillouin zone of a crystal, we would find that the overall sign of the spin-stress obtained from Taylor expanding the cosine term would match the overall sign of the kintetic stress obtained from expanding the sine in the first term. While this gives the same result as our Eq.~(\ref{eq:fflatticestress}) at the $(0,0)$ and $(\pi,\pi)$ points, it differs at the $(0,\pi)$ and $(\pi,0)$ points. 

At first glance, the extra semblance of rotational symmetry gained by this change of index may seem appealing. However, we can ask what form the internal momentum density must take to produce this term in the strain generator. It can be shown that Eq.~(\ref{eq:latticealternativestrain}) implies
\begin{equation}
    g_\mu^\mathrm{int}(\mathbf{R})=\frac{1}{4|\mathbf{a}_\nu|}\tilde{\epsilon}_\mu^{\hphantom{\mu}\nu}\sum_{nm}L_\mathrm{int}^{nm}\left(
c^\dag_{n\mathbf{R+a_\nu+a_\mu}}c_{m\mathbf{R}+a_\mu}
-c^\dag_{n\mathbf{R}}c_{m\mathbf{R+a_\mu}}
+c^\dag_{n\mathbf{R+a_\nu}}c_{m\mathbf{R+a_\nu+a_\mu}}
-c^\dag_{n\mathbf{R+a_\mu}}c_{m\mathbf{R}}\right)\label{eq:latticespinmommodified}
\end{equation}
This discretization of the lattice derivative has the unfortunate features of not being symmetric with respect to the base point of differentiation, and of not including all nearest-neightbor lattice sites to which momentum can flow. For these reasons, we know of no justification for using Eq.~(\ref{eq:latticespinmommodified}) in place of Eq.~(\ref{eq:latticeintmom}).

Nevertheless, we can explore the consequences of Eq.~(\ref{eq:latticespinmommodified}) on the viscosity of the Chern insulator model considered in Sec.~(\ref{sec:chern}). We find for the modified stress tensor

\beq
\bal
T^{\, \lambda}_{\mathrm{B} \rho}({\bf k}) &= t^\lambda \cos(k_\lambda a) \sin(k_\rho a) \sigma^\lambda + r \sin(k_\lambda a) \sin(k_\rho a)\, \sigma^z + \frac{1}{2} \cos(k_\rho a) \left(t^\lambda\sin(k_\lambda a) \sigma^\rho   - t^\rho\sin(k_\rho a) \sigma^\lambda \right)
\eal
\eeq
\end{widetext}
The spin stress tensor here differs from Eq.~(\ref{eq:latticestressci}) by a factor of $\cos k_\rho a / \cos k_\lambda a$.

\textit{$C_4$ case}: For a $C_4$-symmetric system with hoppings $t=t'$, this leads to a total viscosity that is now simple enough to present 
\begin{align}
\eta^\mathrm{H}_{\mathrm{total}}({\bf k}) &= \frac{t^2}{{32 \epsilon_{\bf k}^3}} \biggr(r \cos(k_y a) + \cos(k_x a)\\ \nonumber&\left(r - m \cos(k_y a) \right)\left(\cos(2 k_x a) + \cos(2 k_y a) - 2 \right)\biggr)
\end{align}
\begin{figure}[t]
    \centering
    \includegraphics[width=0.45\textwidth]{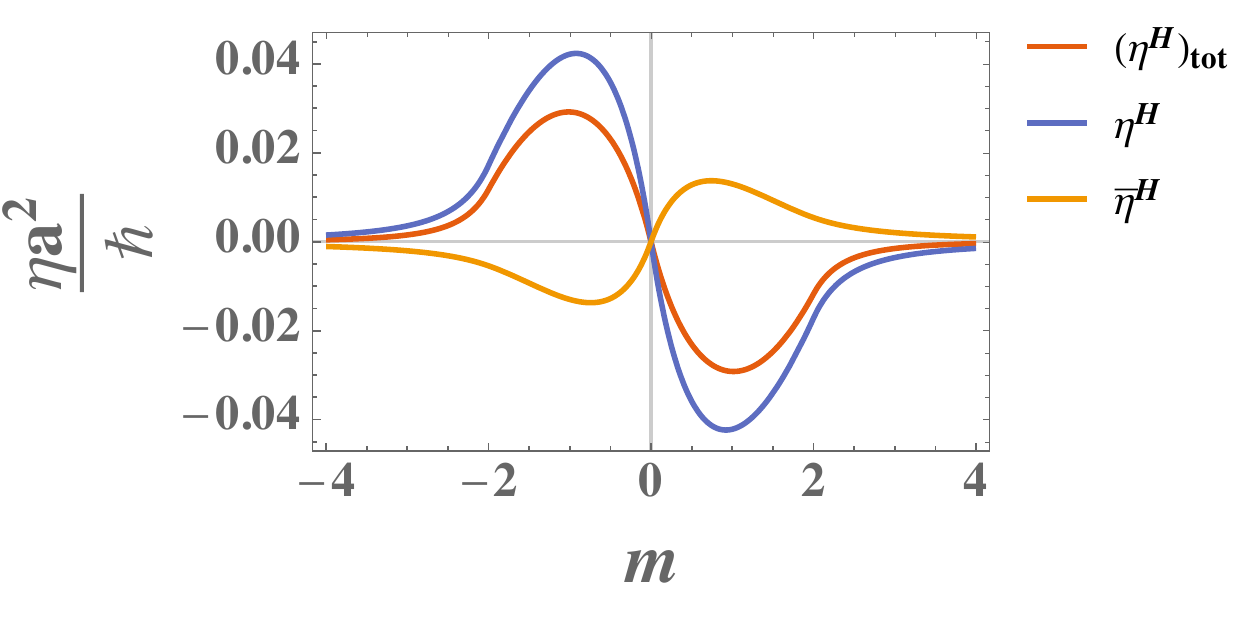}
    \caption{Hall viscosities in $C_4$, $\eta^\mathrm{H}_{\, \mathrm{total}}, \eta^\mathrm{H}, \bar{\eta}^H$, plotted with the alternative form of the strain generator, with parameters $t=t'= r= 1$ and lattice constant $a=1$. We see much less variation in the plots in this case, while the same features of oddness in $m$ and vanishing in the topologically trivial regimes remains.} 
    \label{fig:etatotc42}
\end{figure}
While the expansion of the Hamiltonian around each of the Dirac points is the same as in Eq.~(\ref{eq:expandeddiracham}), the excpanded stress tensor now becomes
The expansion of this tensor around the Dirac points can be written as
\beq
\bal
(T^{\lambda}_{\mathrm{B}, \rho}(\mathbf{k}))^{(i)} &= v_F \left[ \alpha_\lambda^i \alpha_\rho^i  \sigma^\lambda k_\rho + \frac{1}{2} \alpha_\rho^i \left(\alpha_\lambda^i k_\lambda \sigma^\rho - \alpha_\rho^i k_\rho \sigma^\lambda \right) \right]
\\ &=  \frac{v_F}{2} \left[ (2\alpha^i_\lambda - \alpha^i_\rho) \alpha_\rho^i \sigma^\lambda k_\rho + \alpha_\rho^i \alpha_\lambda^i k_\lambda \sigma^\rho\right]
\eal
\eeq

\noindent The resulting viscosity can be computed and takes a similar Dirac-point-dependent form as Eq.~\eqref{eq:viscgeneraldp}. The sum over Dirac points now yields the same expression by magnitude, but this now changes sign across the Dirac points:
\beq
\bal
(\eta^{H}_{\, \mathrm{total}}({\bf k}))^{(i)} &=  \left(\eta^\mathrm{H}({\bf k}) + \bar{\eta}^H({\bf k})\right)^{(i)} \\&= \alpha^i_x \alpha^i_y \frac{v_F^2 M^i}{8 \epsilon_{\bf k}^3} \mathbf{|k|}^2
\eal
\eeq
Which has the same form as the isotropic Hall viscosity in Ref.~\onlinecite{shapourian2015viscoelastic}. Integrating this up to a momentum cuttoff $\Lambda$ and summing over the Dirac points yields the now cutoff-independent viscosity

\beq
\bal
\label{eq:etatotalfinalsumalt}
\eta^{H}_{\, \mathrm{total}} &= -\frac{1}{2}\sum\limits_i \frac{\left(\alpha_x^i  \alpha_y^i \right)}{{4\pi v_F^2}} \left[M^i|M^i|\right]\\
&= \frac{1}{2}
\begin{cases}
\frac{\hbar}{2\pi v_F^2}(-m|m| + 4mr) & |m| < 2r\\
\frac{2 \hbar}{\pi v_F^2} r^2 & |m| \geq 2r
\end{cases}
\eal
\eeq
We find that the cutoff-independent total viscosity obtained with this alternative Belinfante procedure coincides with the cutoff-independent kinetic viscosity computed in Ref.~\onlinecite{shapourian2015viscoelastic}. 

Again separating into the individual barred and unbarred Hall viscosities separately, we see that the end points $(0,0)$ and $(\pi,\pi)$ are again described fully isotropic theories, meaning the unbarred Hall viscosity is the only non-zero contribution $(\eta^\mathrm{H}_{\mathrm{total}})^{(0,3)} = \eta^\mathrm{H}$. At the corner points, however, $(\pi,0)$ and $(0,\pi)$:

\beq
\bal
\eta^\mathrm{H}({\bf k})  =& \frac{-v_F^2 m k^2}{4 \epsilon^3_{\bf k}}\\
\bar{\eta}^H =& \frac{v_F^2 m k^2}{8 \epsilon^3_{\bf k}}
\eal
\eeq

\begin{figure}[ht]
    \includegraphics[width=0.4\textwidth]{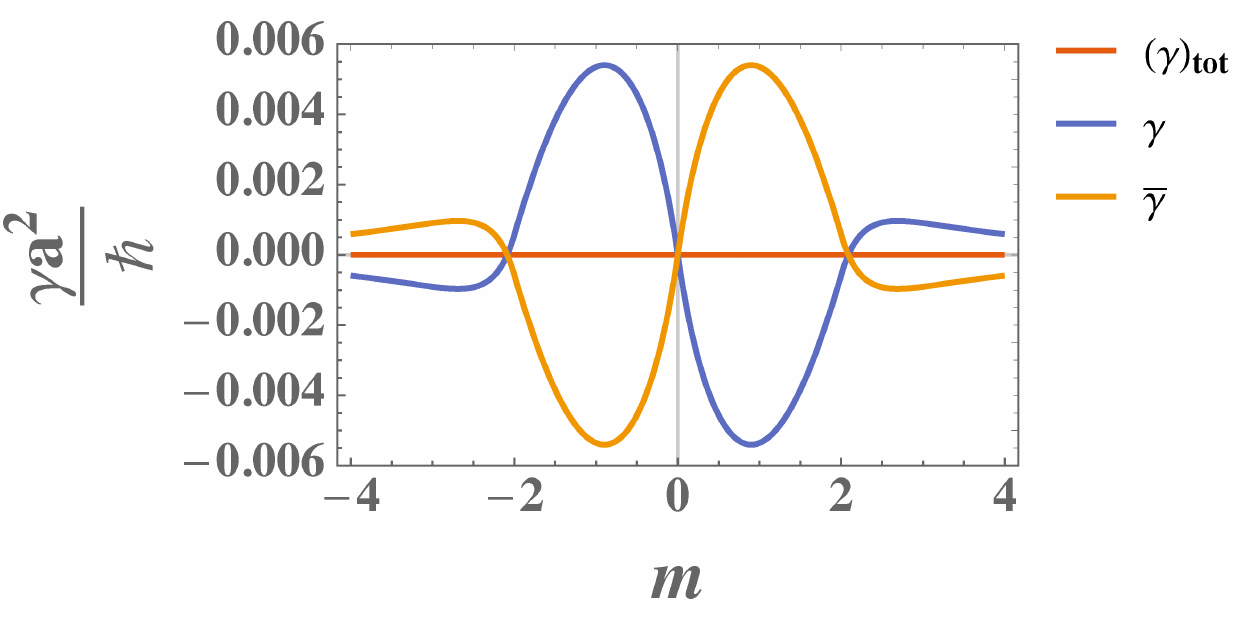}
    \caption{Anisotropic viscosities plotted with the alternative form of the strain generator. We use the same set of parameters $t=1.5$, $t=r=1$, $a=1$ as earlier in the work. Curiously, we see that the total anisotropic viscosity is zero.} 
    \label{fig:altgammatot}
\end{figure}

\textit{$C_2$ case:} We see a similar picture in the general case. The anisotropic viscosities $\gamma_{\, \mathrm{total}}$ are plotted in Fig.~\ref{fig:altgammatot}. We note that for this alternative choice of stress tensor, the total anisotropic viscosity is identically zero.

The continuum expansion is also cutoff-independent with this choice of strain generator, as we see
\beq
\bal
(\eta^\mathrm{H}_{\, \mathrm{total}})^{(i)} &= -\frac{1}{2}\sum\limits_i \frac{(\alpha_x^i \alpha_y^i)}{4\pi\ev{v_F}^2_g} \left[M^i|M^i|\right]\\
&= \frac{1}{2}\begin{cases}
\frac{\hbar}{2\pi \ev{v_F}_g^2}(-m|m| + 4mr) & |m| < 2r\\
\frac{2 \hbar}{\pi \ev{v_F}_g^2} r^2 & |m| \geq 2r
\end{cases}
\eal
\eeq

The anisotropic viscosities are independent of this choice of strain generator (independent of the $\cos(k_\lambda a)$ vs $\cos(k_\rho a)$ prefactor in $(T^{\,\mathrm{spin}})^{\lambda}_{\hphantom{\lambda} \rho}$ ) -- they are zero. 
\vspace{1ex}
\bibliography{refs}

\begin{thebibliography}{82}%
\makeatletter
\providecommand \@ifxundefined [1]{%
 \@ifx{#1\undefined}
}%
\providecommand \@ifnum [1]{%
 \ifnum #1\expandafter \@firstoftwo
 \else \expandafter \@secondoftwo
 \fi
}%
\providecommand \@ifx [1]{%
 \ifx #1\expandafter \@firstoftwo
 \else \expandafter \@secondoftwo
 \fi
}%
\providecommand \natexlab [1]{#1}%
\providecommand \enquote  [1]{``#1''}%
\providecommand \bibnamefont  [1]{#1}%
\providecommand \bibfnamefont [1]{#1}%
\providecommand \citenamefont [1]{#1}%
\providecommand \href@noop [0]{\@secondoftwo}%
\providecommand \href [0]{\begingroup \@sanitize@url \@href}%
\providecommand \@href[1]{\@@startlink{#1}\@@href}%
\providecommand \@@href[1]{\endgroup#1\@@endlink}%
\providecommand \@sanitize@url [0]{\catcode `\\12\catcode `\$12\catcode
  `\&12\catcode `\#12\catcode `\^12\catcode `\_12\catcode `\%12\relax}%
\providecommand \@@startlink[1]{}%
\providecommand \@@endlink[0]{}%
\providecommand \url  [0]{\begingroup\@sanitize@url \@url }%
\providecommand \@url [1]{\endgroup\@href {#1}{\urlprefix }}%
\providecommand \urlprefix  [0]{URL }%
\providecommand \Eprint [0]{\href }%
\providecommand \doibase [0]{http://dx.doi.org/}%
\providecommand \selectlanguage [0]{\@gobble}%
\providecommand \bibinfo  [0]{\@secondoftwo}%
\providecommand \bibfield  [0]{\@secondoftwo}%
\providecommand \translation [1]{[#1]}%
\providecommand \BibitemOpen [0]{}%
\providecommand \bibitemStop [0]{}%
\providecommand \bibitemNoStop [0]{.\EOS\space}%
\providecommand \EOS [0]{\spacefactor3000\relax}%
\providecommand \BibitemShut  [1]{\csname bibitem#1\endcsname}%
\let\auto@bib@innerbib\@empty
\bibitem [{\citenamefont {Klitzing}\ \emph {et~al.}(1980)\citenamefont
  {Klitzing}, \citenamefont {Dorda},\ and\ \citenamefont
  {Pepper}}]{klitzing1980new}%
  \BibitemOpen
  \bibfield  {author} {\bibinfo {author} {\bibfnamefont {K~v}\ \bibnamefont
  {Klitzing}}, \bibinfo {author} {\bibfnamefont {Gerhard}\ \bibnamefont
  {Dorda}}, \ and\ \bibinfo {author} {\bibfnamefont {Michael}\ \bibnamefont
  {Pepper}},\ }\bibfield  {title} {\enquote {\bibinfo {title} {New method for
  high-accuracy determination of the fine-structure constant based on quantized
  hall resistance},}\ }\href@noop {} {\bibfield  {journal} {\bibinfo  {journal}
  {Physical Review Letters}\ }\textbf {\bibinfo {volume} {45}},\ \bibinfo
  {pages} {494} (\bibinfo {year} {1980})}\BibitemShut {NoStop}%
\bibitem [{\citenamefont {Laughlin}(1981)}]{laughlin1981quantized}%
  \BibitemOpen
  \bibfield  {author} {\bibinfo {author} {\bibfnamefont {Robert~B}\
  \bibnamefont {Laughlin}},\ }\bibfield  {title} {\enquote {\bibinfo {title}
  {Quantized hall conductivity in two dimensions},}\ }\href@noop {} {\bibfield
  {journal} {\bibinfo  {journal} {Physical Review B}\ }\textbf {\bibinfo
  {volume} {23}},\ \bibinfo {pages} {5632} (\bibinfo {year}
  {1981})}\BibitemShut {NoStop}%
\bibitem [{\citenamefont {Thouless}\ \emph {et~al.}(1982)\citenamefont
  {Thouless}, \citenamefont {Kohmoto}, \citenamefont {Nightingale},\ and\
  \citenamefont {den Nijs}}]{thouless1982quantized}%
  \BibitemOpen
  \bibfield  {author} {\bibinfo {author} {\bibfnamefont {David~J}\ \bibnamefont
  {Thouless}}, \bibinfo {author} {\bibfnamefont {Mahito}\ \bibnamefont
  {Kohmoto}}, \bibinfo {author} {\bibfnamefont {M~Peter}\ \bibnamefont
  {Nightingale}}, \ and\ \bibinfo {author} {\bibfnamefont {Md}~\bibnamefont
  {den Nijs}},\ }\bibfield  {title} {\enquote {\bibinfo {title} {Quantized hall
  conductance in a two-dimensional periodic potential},}\ }\href@noop {}
  {\bibfield  {journal} {\bibinfo  {journal} {Physical review letters}\
  }\textbf {\bibinfo {volume} {49}},\ \bibinfo {pages} {405} (\bibinfo {year}
  {1982})}\BibitemShut {NoStop}%
\bibitem [{\citenamefont {Avron}\ \emph {et~al.}(1995)\citenamefont {Avron},
  \citenamefont {Seiler},\ and\ \citenamefont
  {Zograf}}]{1995-AvronSeilerZograf}%
  \BibitemOpen
  \bibfield  {author} {\bibinfo {author} {\bibfnamefont {J~E}\ \bibnamefont
  {Avron}}, \bibinfo {author} {\bibfnamefont {R}~\bibnamefont {Seiler}}, \ and\
  \bibinfo {author} {\bibfnamefont {P~G}\ \bibnamefont {Zograf}},\ }\bibfield
  {title} {\enquote {\bibinfo {title} {Viscosity of quantum hall fluids},}\
  }\href
  {http://www.google.com/search?client=safari&rls=en-us&q=VISCOSITY+OF+QUANTUM+HALL+FLUIDS&ie=UTF-8&oe=UTF-8}
  {\bibfield  {journal} {\bibinfo  {journal} {Phys Rev Lett}\ }\textbf
  {\bibinfo {volume} {75}},\ \bibinfo {pages} {697--700} (\bibinfo {year}
  {1995})}\BibitemShut {NoStop}%
\bibitem [{\citenamefont {Tokatly}\ and\ \citenamefont
  {Vignale}(2007)}]{2007-TokatlyVignale}%
  \BibitemOpen
  \bibfield  {author} {\bibinfo {author} {\bibfnamefont {I.~V}\ \bibnamefont
  {Tokatly}}\ and\ \bibinfo {author} {\bibfnamefont {G}~\bibnamefont
  {Vignale}},\ }\bibfield  {title} {\enquote {\bibinfo {title} {Lorentz shear
  modulus of a two-dimensional electron gas at high magnetic field},}\ }\href
  {\doibase 10.1103/PhysRevB.76.161305} {\bibfield  {journal} {\bibinfo
  {journal} {Phys Rev B}\ }\textbf {\bibinfo {volume} {76}},\ \bibinfo {pages}
  {161305} (\bibinfo {year} {2007})}\BibitemShut {NoStop}%
\bibitem [{\citenamefont {Read}(2009)}]{read2009non}%
  \BibitemOpen
  \bibfield  {author} {\bibinfo {author} {\bibfnamefont {N}~\bibnamefont
  {Read}},\ }\bibfield  {title} {\enquote {\bibinfo {title} {Non-abelian
  adiabatic statistics and hall viscosity in quantum hall states and p x+ i p y
  paired superfluids},}\ }\href@noop {} {\bibfield  {journal} {\bibinfo
  {journal} {Physical Review B}\ }\textbf {\bibinfo {volume} {79}},\ \bibinfo
  {pages} {045308} (\bibinfo {year} {2009})}\BibitemShut {NoStop}%
\bibitem [{\citenamefont {Read}\ and\ \citenamefont
  {Rezayi}(2011)}]{read2011hall}%
  \BibitemOpen
  \bibfield  {author} {\bibinfo {author} {\bibfnamefont {N}~\bibnamefont
  {Read}}\ and\ \bibinfo {author} {\bibfnamefont {EH}~\bibnamefont {Rezayi}},\
  }\bibfield  {title} {\enquote {\bibinfo {title} {Hall viscosity, orbital
  spin, and geometry: paired superfluids and quantum hall systems},}\
  }\href@noop {} {\bibfield  {journal} {\bibinfo  {journal} {Physical Review
  B}\ }\textbf {\bibinfo {volume} {84}},\ \bibinfo {pages} {085316} (\bibinfo
  {year} {2011})}\BibitemShut {NoStop}%
\bibitem [{\citenamefont {Haldane}(1983)}]{Haldane1983}%
  \BibitemOpen
  \bibfield  {author} {\bibinfo {author} {\bibfnamefont {F.~D.~M.}\
  \bibnamefont {Haldane}},\ }\bibfield  {title} {\enquote {\bibinfo {title}
  {Fractional quantization of the hall effect: a hierarchy of incompressible
  quantum fluid states},}\ }\href {\doibase 10.1103/PhysRevLett.51.605}
  {\bibfield  {journal} {\bibinfo  {journal} {Phys. Rev. Lett.}\ }\textbf
  {\bibinfo {volume} {51}},\ \bibinfo {pages} {605} (\bibinfo {year}
  {1983})}\BibitemShut {NoStop}%
\bibitem [{\citenamefont {Zaletel}\ \emph {et~al.}(2013)\citenamefont
  {Zaletel}, \citenamefont {Mong},\ and\ \citenamefont
  {Pollmann}}]{Zaletel-PhysRevLett.110.236801}%
  \BibitemOpen
  \bibfield  {author} {\bibinfo {author} {\bibfnamefont {Michael~P.}\
  \bibnamefont {Zaletel}}, \bibinfo {author} {\bibfnamefont {Roger S.~K.}\
  \bibnamefont {Mong}}, \ and\ \bibinfo {author} {\bibfnamefont {Frank}\
  \bibnamefont {Pollmann}},\ }\bibfield  {title} {\enquote {\bibinfo {title}
  {Topological characterization of fractional quantum hall ground states from
  microscopic hamiltonians},}\ }\href {\doibase 10.1103/PhysRevLett.110.236801}
  {\bibfield  {journal} {\bibinfo  {journal} {Phys. Rev. Lett.}\ }\textbf
  {\bibinfo {volume} {110}},\ \bibinfo {pages} {236801} (\bibinfo {year}
  {2013})}\BibitemShut {NoStop}%
\bibitem [{\citenamefont {Haldane}\ and\ \citenamefont
  {Shen}(2015)}]{haldane2015geometry}%
  \BibitemOpen
  \bibfield  {author} {\bibinfo {author} {\bibfnamefont {F.~D.~M.}\
  \bibnamefont {Haldane}}\ and\ \bibinfo {author} {\bibfnamefont
  {Yu}~\bibnamefont {Shen}},\ }\bibfield  {title} {\enquote {\bibinfo {title}
  {Geometry of landau orbits in the absence of rotational symmetry},}\
  }\href@noop {} {\bibfield  {journal} {\bibinfo  {journal} {arXiv preprint
  arXiv:1512.04502}\ } (\bibinfo {year} {2015})}\BibitemShut {NoStop}%
\bibitem [{\citenamefont {Gromov}\ \emph {et~al.}(2017)\citenamefont {Gromov},
  \citenamefont {Geraedts},\ and\ \citenamefont
  {Bradlyn}}]{gromov2017investigating}%
  \BibitemOpen
  \bibfield  {author} {\bibinfo {author} {\bibfnamefont {Andrey}\ \bibnamefont
  {Gromov}}, \bibinfo {author} {\bibfnamefont {Scott~D}\ \bibnamefont
  {Geraedts}}, \ and\ \bibinfo {author} {\bibfnamefont {Barry}\ \bibnamefont
  {Bradlyn}},\ }\bibfield  {title} {\enquote {\bibinfo {title} {Investigating
  anisotropic quantum hall states with bimetric geometry},}\ }\href@noop {}
  {\bibfield  {journal} {\bibinfo  {journal} {Physical review letters}\
  }\textbf {\bibinfo {volume} {119}},\ \bibinfo {pages} {146602} (\bibinfo
  {year} {2017})}\BibitemShut {NoStop}%
\bibitem [{\citenamefont {Offertaler}\ and\ \citenamefont
  {Bradlyn}(2019)}]{offertaler2019viscoelastic}%
  \BibitemOpen
  \bibfield  {author} {\bibinfo {author} {\bibfnamefont {Bendeguz}\
  \bibnamefont {Offertaler}}\ and\ \bibinfo {author} {\bibfnamefont {Barry}\
  \bibnamefont {Bradlyn}},\ }\bibfield  {title} {\enquote {\bibinfo {title}
  {Viscoelastic response of quantum hall fluids in a tilted field},}\
  }\href@noop {} {\bibfield  {journal} {\bibinfo  {journal} {Physical Review
  B}\ }\textbf {\bibinfo {volume} {99}},\ \bibinfo {pages} {035427} (\bibinfo
  {year} {2019})}\BibitemShut {NoStop}%
\bibitem [{\citenamefont {Souslov}\ \emph {et~al.}(2019)\citenamefont
  {Souslov}, \citenamefont {Gromov},\ and\ \citenamefont
  {Vitelli}}]{souslov2019anisotropic}%
  \BibitemOpen
  \bibfield  {author} {\bibinfo {author} {\bibfnamefont {Anton}\ \bibnamefont
  {Souslov}}, \bibinfo {author} {\bibfnamefont {Andrey}\ \bibnamefont
  {Gromov}}, \ and\ \bibinfo {author} {\bibfnamefont {Vincenzo}\ \bibnamefont
  {Vitelli}},\ }\bibfield  {title} {\enquote {\bibinfo {title} {Anisotropic odd
  viscosity via time-modulated drive},}\ }\href@noop {} {\bibfield  {journal}
  {\bibinfo  {journal} {arXiv preprint arXiv:1909.08505}\ } (\bibinfo {year}
  {2019})}\BibitemShut {NoStop}%
\bibitem [{\citenamefont {Karplus}\ and\ \citenamefont
  {Luttinger}(1954)}]{karplus1954hall}%
  \BibitemOpen
  \bibfield  {author} {\bibinfo {author} {\bibfnamefont {Robert}\ \bibnamefont
  {Karplus}}\ and\ \bibinfo {author} {\bibfnamefont {JM}~\bibnamefont
  {Luttinger}},\ }\bibfield  {title} {\enquote {\bibinfo {title} {Hall effect
  in ferromagnetics},}\ }\href@noop {} {\bibfield  {journal} {\bibinfo
  {journal} {Physical Review}\ }\textbf {\bibinfo {volume} {95}},\ \bibinfo
  {pages} {1154} (\bibinfo {year} {1954})}\BibitemShut {NoStop}%
\bibitem [{\citenamefont {Haldane}(2004)}]{haldaneahe}%
  \BibitemOpen
  \bibfield  {author} {\bibinfo {author} {\bibfnamefont {F.~D.~M.}\
  \bibnamefont {Haldane}},\ }\bibfield  {title} {\enquote {\bibinfo {title}
  {Berry curvature on the fermi surface: Anomalous hall effect as a topological
  fermi-liquid property},}\ }\href {\doibase 10.1103/PhysRevLett.93.206602}
  {\bibfield  {journal} {\bibinfo  {journal} {Phys. Rev. Lett.}\ }\textbf
  {\bibinfo {volume} {93}},\ \bibinfo {pages} {206602} (\bibinfo {year}
  {2004})}\BibitemShut {NoStop}%
\bibitem [{\citenamefont {Lucas}\ and\ \citenamefont
  {Fong}(2018)}]{lucas2018hydrodynamics}%
  \BibitemOpen
  \bibfield  {author} {\bibinfo {author} {\bibfnamefont {Andrew}\ \bibnamefont
  {Lucas}}\ and\ \bibinfo {author} {\bibfnamefont {Kin~Chung}\ \bibnamefont
  {Fong}},\ }\bibfield  {title} {\enquote {\bibinfo {title} {Hydrodynamics of
  electrons in graphene},}\ }\href@noop {} {\bibfield  {journal} {\bibinfo
  {journal} {Journal of Physics: Condensed Matter}\ }\textbf {\bibinfo {volume}
  {30}},\ \bibinfo {pages} {053001} (\bibinfo {year} {2018})}\BibitemShut
  {NoStop}%
\bibitem [{\citenamefont {Scaffidi}\ \emph {et~al.}(2017)\citenamefont
  {Scaffidi}, \citenamefont {Nandi}, \citenamefont {Schmidt}, \citenamefont
  {Mackenzie},\ and\ \citenamefont {Moore}}]{scaffidi2017hydrodynamic}%
  \BibitemOpen
  \bibfield  {author} {\bibinfo {author} {\bibfnamefont {Thomas}\ \bibnamefont
  {Scaffidi}}, \bibinfo {author} {\bibfnamefont {Nabhanila}\ \bibnamefont
  {Nandi}}, \bibinfo {author} {\bibfnamefont {Burkhard}\ \bibnamefont
  {Schmidt}}, \bibinfo {author} {\bibfnamefont {Andrew~P}\ \bibnamefont
  {Mackenzie}}, \ and\ \bibinfo {author} {\bibfnamefont {Joel~E}\ \bibnamefont
  {Moore}},\ }\bibfield  {title} {\enquote {\bibinfo {title} {Hydrodynamic
  electron flow and hall viscosity},}\ }\href@noop {} {\bibfield  {journal}
  {\bibinfo  {journal} {Physical review letters}\ }\textbf {\bibinfo {volume}
  {118}},\ \bibinfo {pages} {226601} (\bibinfo {year} {2017})}\BibitemShut
  {NoStop}%
\bibitem [{\citenamefont {Alekseev}(2016)}]{alekseev2016negative}%
  \BibitemOpen
  \bibfield  {author} {\bibinfo {author} {\bibfnamefont {PS}~\bibnamefont
  {Alekseev}},\ }\bibfield  {title} {\enquote {\bibinfo {title} {Negative
  magnetoresistance in viscous flow of two-dimensional electrons},}\
  }\href@noop {} {\bibfield  {journal} {\bibinfo  {journal} {Physical review
  letters}\ }\textbf {\bibinfo {volume} {117}},\ \bibinfo {pages} {166601}
  (\bibinfo {year} {2016})}\BibitemShut {NoStop}%
\bibitem [{\citenamefont {Delacr{\'e}taz}\ and\ \citenamefont
  {Gromov}(2017)}]{delacretaz2017transport}%
  \BibitemOpen
  \bibfield  {author} {\bibinfo {author} {\bibfnamefont {Luca~V}\ \bibnamefont
  {Delacr{\'e}taz}}\ and\ \bibinfo {author} {\bibfnamefont {Andrey}\
  \bibnamefont {Gromov}},\ }\bibfield  {title} {\enquote {\bibinfo {title}
  {Transport signatures of the hall viscosity},}\ }\href@noop {} {\bibfield
  {journal} {\bibinfo  {journal} {Physical review letters}\ }\textbf {\bibinfo
  {volume} {119}},\ \bibinfo {pages} {226602} (\bibinfo {year}
  {2017})}\BibitemShut {NoStop}%
\bibitem [{\citenamefont {Pellegrino}\ \emph {et~al.}(2017)\citenamefont
  {Pellegrino}, \citenamefont {Torre},\ and\ \citenamefont
  {Polini}}]{pellegrino2017nonlocal}%
  \BibitemOpen
  \bibfield  {author} {\bibinfo {author} {\bibfnamefont {Francesco~MD}\
  \bibnamefont {Pellegrino}}, \bibinfo {author} {\bibfnamefont {Iacopo}\
  \bibnamefont {Torre}}, \ and\ \bibinfo {author} {\bibfnamefont {Marco}\
  \bibnamefont {Polini}},\ }\bibfield  {title} {\enquote {\bibinfo {title}
  {Nonlocal transport and the hall viscosity of two-dimensional hydrodynamic
  electron liquids},}\ }\href@noop {} {\bibfield  {journal} {\bibinfo
  {journal} {Physical Review B}\ }\textbf {\bibinfo {volume} {96}},\ \bibinfo
  {pages} {195401} (\bibinfo {year} {2017})}\BibitemShut {NoStop}%
\bibitem [{\citenamefont {Holder}\ \emph {et~al.}(2019)\citenamefont {Holder},
  \citenamefont {Queiroz},\ and\ \citenamefont {Stern}}]{holder2019unified}%
  \BibitemOpen
  \bibfield  {author} {\bibinfo {author} {\bibfnamefont {Tobias}\ \bibnamefont
  {Holder}}, \bibinfo {author} {\bibfnamefont {Raquel}\ \bibnamefont
  {Queiroz}}, \ and\ \bibinfo {author} {\bibfnamefont {Ady}\ \bibnamefont
  {Stern}},\ }\bibfield  {title} {\enquote {\bibinfo {title} {Unified
  description of the classical hall viscosity},}\ }\href {\doibase
  10.1103/PhysRevLett.123.106801} {\bibfield  {journal} {\bibinfo  {journal}
  {Phys. Rev. Lett.}\ }\textbf {\bibinfo {volume} {123}},\ \bibinfo {pages}
  {106801} (\bibinfo {year} {2019})}\BibitemShut {NoStop}%
\bibitem [{\citenamefont {Berdyugin}\ \emph {et~al.}(2019)\citenamefont
  {Berdyugin}, \citenamefont {Xu}, \citenamefont {Pellegrino}, \citenamefont
  {Kumar}, \citenamefont {Principi}, \citenamefont {Torre}, \citenamefont
  {Shalom}, \citenamefont {Taniguchi}, \citenamefont {Watanabe}, \citenamefont
  {Grigorieva} \emph {et~al.}}]{berdyugin2019measuring}%
  \BibitemOpen
  \bibfield  {author} {\bibinfo {author} {\bibfnamefont {Alexey~I}\
  \bibnamefont {Berdyugin}}, \bibinfo {author} {\bibfnamefont {SG}~\bibnamefont
  {Xu}}, \bibinfo {author} {\bibfnamefont {FMD}\ \bibnamefont {Pellegrino}},
  \bibinfo {author} {\bibfnamefont {R~Krishna}\ \bibnamefont {Kumar}}, \bibinfo
  {author} {\bibfnamefont {Alessandro}\ \bibnamefont {Principi}}, \bibinfo
  {author} {\bibfnamefont {Iacopo}\ \bibnamefont {Torre}}, \bibinfo {author}
  {\bibfnamefont {M~Ben}\ \bibnamefont {Shalom}}, \bibinfo {author}
  {\bibfnamefont {Takashi}\ \bibnamefont {Taniguchi}}, \bibinfo {author}
  {\bibfnamefont {Kenji}\ \bibnamefont {Watanabe}}, \bibinfo {author}
  {\bibfnamefont {IV}~\bibnamefont {Grigorieva}},  \emph {et~al.},\ }\bibfield
  {title} {\enquote {\bibinfo {title} {Measuring hall viscosity of graphene’s
  electron fluid},}\ }\href@noop {} {\bibfield  {journal} {\bibinfo  {journal}
  {Science}\ }\textbf {\bibinfo {volume} {364}},\ \bibinfo {pages} {162--165}
  (\bibinfo {year} {2019})}\BibitemShut {NoStop}%
\bibitem [{\citenamefont {Narozhny}\ and\ \citenamefont
  {Sch{\"u}tt}(2019)}]{narozhny2019magnetohydrodynamics}%
  \BibitemOpen
  \bibfield  {author} {\bibinfo {author} {\bibfnamefont {Boris~N}\ \bibnamefont
  {Narozhny}}\ and\ \bibinfo {author} {\bibfnamefont {Michael}\ \bibnamefont
  {Sch{\"u}tt}},\ }\bibfield  {title} {\enquote {\bibinfo {title}
  {Magnetohydrodynamics in graphene: shear and hall viscosities},}\ }\href@noop
  {} {\bibfield  {journal} {\bibinfo  {journal} {Physical Review B}\ }\textbf
  {\bibinfo {volume} {100}},\ \bibinfo {pages} {035125} (\bibinfo {year}
  {2019})}\BibitemShut {NoStop}%
\bibitem [{\citenamefont {Imran}(2019)}]{imran2019quantizing}%
  \BibitemOpen
  \bibfield  {author} {\bibinfo {author} {\bibfnamefont {Muhammad}\
  \bibnamefont {Imran}},\ }\bibfield  {title} {\enquote {\bibinfo {title}
  {Quantizing momentum transport in bilayer graphene},}\ }\href@noop {}
  {\bibfield  {journal} {\bibinfo  {journal} {arXiv preprint arXiv:1909.09608}\
  } (\bibinfo {year} {2019})}\BibitemShut {NoStop}%
\bibitem [{\citenamefont {Narozhny}(2019)}]{narozhny2019electronic}%
  \BibitemOpen
  \bibfield  {author} {\bibinfo {author} {\bibfnamefont {Boris~N}\ \bibnamefont
  {Narozhny}},\ }\bibfield  {title} {\enquote {\bibinfo {title} {Electronic
  hydrodynamics in graphene},}\ }\href@noop {} {\bibfield  {journal} {\bibinfo
  {journal} {Annals of Physics}\ }\textbf {\bibinfo {volume} {411}},\ \bibinfo
  {pages} {167979} (\bibinfo {year} {2019})}\BibitemShut {NoStop}%
\bibitem [{\citenamefont {Son}(2019)}]{son2019chiral}%
  \BibitemOpen
  \bibfield  {author} {\bibinfo {author} {\bibfnamefont {Dam~Thanh}\
  \bibnamefont {Son}},\ }\bibfield  {title} {\enquote {\bibinfo {title} {Chiral
  metric hydrodynamics, kelvin circulation theorem, and the fractional quantum
  hall effect},}\ }\href@noop {} {\bibfield  {journal} {\bibinfo  {journal}
  {arXiv preprint arXiv:1907.07187}\ } (\bibinfo {year} {2019})}\BibitemShut
  {NoStop}%
\bibitem [{\citenamefont {Pu}\ \emph {et~al.}(2020)\citenamefont {Pu},
  \citenamefont {Fremling},\ and\ \citenamefont {Jain}}]{pu2020hall}%
  \BibitemOpen
  \bibfield  {author} {\bibinfo {author} {\bibfnamefont {Songyang}\
  \bibnamefont {Pu}}, \bibinfo {author} {\bibfnamefont {Mikael}\ \bibnamefont
  {Fremling}}, \ and\ \bibinfo {author} {\bibfnamefont {JK}~\bibnamefont
  {Jain}},\ }\bibfield  {title} {\enquote {\bibinfo {title} {Hall viscosity of
  composite fermions},}\ }\href@noop {} {\bibfield  {journal} {\bibinfo
  {journal} {Physical Review Research}\ }\textbf {\bibinfo {volume} {2}},\
  \bibinfo {pages} {013139} (\bibinfo {year} {2020})}\BibitemShut {NoStop}%
\bibitem [{\citenamefont {Buchel}\ and\ \citenamefont
  {Baggioli}(2019)}]{buchel2019holographic}%
  \BibitemOpen
  \bibfield  {author} {\bibinfo {author} {\bibfnamefont {Alex}\ \bibnamefont
  {Buchel}}\ and\ \bibinfo {author} {\bibfnamefont {Matteo}\ \bibnamefont
  {Baggioli}},\ }\bibfield  {title} {\enquote {\bibinfo {title} {Holographic
  viscoelastic hydrodynamics},}\ }\href@noop {} {\bibfield  {journal} {\bibinfo
   {journal} {Journal of High Energy Physics}\ }\textbf {\bibinfo {volume}
  {2019}},\ \bibinfo {pages} {146} (\bibinfo {year} {2019})}\BibitemShut
  {NoStop}%
\bibitem [{\citenamefont {Apostolov}\ \emph {et~al.}(2019)\citenamefont
  {Apostolov}, \citenamefont {Pesin},\ and\ \citenamefont
  {Levchenko}}]{apostolov2019magnetodrag}%
  \BibitemOpen
  \bibfield  {author} {\bibinfo {author} {\bibfnamefont {SS}~\bibnamefont
  {Apostolov}}, \bibinfo {author} {\bibfnamefont {DA}~\bibnamefont {Pesin}}, \
  and\ \bibinfo {author} {\bibfnamefont {A}~\bibnamefont {Levchenko}},\
  }\bibfield  {title} {\enquote {\bibinfo {title} {Magnetodrag in the
  hydrodynamic regime: Effects of magnetoplasmon resonance and hall
  viscosity},}\ }\href@noop {} {\bibfield  {journal} {\bibinfo  {journal}
  {Physical Review B}\ }\textbf {\bibinfo {volume} {100}},\ \bibinfo {pages}
  {115401} (\bibinfo {year} {2019})}\BibitemShut {NoStop}%
\bibitem [{\citenamefont {Shapourian}\ \emph {et~al.}(2015)\citenamefont
  {Shapourian}, \citenamefont {Hughes},\ and\ \citenamefont
  {Ryu}}]{shapourian2015viscoelastic}%
  \BibitemOpen
  \bibfield  {author} {\bibinfo {author} {\bibfnamefont {Hassan}\ \bibnamefont
  {Shapourian}}, \bibinfo {author} {\bibfnamefont {Taylor~L}\ \bibnamefont
  {Hughes}}, \ and\ \bibinfo {author} {\bibfnamefont {Shinsei}\ \bibnamefont
  {Ryu}},\ }\bibfield  {title} {\enquote {\bibinfo {title} {Viscoelastic
  response of topological tight-binding models in two and three dimensions},}\
  }\href@noop {} {\bibfield  {journal} {\bibinfo  {journal} {Physical Review
  B}\ }\textbf {\bibinfo {volume} {92}},\ \bibinfo {pages} {165131} (\bibinfo
  {year} {2015})}\BibitemShut {NoStop}%
\bibitem [{\citenamefont {Tuegel}\ and\ \citenamefont
  {Hughes}(2015)}]{tuegel2015hall}%
  \BibitemOpen
  \bibfield  {author} {\bibinfo {author} {\bibfnamefont {Thomas~I}\
  \bibnamefont {Tuegel}}\ and\ \bibinfo {author} {\bibfnamefont {Taylor~L}\
  \bibnamefont {Hughes}},\ }\bibfield  {title} {\enquote {\bibinfo {title}
  {Hall viscosity and momentum transport in lattice and continuum models of the
  integer quantum hall effect in strong magnetic fields},}\ }\href@noop {}
  {\bibfield  {journal} {\bibinfo  {journal} {Physical Review B}\ }\textbf
  {\bibinfo {volume} {92}},\ \bibinfo {pages} {165127} (\bibinfo {year}
  {2015})}\BibitemShut {NoStop}%
\bibitem [{\citenamefont {Dong}\ and\ \citenamefont
  {Niu}(2018)}]{niu2018geometry}%
  \BibitemOpen
  \bibfield  {author} {\bibinfo {author} {\bibfnamefont {Liang}\ \bibnamefont
  {Dong}}\ and\ \bibinfo {author} {\bibfnamefont {Qian}\ \bibnamefont {Niu}},\
  }\bibfield  {title} {\enquote {\bibinfo {title} {Geometrodynamics of
  electrons in a crystal under position and time-dependent deformation},}\
  }\href {\doibase 10.1103/PhysRevB.98.115162} {\bibfield  {journal} {\bibinfo
  {journal} {Phys. Rev. B}\ }\textbf {\bibinfo {volume} {98}},\ \bibinfo
  {pages} {115162} (\bibinfo {year} {2018})}\BibitemShut {NoStop}%
\bibitem [{\citenamefont {Burmistrov}\ \emph {et~al.}(2019)\citenamefont
  {Burmistrov}, \citenamefont {Goldstein}, \citenamefont {Kot}, \citenamefont
  {Kurilovich},\ and\ \citenamefont {Kurilovich}}]{hallvisc-disorder-numerics}%
  \BibitemOpen
  \bibfield  {author} {\bibinfo {author} {\bibfnamefont {Igor~S.}\ \bibnamefont
  {Burmistrov}}, \bibinfo {author} {\bibfnamefont {Moshe}\ \bibnamefont
  {Goldstein}}, \bibinfo {author} {\bibfnamefont {Mordecai}\ \bibnamefont
  {Kot}}, \bibinfo {author} {\bibfnamefont {Vladislav~D.}\ \bibnamefont
  {Kurilovich}}, \ and\ \bibinfo {author} {\bibfnamefont {Pavel~D.}\
  \bibnamefont {Kurilovich}},\ }\bibfield  {title} {\enquote {\bibinfo {title}
  {Dissipative and hall viscosity of a disordered 2d electron gas},}\ }\href
  {\doibase 10.1103/PhysRevLett.123.026804} {\bibfield  {journal} {\bibinfo
  {journal} {Phys. Rev. Lett.}\ }\textbf {\bibinfo {volume} {123}},\ \bibinfo
  {pages} {026804} (\bibinfo {year} {2019})}\BibitemShut {NoStop}%
\bibitem [{\citenamefont {Landau}\ and\ \citenamefont
  {Lifshitz}(1987)}]{landau1987fluid}%
  \BibitemOpen
  \bibfield  {author} {\bibinfo {author} {\bibfnamefont {Lev~D}\ \bibnamefont
  {Landau}}\ and\ \bibinfo {author} {\bibfnamefont {Evgeny~Mikhailovich}\
  \bibnamefont {Lifshitz}},\ }\bibfield  {title} {\enquote {\bibinfo {title}
  {Fluid mechanics},}\ }\href@noop {} {\bibfield  {journal} {\bibinfo
  {journal} {Fluid Mechanics. Second Edition. 1987. Pergamon, Oxford}\ }
  (\bibinfo {year} {1987})}\BibitemShut {NoStop}%
\bibitem [{\citenamefont {Bradlyn}\ and\ \citenamefont
  {Read}(2015)}]{bradlyn2014low}%
  \BibitemOpen
  \bibfield  {author} {\bibinfo {author} {\bibfnamefont {Barry}\ \bibnamefont
  {Bradlyn}}\ and\ \bibinfo {author} {\bibfnamefont {N}~\bibnamefont {Read}},\
  }\bibfield  {title} {\enquote {\bibinfo {title} {Low-energy effective theory
  in the bulk for transport in a topological phase},}\ }\href@noop {}
  {\bibfield  {journal} {\bibinfo  {journal} {Phys. Rev. B}\ }\textbf {\bibinfo
  {volume} {91}},\ \bibinfo {pages} {125303} (\bibinfo {year}
  {2015})}\BibitemShut {NoStop}%
\bibitem [{\citenamefont {Link}\ \emph {et~al.}(2018)\citenamefont {Link},
  \citenamefont {Sheehy}, \citenamefont {Narozhny},\ and\ \citenamefont
  {Schmalian}}]{link2018elastic}%
  \BibitemOpen
  \bibfield  {author} {\bibinfo {author} {\bibfnamefont {Julia~M}\ \bibnamefont
  {Link}}, \bibinfo {author} {\bibfnamefont {Daniel~E}\ \bibnamefont {Sheehy}},
  \bibinfo {author} {\bibfnamefont {Boris~N}\ \bibnamefont {Narozhny}}, \ and\
  \bibinfo {author} {\bibfnamefont {J{\"o}rg}\ \bibnamefont {Schmalian}},\
  }\bibfield  {title} {\enquote {\bibinfo {title} {Elastic response of the
  electron fluid in intrinsic graphene: The collisionless regime},}\
  }\href@noop {} {\bibfield  {journal} {\bibinfo  {journal} {Physical Review
  B}\ }\textbf {\bibinfo {volume} {98}},\ \bibinfo {pages} {195103} (\bibinfo
  {year} {2018})}\BibitemShut {NoStop}%
\bibitem [{\citenamefont {Bradlyn}\ \emph {et~al.}(2012)\citenamefont
  {Bradlyn}, \citenamefont {Goldstein},\ and\ \citenamefont
  {Read}}]{bradlyn2012kubo}%
  \BibitemOpen
  \bibfield  {author} {\bibinfo {author} {\bibfnamefont {Barry}\ \bibnamefont
  {Bradlyn}}, \bibinfo {author} {\bibfnamefont {Moshe}\ \bibnamefont
  {Goldstein}}, \ and\ \bibinfo {author} {\bibfnamefont {N}~\bibnamefont
  {Read}},\ }\bibfield  {title} {\enquote {\bibinfo {title} {Kubo formulas for
  viscosity: Hall viscosity, ward identities, and the relation with
  conductivity},}\ }\href@noop {} {\bibfield  {journal} {\bibinfo  {journal}
  {Physical Review B}\ }\textbf {\bibinfo {volume} {86}},\ \bibinfo {pages}
  {245309} (\bibinfo {year} {2012})}\BibitemShut {NoStop}%
\bibitem [{\citenamefont {Belinfante}(1940)}]{belinfante1940current}%
  \BibitemOpen
  \bibfield  {author} {\bibinfo {author} {\bibfnamefont {Frederik~J}\
  \bibnamefont {Belinfante}},\ }\bibfield  {title} {\enquote {\bibinfo {title}
  {On the current and the density of the electric charge, the energy, the
  linear momentum and the angular momentum of arbitrary fields},}\ }\href@noop
  {} {\bibfield  {journal} {\bibinfo  {journal} {Physica}\ }\textbf {\bibinfo
  {volume} {7}},\ \bibinfo {pages} {449--474} (\bibinfo {year}
  {1940})}\BibitemShut {NoStop}%
\bibitem [{\citenamefont {Nakahara}(2003)}]{nakahara2003geometry}%
  \BibitemOpen
  \bibfield  {author} {\bibinfo {author} {\bibfnamefont {Mikio}\ \bibnamefont
  {Nakahara}},\ }\href@noop {} {\emph {\bibinfo {title} {Geometry, topology and
  physics}}}\ (\bibinfo  {publisher} {CRC Press},\ \bibinfo {year}
  {2003})\BibitemShut {NoStop}%
\bibitem [{\citenamefont {Rosenfeld}(1940)}]{rosenfeld1940tenseur}%
  \BibitemOpen
  \bibfield  {author} {\bibinfo {author} {\bibfnamefont {L{\'e}on Jacques
  Henri~Constant}\ \bibnamefont {Rosenfeld}},\ }\href@noop {} {\emph {\bibinfo
  {title} {Sur le tenseur d'impulsion-{\'e}nergie}}}\ (\bibinfo  {publisher}
  {Palais des Acad{\'e}mies},\ \bibinfo {year} {1940})\BibitemShut {NoStop}%
\bibitem [{\citenamefont {Abanov}\ and\ \citenamefont
  {Gromov}(2014)}]{Gromov20141}%
  \BibitemOpen
  \bibfield  {author} {\bibinfo {author} {\bibfnamefont {A.}~\bibnamefont
  {Abanov}}\ and\ \bibinfo {author} {\bibfnamefont {A.}~\bibnamefont
  {Gromov}},\ }\bibfield  {title} {\enquote {\bibinfo {title} {Electromagnetic
  and gravitational responses of two-dimensional noninteracting electrons in a
  background magnetic field},}\ }\href {\doibase 10.1103/PhysRevB.90.014435}
  {\bibfield  {journal} {\bibinfo  {journal} {Phys. Rev. B}\ }\textbf {\bibinfo
  {volume} {90}},\ \bibinfo {pages} {014435} (\bibinfo {year}
  {2014})}\BibitemShut {NoStop}%
\bibitem [{\citenamefont {Gromov}\ and\ \citenamefont
  {Abanov}(2014)}]{Abanov2014}%
  \BibitemOpen
  \bibfield  {author} {\bibinfo {author} {\bibfnamefont {A.}~\bibnamefont
  {Gromov}}\ and\ \bibinfo {author} {\bibfnamefont {A.}~\bibnamefont
  {Abanov}},\ }\bibfield  {title} {\enquote {\bibinfo {title}
  {Density-curvature response and gravitational anomaly},}\ }\href {\doibase
  10.1103/PhysRevLett.113.046803} {\bibfield  {journal} {\bibinfo  {journal}
  {Phys. Rev. Lett.}\ }\textbf {\bibinfo {volume} {113}},\ \bibinfo {pages}
  {266802} (\bibinfo {year} {2014})}\BibitemShut {NoStop}%
\bibitem [{\citenamefont {Lipkin}(2002)}]{lipkin2002lie}%
  \BibitemOpen
  \bibfield  {author} {\bibinfo {author} {\bibfnamefont {Harry~J}\ \bibnamefont
  {Lipkin}},\ }\href@noop {} {\emph {\bibinfo {title} {Lie groups for
  pedestrians}}}\ (\bibinfo  {publisher} {Courier Corporation},\ \bibinfo
  {year} {2002})\BibitemShut {NoStop}%
\bibitem [{\citenamefont {Kogut}\ and\ \citenamefont
  {Susskind}(1975)}]{kogut1975hamiltonian}%
  \BibitemOpen
  \bibfield  {author} {\bibinfo {author} {\bibfnamefont {John}\ \bibnamefont
  {Kogut}}\ and\ \bibinfo {author} {\bibfnamefont {Leonard}\ \bibnamefont
  {Susskind}},\ }\bibfield  {title} {\enquote {\bibinfo {title} {Hamiltonian
  formulation of wilson's lattice gauge theories},}\ }\href@noop {} {\bibfield
  {journal} {\bibinfo  {journal} {Physical Review D}\ }\textbf {\bibinfo
  {volume} {11}},\ \bibinfo {pages} {395} (\bibinfo {year} {1975})}\BibitemShut
  {NoStop}%
\bibitem [{\citenamefont {Fetter}\ and\ \citenamefont
  {Walecka}(2012)}]{fetter2012quantum}%
  \BibitemOpen
  \bibfield  {author} {\bibinfo {author} {\bibfnamefont {Alexander~L}\
  \bibnamefont {Fetter}}\ and\ \bibinfo {author} {\bibfnamefont {John~Dirk}\
  \bibnamefont {Walecka}},\ }\href@noop {} {\emph {\bibinfo {title} {Quantum
  theory of many-particle systems}}}\ (\bibinfo  {publisher} {Courier
  Corporation, Mineola NY},\ \bibinfo {year} {2012})\BibitemShut {NoStop}%
\bibitem [{\citenamefont {Ashcroft}\ and\ \citenamefont
  {Mermin}(2005)}]{ashcroft2005solid}%
  \BibitemOpen
  \bibfield  {author} {\bibinfo {author} {\bibfnamefont {Neil~W}\ \bibnamefont
  {Ashcroft}}\ and\ \bibinfo {author} {\bibfnamefont {N~David}\ \bibnamefont
  {Mermin}},\ }\bibfield  {title} {\enquote {\bibinfo {title} {Solid state
  physics (holt, rinehart and winston, new york, 1976)},}\ }\href@noop {}
  {\bibfield  {journal} {\bibinfo  {journal} {Google Scholar}\ }\textbf
  {\bibinfo {volume} {403}} (\bibinfo {year} {2005})}\BibitemShut {NoStop}%
\bibitem [{\citenamefont {Irving}\ and\ \citenamefont
  {Kirkwood}(1950)}]{irving1950statistical}%
  \BibitemOpen
  \bibfield  {author} {\bibinfo {author} {\bibfnamefont {JH}~\bibnamefont
  {Irving}}\ and\ \bibinfo {author} {\bibfnamefont {John~G}\ \bibnamefont
  {Kirkwood}},\ }\bibfield  {title} {\enquote {\bibinfo {title} {The
  statistical mechanical theory of transport processes. iv. the equations of
  hydrodynamics},}\ }\href@noop {} {\bibfield  {journal} {\bibinfo  {journal}
  {The Journal of chemical physics}\ }\textbf {\bibinfo {volume} {18}},\
  \bibinfo {pages} {817--829} (\bibinfo {year} {1950})}\BibitemShut {NoStop}%
\bibitem [{Note1()}]{Note1}%
  \BibitemOpen
  \bibinfo {note} {For a model with a symmetric stress tensor and no spin, our
  stress tensor matches Ref~.\protect \rev@citealpnum
  {shapourian2015viscoelastic}.}\BibitemShut {Stop}%
\bibitem [{\citenamefont {Yu}\ and\ \citenamefont
  {Bradley}(2017)}]{yu2017emergent}%
  \BibitemOpen
  \bibfield  {author} {\bibinfo {author} {\bibfnamefont {Xiaoquan}\
  \bibnamefont {Yu}}\ and\ \bibinfo {author} {\bibfnamefont {Ashton~S}\
  \bibnamefont {Bradley}},\ }\bibfield  {title} {\enquote {\bibinfo {title}
  {Emergent non-eulerian hydrodynamics of quantum vortices in two
  dimensions},}\ }\href@noop {} {\bibfield  {journal} {\bibinfo  {journal}
  {Physical review letters}\ }\textbf {\bibinfo {volume} {119}},\ \bibinfo
  {pages} {185301} (\bibinfo {year} {2017})}\BibitemShut {NoStop}%
\bibitem [{\citenamefont {Haldane}(2009)}]{haldane2009hall}%
  \BibitemOpen
  \bibfield  {author} {\bibinfo {author} {\bibfnamefont {F.~D.~M.}\
  \bibnamefont {Haldane}},\ }\bibfield  {title} {\enquote {\bibinfo {title}
  {Hall viscosity and intrinsic metric of incompressible fractional hall
  fluids},}\ }\href@noop {} {\bibfield  {journal} {\bibinfo  {journal} {arXiv
  preprint arXiv:0906.1854}\ } (\bibinfo {year} {2009})}\BibitemShut {NoStop}%
\bibitem [{\citenamefont {Haldane}(2011)}]{haldane2011geometrical}%
  \BibitemOpen
  \bibfield  {author} {\bibinfo {author} {\bibfnamefont {F.~D.~M.}\
  \bibnamefont {Haldane}},\ }\bibfield  {title} {\enquote {\bibinfo {title}
  {Geometrical description of the fractional quantum hall effect},}\
  }\href@noop {} {\bibfield  {journal} {\bibinfo  {journal} {Physical review
  letters}\ }\textbf {\bibinfo {volume} {107}},\ \bibinfo {pages} {116801}
  (\bibinfo {year} {2011})}\BibitemShut {NoStop}%
\bibitem [{\citenamefont {Aroyo}\ \emph {et~al.}(2011)\citenamefont {Aroyo},
  \citenamefont {Perez-Mato}, \citenamefont {Orobengoa}, \citenamefont {Tasci},
  \citenamefont {de~la Flor},\ and\ \citenamefont {Kirov}}]{Bilbao1}%
  \BibitemOpen
  \bibfield  {author} {\bibinfo {author} {\bibfnamefont {M.~I.}\ \bibnamefont
  {Aroyo}}, \bibinfo {author} {\bibfnamefont {J.~M.}\ \bibnamefont
  {Perez-Mato}}, \bibinfo {author} {\bibfnamefont {D.}~\bibnamefont
  {Orobengoa}}, \bibinfo {author} {\bibfnamefont {E.}~\bibnamefont {Tasci}},
  \bibinfo {author} {\bibfnamefont {G.}~\bibnamefont {de~la Flor}}, \ and\
  \bibinfo {author} {\bibfnamefont {A.}~\bibnamefont {Kirov}},\ }\bibfield
  {title} {\enquote {\bibinfo {title} {Crystallography online: Bilbao
  crystallographic server},}\ }\href@noop {} {\bibfield  {journal} {\bibinfo
  {journal} {Bulg. Chem. Commun.}\ }\textbf {\bibinfo {volume} {43(2)}},\
  \bibinfo {pages} {183} (\bibinfo {year} {2011})}\BibitemShut {NoStop}%
\bibitem [{\citenamefont {Aroyo}\ \emph
  {et~al.}(2006{\natexlab{a}})\citenamefont {Aroyo}, \citenamefont
  {Perez-Mato}, \citenamefont {Capillas}, \citenamefont {Kroumova},
  \citenamefont {Ivantchev}, \citenamefont {Madariaga}, \citenamefont {Kirov},\
  and\ \citenamefont {Wondratschek}}]{Bilbao2}%
  \BibitemOpen
  \bibfield  {author} {\bibinfo {author} {\bibfnamefont {M.~I.}\ \bibnamefont
  {Aroyo}}, \bibinfo {author} {\bibfnamefont {J.~M.}\ \bibnamefont
  {Perez-Mato}}, \bibinfo {author} {\bibfnamefont {C.}~\bibnamefont
  {Capillas}}, \bibinfo {author} {\bibfnamefont {E.}~\bibnamefont {Kroumova}},
  \bibinfo {author} {\bibfnamefont {S.}~\bibnamefont {Ivantchev}}, \bibinfo
  {author} {\bibfnamefont {G.}~\bibnamefont {Madariaga}}, \bibinfo {author}
  {\bibfnamefont {A.}~\bibnamefont {Kirov}}, \ and\ \bibinfo {author}
  {\bibfnamefont {H.}~\bibnamefont {Wondratschek}},\ }\bibfield  {title}
  {\enquote {\bibinfo {title} {Bilbao crystallographic server i: Databases and
  crystallographic computing programs},}\ }\href@noop {} {\bibfield  {journal}
  {\bibinfo  {journal} {Z. Krist.}\ }\textbf {\bibinfo {volume} {221}},\
  \bibinfo {pages} {15} (\bibinfo {year} {2006}{\natexlab{a}})}\BibitemShut
  {NoStop}%
\bibitem [{\citenamefont {Aroyo}\ \emph
  {et~al.}(2006{\natexlab{b}})\citenamefont {Aroyo}, \citenamefont {Kirov},
  \citenamefont {Capillas}, \citenamefont {Perez-Mato},\ and\ \citenamefont
  {Wondratschek}}]{Bilbao3}%
  \BibitemOpen
  \bibfield  {author} {\bibinfo {author} {\bibfnamefont {M.~I.}\ \bibnamefont
  {Aroyo}}, \bibinfo {author} {\bibfnamefont {A.}~\bibnamefont {Kirov}},
  \bibinfo {author} {\bibfnamefont {C.}~\bibnamefont {Capillas}}, \bibinfo
  {author} {\bibfnamefont {J.~M.}\ \bibnamefont {Perez-Mato}}, \ and\ \bibinfo
  {author} {\bibfnamefont {H.}~\bibnamefont {Wondratschek}},\ }\bibfield
  {title} {\enquote {\bibinfo {title} {Bilbao crystallographic server ii:
  Representations of crystallographic point groups and space groups},}\
  }\href@noop {} {\bibfield  {journal} {\bibinfo  {journal} {Acta Cryst.}\
  }\textbf {\bibinfo {volume} {A62}},\ \bibinfo {pages} {115} (\bibinfo {year}
  {2006}{\natexlab{b}})}\BibitemShut {NoStop}%
\bibitem [{\citenamefont {Golan}\ \emph {et~al.}(2019)\citenamefont {Golan},
  \citenamefont {Hoyos},\ and\ \citenamefont {Moroz}}]{golan2019boundary}%
  \BibitemOpen
  \bibfield  {author} {\bibinfo {author} {\bibfnamefont {Omri}\ \bibnamefont
  {Golan}}, \bibinfo {author} {\bibfnamefont {Carlos}\ \bibnamefont {Hoyos}}, \
  and\ \bibinfo {author} {\bibfnamefont {Sergej}\ \bibnamefont {Moroz}},\
  }\bibfield  {title} {\enquote {\bibinfo {title} {Boundary central charge from
  bulk odd viscosity: Chiral superfluids},}\ }\href@noop {} {\bibfield
  {journal} {\bibinfo  {journal} {Physical Review B}\ }\textbf {\bibinfo
  {volume} {100}},\ \bibinfo {pages} {104512} (\bibinfo {year}
  {2019})}\BibitemShut {NoStop}%
\bibitem [{\citenamefont {Forster}(1975)}]{forster1975hydrodynamic}%
  \BibitemOpen
  \bibfield  {author} {\bibinfo {author} {\bibfnamefont {Dieter}\ \bibnamefont
  {Forster}},\ }\bibfield  {title} {\enquote {\bibinfo {title} {Hydrodynamic
  fluctuations, broken symmetry, and correlation functions},}\ }in\ \href@noop
  {} {\emph {\bibinfo {booktitle} {Reading, Mass., WA Benjamin, Inc.(Frontiers
  in Physics. Volume 47), 1975. 343 p.}}},\ Vol.~\bibinfo {volume} {47}\
  (\bibinfo {year} {1975})\BibitemShut {NoStop}%
\bibitem [{\citenamefont {Scheibner}\ \emph {et~al.}(2019)\citenamefont
  {Scheibner}, \citenamefont {Souslov}, \citenamefont {Banerjee}, \citenamefont
  {Surowka}, \citenamefont {Irvine},\ and\ \citenamefont
  {Vitelli}}]{scheibner2019odd}%
  \BibitemOpen
  \bibfield  {author} {\bibinfo {author} {\bibfnamefont {Colin}\ \bibnamefont
  {Scheibner}}, \bibinfo {author} {\bibfnamefont {Anton}\ \bibnamefont
  {Souslov}}, \bibinfo {author} {\bibfnamefont {Debarghya}\ \bibnamefont
  {Banerjee}}, \bibinfo {author} {\bibfnamefont {Piotr}\ \bibnamefont
  {Surowka}}, \bibinfo {author} {\bibfnamefont {William T.~M.}\ \bibnamefont
  {Irvine}}, \ and\ \bibinfo {author} {\bibfnamefont {Vincenzo}\ \bibnamefont
  {Vitelli}},\ }\href@noop {} {\enquote {\bibinfo {title} {Odd elasticity},}\ }
  (\bibinfo {year} {2019}),\ \Eprint {http://arxiv.org/abs/1902.07760}
  {arXiv:1902.07760 [cond-mat.soft]} \BibitemShut {NoStop}%
\bibitem [{\citenamefont {Avron}\ \emph {et~al.}(1987)\citenamefont {Avron},
  \citenamefont {Seiler},\ and\ \citenamefont {Yaffe}}]{avron1987adiabatic}%
  \BibitemOpen
  \bibfield  {author} {\bibinfo {author} {\bibfnamefont {JE}~\bibnamefont
  {Avron}}, \bibinfo {author} {\bibfnamefont {Rued}\ \bibnamefont {Seiler}}, \
  and\ \bibinfo {author} {\bibfnamefont {LG}~\bibnamefont {Yaffe}},\ }\bibfield
   {title} {\enquote {\bibinfo {title} {Adiabatic theorems and applications to
  the quantum hall effect},}\ }\href@noop {} {\bibfield  {journal} {\bibinfo
  {journal} {Communications in Mathematical Physics}\ }\textbf {\bibinfo
  {volume} {110}},\ \bibinfo {pages} {33--49} (\bibinfo {year}
  {1987})}\BibitemShut {NoStop}%
\bibitem [{\citenamefont {Hughes}\ \emph {et~al.}(2011)\citenamefont {Hughes},
  \citenamefont {Leigh},\ and\ \citenamefont {Fradkin}}]{hughesleighfradkin}%
  \BibitemOpen
  \bibfield  {author} {\bibinfo {author} {\bibfnamefont {Taylor~L}\
  \bibnamefont {Hughes}}, \bibinfo {author} {\bibfnamefont {Robert~G}\
  \bibnamefont {Leigh}}, \ and\ \bibinfo {author} {\bibfnamefont {Eduardo}\
  \bibnamefont {Fradkin}},\ }\bibfield  {title} {\enquote {\bibinfo {title}
  {Torsional response and dissipationless viscosity in topological
  insulators},}\ }\href@noop {} {\bibfield  {journal} {\bibinfo  {journal}
  {Physical review letters}\ }\textbf {\bibinfo {volume} {107}},\ \bibinfo
  {pages} {075502} (\bibinfo {year} {2011})}\BibitemShut {NoStop}%
\bibitem [{\citenamefont {Read}\ and\ \citenamefont
  {Green}(2000)}]{read2000paired}%
  \BibitemOpen
  \bibfield  {author} {\bibinfo {author} {\bibfnamefont {Nicholas}\
  \bibnamefont {Read}}\ and\ \bibinfo {author} {\bibfnamefont {Dmitry}\
  \bibnamefont {Green}},\ }\bibfield  {title} {\enquote {\bibinfo {title}
  {Paired states of fermions in two dimensions with breaking of parity and
  time-reversal symmetries and the fractional quantum hall effect},}\
  }\href@noop {} {\bibfield  {journal} {\bibinfo  {journal} {Physical Review
  B}\ }\textbf {\bibinfo {volume} {61}},\ \bibinfo {pages} {10267} (\bibinfo
  {year} {2000})}\BibitemShut {NoStop}%
\bibitem [{\citenamefont {Xia}\ \emph {et~al.}(2011)\citenamefont {Xia},
  \citenamefont {Eisenstein}, \citenamefont {Pfeiffer},\ and\ \citenamefont
  {West}}]{xia2011evidence}%
  \BibitemOpen
  \bibfield  {author} {\bibinfo {author} {\bibfnamefont {Jing}\ \bibnamefont
  {Xia}}, \bibinfo {author} {\bibfnamefont {JP}~\bibnamefont {Eisenstein}},
  \bibinfo {author} {\bibfnamefont {Loren~N}\ \bibnamefont {Pfeiffer}}, \ and\
  \bibinfo {author} {\bibfnamefont {Ken~W}\ \bibnamefont {West}},\ }\bibfield
  {title} {\enquote {\bibinfo {title} {Evidence for a fractionally quantized
  hall state with anisotropic longitudinal transport},}\ }\href@noop {}
  {\bibfield  {journal} {\bibinfo  {journal} {Nature Physics}\ }\textbf
  {\bibinfo {volume} {7}},\ \bibinfo {pages} {845--848} (\bibinfo {year}
  {2011})}\BibitemShut {NoStop}%
\bibitem [{\citenamefont {Ganeshan}\ and\ \citenamefont
  {Abanov}(2017)}]{sriramhydro}%
  \BibitemOpen
  \bibfield  {author} {\bibinfo {author} {\bibfnamefont {Sriram}\ \bibnamefont
  {Ganeshan}}\ and\ \bibinfo {author} {\bibfnamefont {Alexander~G.}\
  \bibnamefont {Abanov}},\ }\bibfield  {title} {\enquote {\bibinfo {title} {Odd
  viscosity in two-dimensional incompressible fluids},}\ }\href {\doibase
  10.1103/PhysRevFluids.2.094101} {\bibfield  {journal} {\bibinfo  {journal}
  {Phys. Rev. Fluids}\ }\textbf {\bibinfo {volume} {2}},\ \bibinfo {pages}
  {094101} (\bibinfo {year} {2017})}\BibitemShut {NoStop}%
\bibitem [{\citenamefont {Abanov}\ \emph {et~al.}(2018)\citenamefont {Abanov},
  \citenamefont {Can},\ and\ \citenamefont {Ganeshan}}]{abanov2018odd}%
  \BibitemOpen
  \bibfield  {author} {\bibinfo {author} {\bibfnamefont {Alexander}\
  \bibnamefont {Abanov}}, \bibinfo {author} {\bibfnamefont {Tankut}\
  \bibnamefont {Can}}, \ and\ \bibinfo {author} {\bibfnamefont {Sriram}\
  \bibnamefont {Ganeshan}},\ }\bibfield  {title} {\enquote {\bibinfo {title}
  {Odd surface waves in two-dimensional incompressible fluids},}\ }\href@noop
  {} {\bibfield  {journal} {\bibinfo  {journal} {SciPost Physics}\ }\textbf
  {\bibinfo {volume} {5}} (\bibinfo {year} {2018})}\BibitemShut {NoStop}%
\bibitem [{\citenamefont {Abanov}\ and\ \citenamefont
  {Monteiro}(2019)}]{abanov2019free}%
  \BibitemOpen
  \bibfield  {author} {\bibinfo {author} {\bibfnamefont {Alexander~G}\
  \bibnamefont {Abanov}}\ and\ \bibinfo {author} {\bibfnamefont {Gustavo~M}\
  \bibnamefont {Monteiro}},\ }\bibfield  {title} {\enquote {\bibinfo {title}
  {Free-surface variational principle for an incompressible fluid with odd
  viscosity},}\ }\href@noop {} {\bibfield  {journal} {\bibinfo  {journal}
  {Physical review letters}\ }\textbf {\bibinfo {volume} {122}},\ \bibinfo
  {pages} {154501} (\bibinfo {year} {2019})}\BibitemShut {NoStop}%
\bibitem [{\citenamefont {Abanov}\ \emph {et~al.}(2019)\citenamefont {Abanov},
  \citenamefont {Can}, \citenamefont {Ganeshan},\ and\ \citenamefont
  {Monteiro}}]{abanov2019hydrodynamics}%
  \BibitemOpen
  \bibfield  {author} {\bibinfo {author} {\bibfnamefont {Alexander~G}\
  \bibnamefont {Abanov}}, \bibinfo {author} {\bibfnamefont {Tankut}\
  \bibnamefont {Can}}, \bibinfo {author} {\bibfnamefont {Sriram}\ \bibnamefont
  {Ganeshan}}, \ and\ \bibinfo {author} {\bibfnamefont {Gustavo~M}\
  \bibnamefont {Monteiro}},\ }\bibfield  {title} {\enquote {\bibinfo {title}
  {Hydrodynamics of two-dimensional compressible fluid with broken parity:
  variational principle and free surface dynamics in the absence of
  dissipation},}\ }\href@noop {} {\bibfield  {journal} {\bibinfo  {journal}
  {arXiv preprint arXiv:1907.11196}\ } (\bibinfo {year} {2019})}\BibitemShut
  {NoStop}%
\bibitem [{\citenamefont {Banerjee}\ \emph {et~al.}(2017)\citenamefont
  {Banerjee}, \citenamefont {Souslov}, \citenamefont {Abanov},\ and\
  \citenamefont {Vitelli}}]{vitelli2017odd}%
  \BibitemOpen
  \bibfield  {author} {\bibinfo {author} {\bibfnamefont {Debarghya}\
  \bibnamefont {Banerjee}}, \bibinfo {author} {\bibfnamefont {Anton}\
  \bibnamefont {Souslov}}, \bibinfo {author} {\bibfnamefont {Alexander~G}\
  \bibnamefont {Abanov}}, \ and\ \bibinfo {author} {\bibfnamefont {Vincenzo}\
  \bibnamefont {Vitelli}},\ }\bibfield  {title} {\enquote {\bibinfo {title}
  {Odd viscosity in chiral active fluids},}\ }\href@noop {} {\bibfield
  {journal} {\bibinfo  {journal} {Nature Communications}\ }\textbf {\bibinfo
  {volume} {8}},\ \bibinfo {pages} {1573} (\bibinfo {year} {2017})}\BibitemShut
  {NoStop}%
\bibitem [{\citenamefont {Harris}\ \emph {et~al.}(2000)\citenamefont {Harris},
  \citenamefont {Awschalom}, \citenamefont {Maranowski},\ and\ \citenamefont
  {Gossard}}]{harris2000magnetization}%
  \BibitemOpen
  \bibfield  {author} {\bibinfo {author} {\bibfnamefont {JGE}\ \bibnamefont
  {Harris}}, \bibinfo {author} {\bibfnamefont {DD}~\bibnamefont {Awschalom}},
  \bibinfo {author} {\bibfnamefont {KD}~\bibnamefont {Maranowski}}, \ and\
  \bibinfo {author} {\bibfnamefont {AC}~\bibnamefont {Gossard}},\ }\bibfield
  {title} {\enquote {\bibinfo {title} {Magnetization and dissipation
  measurements in the quantum hall regime using an integrated micromechanical
  magnetometer},}\ }\href@noop {} {\bibfield  {journal} {\bibinfo  {journal}
  {Journal of Applied Physics}\ }\textbf {\bibinfo {volume} {87}},\ \bibinfo
  {pages} {5102--5104} (\bibinfo {year} {2000})}\BibitemShut {NoStop}%
\bibitem [{\citenamefont {Mumford}\ \emph {et~al.}(2019)\citenamefont
  {Mumford}, \citenamefont {Paul}, \citenamefont {Lee}, \citenamefont
  {Yacoby},\ and\ \citenamefont {Kapitulnik}}]{mumford2019cantilever}%
  \BibitemOpen
  \bibfield  {author} {\bibinfo {author} {\bibfnamefont {Samuel}\ \bibnamefont
  {Mumford}}, \bibinfo {author} {\bibfnamefont {Tiffany}\ \bibnamefont {Paul}},
  \bibinfo {author} {\bibfnamefont {Seung~Hwan}\ \bibnamefont {Lee}}, \bibinfo
  {author} {\bibfnamefont {Amir}\ \bibnamefont {Yacoby}}, \ and\ \bibinfo
  {author} {\bibfnamefont {Aharon}\ \bibnamefont {Kapitulnik}},\ }\bibfield
  {title} {\enquote {\bibinfo {title} {A cantilever torque magnetometry method
  for the measurement of hall conductivity of highly resistive samples},}\
  }\href@noop {} {\bibfield  {journal} {\bibinfo  {journal} {arXiv preprint
  arXiv:1908.10857}\ } (\bibinfo {year} {2019})}\BibitemShut {NoStop}%
\bibitem [{\citenamefont {Wiegers}\ \emph {et~al.}(1997)\citenamefont
  {Wiegers}, \citenamefont {Specht}, \citenamefont {L\'evy}, \citenamefont
  {Simmons}, \citenamefont {Ritchie}, \citenamefont {Cavanna}, \citenamefont
  {Etienne}, \citenamefont {Martinez},\ and\ \citenamefont
  {Wyder}}]{PhysRevLett.79.3238}%
  \BibitemOpen
  \bibfield  {author} {\bibinfo {author} {\bibfnamefont {S.~A.~J.}\
  \bibnamefont {Wiegers}}, \bibinfo {author} {\bibfnamefont {M.}~\bibnamefont
  {Specht}}, \bibinfo {author} {\bibfnamefont {L.~P.}\ \bibnamefont {L\'evy}},
  \bibinfo {author} {\bibfnamefont {M.~Y.}\ \bibnamefont {Simmons}}, \bibinfo
  {author} {\bibfnamefont {D.~A.}\ \bibnamefont {Ritchie}}, \bibinfo {author}
  {\bibfnamefont {A.}~\bibnamefont {Cavanna}}, \bibinfo {author} {\bibfnamefont
  {B.}~\bibnamefont {Etienne}}, \bibinfo {author} {\bibfnamefont
  {G.}~\bibnamefont {Martinez}}, \ and\ \bibinfo {author} {\bibfnamefont
  {P.}~\bibnamefont {Wyder}},\ }\bibfield  {title} {\enquote {\bibinfo {title}
  {Magnetization and energy gaps of a high-mobility 2d electron gas in the
  quantum limit},}\ }\href {\doibase 10.1103/PhysRevLett.79.3238} {\bibfield
  {journal} {\bibinfo  {journal} {Phys. Rev. Lett.}\ }\textbf {\bibinfo
  {volume} {79}},\ \bibinfo {pages} {3238--3241} (\bibinfo {year}
  {1997})}\BibitemShut {NoStop}%
\bibitem [{\citenamefont {Wang}\ \emph {et~al.}(2015)\citenamefont {Wang},
  \citenamefont {Jones}, \citenamefont {Seyler}, \citenamefont {Tran},
  \citenamefont {Jia}, \citenamefont {Zhao}, \citenamefont {Wang},
  \citenamefont {Yang}, \citenamefont {Xu},\ and\ \citenamefont
  {Xia}}]{wang2015highly}%
  \BibitemOpen
  \bibfield  {author} {\bibinfo {author} {\bibfnamefont {Xiaomu}\ \bibnamefont
  {Wang}}, \bibinfo {author} {\bibfnamefont {Aaron~M}\ \bibnamefont {Jones}},
  \bibinfo {author} {\bibfnamefont {Kyle~L}\ \bibnamefont {Seyler}}, \bibinfo
  {author} {\bibfnamefont {Vy}~\bibnamefont {Tran}}, \bibinfo {author}
  {\bibfnamefont {Yichen}\ \bibnamefont {Jia}}, \bibinfo {author}
  {\bibfnamefont {Huan}\ \bibnamefont {Zhao}}, \bibinfo {author} {\bibfnamefont
  {Han}\ \bibnamefont {Wang}}, \bibinfo {author} {\bibfnamefont
  {Li}~\bibnamefont {Yang}}, \bibinfo {author} {\bibfnamefont {Xiaodong}\
  \bibnamefont {Xu}}, \ and\ \bibinfo {author} {\bibfnamefont {Fengnian}\
  \bibnamefont {Xia}},\ }\bibfield  {title} {\enquote {\bibinfo {title} {Highly
  anisotropic and robust excitons in monolayer black phosphorus},}\ }\href@noop
  {} {\bibfield  {journal} {\bibinfo  {journal} {Nature nanotechnology}\
  }\textbf {\bibinfo {volume} {10}},\ \bibinfo {pages} {517} (\bibinfo {year}
  {2015})}\BibitemShut {NoStop}%
\bibitem [{\citenamefont {Li}\ \emph {et~al.}(2016)\citenamefont {Li},
  \citenamefont {Yang}, \citenamefont {Ye}, \citenamefont {Zhang},
  \citenamefont {Zhu}, \citenamefont {Lou}, \citenamefont {Zhou}, \citenamefont
  {Li}, \citenamefont {Watanabe}, \citenamefont {Taniguchi} \emph
  {et~al.}}]{li2016quantum}%
  \BibitemOpen
  \bibfield  {author} {\bibinfo {author} {\bibfnamefont {Likai}\ \bibnamefont
  {Li}}, \bibinfo {author} {\bibfnamefont {Fangyuan}\ \bibnamefont {Yang}},
  \bibinfo {author} {\bibfnamefont {Guo~Jun}\ \bibnamefont {Ye}}, \bibinfo
  {author} {\bibfnamefont {Zuocheng}\ \bibnamefont {Zhang}}, \bibinfo {author}
  {\bibfnamefont {Zengwei}\ \bibnamefont {Zhu}}, \bibinfo {author}
  {\bibfnamefont {Wenkai}\ \bibnamefont {Lou}}, \bibinfo {author}
  {\bibfnamefont {Xiaoying}\ \bibnamefont {Zhou}}, \bibinfo {author}
  {\bibfnamefont {Liang}\ \bibnamefont {Li}}, \bibinfo {author} {\bibfnamefont
  {Kenji}\ \bibnamefont {Watanabe}}, \bibinfo {author} {\bibfnamefont
  {Takashi}\ \bibnamefont {Taniguchi}},  \emph {et~al.},\ }\bibfield  {title}
  {\enquote {\bibinfo {title} {Quantum hall effect in black phosphorus
  two-dimensional electron system},}\ }\href@noop {} {\bibfield  {journal}
  {\bibinfo  {journal} {Nature nanotechnology}\ }\textbf {\bibinfo {volume}
  {11}},\ \bibinfo {pages} {593} (\bibinfo {year} {2016})}\BibitemShut
  {NoStop}%
\bibitem [{\citenamefont {Moll}\ \emph {et~al.}(2016)\citenamefont {Moll},
  \citenamefont {Kushwaha}, \citenamefont {Nandi}, \citenamefont {Schmidt},\
  and\ \citenamefont {Mackenzie}}]{moll2016evidence}%
  \BibitemOpen
  \bibfield  {author} {\bibinfo {author} {\bibfnamefont {Philip~JW}\
  \bibnamefont {Moll}}, \bibinfo {author} {\bibfnamefont {Pallavi}\
  \bibnamefont {Kushwaha}}, \bibinfo {author} {\bibfnamefont {Nabhanila}\
  \bibnamefont {Nandi}}, \bibinfo {author} {\bibfnamefont {Burkhard}\
  \bibnamefont {Schmidt}}, \ and\ \bibinfo {author} {\bibfnamefont {Andrew~P}\
  \bibnamefont {Mackenzie}},\ }\bibfield  {title} {\enquote {\bibinfo {title}
  {Evidence for hydrodynamic electron flow in pdcoo2},}\ }\href@noop {}
  {\bibfield  {journal} {\bibinfo  {journal} {Science}\ }\textbf {\bibinfo
  {volume} {351}},\ \bibinfo {pages} {1061--1064} (\bibinfo {year}
  {2016})}\BibitemShut {NoStop}%
\bibitem [{\citenamefont {Spanton}\ \emph {et~al.}(2018)\citenamefont
  {Spanton}, \citenamefont {Zibrov}, \citenamefont {Zhou}, \citenamefont
  {Taniguchi}, \citenamefont {Watanabe}, \citenamefont {Zaletel},\ and\
  \citenamefont {Young}}]{spanton2018observation}%
  \BibitemOpen
  \bibfield  {author} {\bibinfo {author} {\bibfnamefont {Eric~M}\ \bibnamefont
  {Spanton}}, \bibinfo {author} {\bibfnamefont {Alexander~A}\ \bibnamefont
  {Zibrov}}, \bibinfo {author} {\bibfnamefont {Haoxin}\ \bibnamefont {Zhou}},
  \bibinfo {author} {\bibfnamefont {Takashi}\ \bibnamefont {Taniguchi}},
  \bibinfo {author} {\bibfnamefont {Kenji}\ \bibnamefont {Watanabe}}, \bibinfo
  {author} {\bibfnamefont {Michael~P}\ \bibnamefont {Zaletel}}, \ and\ \bibinfo
  {author} {\bibfnamefont {Andrea~F}\ \bibnamefont {Young}},\ }\bibfield
  {title} {\enquote {\bibinfo {title} {Observation of fractional chern
  insulators in a van der waals heterostructure},}\ }\href@noop {} {\bibfield
  {journal} {\bibinfo  {journal} {Science}\ }\textbf {\bibinfo {volume}
  {360}},\ \bibinfo {pages} {62--66} (\bibinfo {year} {2018})}\BibitemShut
  {NoStop}%
\bibitem [{\citenamefont {Zibrov}\ \emph {et~al.}(2017)\citenamefont {Zibrov},
  \citenamefont {Kometter}, \citenamefont {Zhou}, \citenamefont {Spanton},
  \citenamefont {Taniguchi}, \citenamefont {Watanabe}, \citenamefont
  {Zaletel},\ and\ \citenamefont {Young}}]{zibrov2017tunable}%
  \BibitemOpen
  \bibfield  {author} {\bibinfo {author} {\bibfnamefont {Alexander~A}\
  \bibnamefont {Zibrov}}, \bibinfo {author} {\bibfnamefont {C}~\bibnamefont
  {Kometter}}, \bibinfo {author} {\bibfnamefont {H}~\bibnamefont {Zhou}},
  \bibinfo {author} {\bibfnamefont {EM}~\bibnamefont {Spanton}}, \bibinfo
  {author} {\bibfnamefont {T}~\bibnamefont {Taniguchi}}, \bibinfo {author}
  {\bibfnamefont {K}~\bibnamefont {Watanabe}}, \bibinfo {author} {\bibfnamefont
  {MP}~\bibnamefont {Zaletel}}, \ and\ \bibinfo {author} {\bibfnamefont
  {AF}~\bibnamefont {Young}},\ }\bibfield  {title} {\enquote {\bibinfo {title}
  {Tunable interacting composite fermion phases in a half-filled
  bilayer-graphene landau level},}\ }\href@noop {} {\bibfield  {journal}
  {\bibinfo  {journal} {Nature}\ }\textbf {\bibinfo {volume} {549}},\ \bibinfo
  {pages} {360--364} (\bibinfo {year} {2017})}\BibitemShut {NoStop}%
\bibitem [{\citenamefont {Serlin}\ \emph {et~al.}(2019)\citenamefont {Serlin},
  \citenamefont {Tschirhart}, \citenamefont {Polshyn}, \citenamefont {Zhang},
  \citenamefont {Zhu}, \citenamefont {Watanabe}, \citenamefont {Taniguchi},
  \citenamefont {Balents},\ and\ \citenamefont {Young}}]{serlin2019intrinsic}%
  \BibitemOpen
  \bibfield  {author} {\bibinfo {author} {\bibfnamefont {M}~\bibnamefont
  {Serlin}}, \bibinfo {author} {\bibfnamefont {CL}~\bibnamefont {Tschirhart}},
  \bibinfo {author} {\bibfnamefont {H}~\bibnamefont {Polshyn}}, \bibinfo
  {author} {\bibfnamefont {Y}~\bibnamefont {Zhang}}, \bibinfo {author}
  {\bibfnamefont {J}~\bibnamefont {Zhu}}, \bibinfo {author} {\bibfnamefont
  {K}~\bibnamefont {Watanabe}}, \bibinfo {author} {\bibfnamefont
  {T}~\bibnamefont {Taniguchi}}, \bibinfo {author} {\bibfnamefont
  {L}~\bibnamefont {Balents}}, \ and\ \bibinfo {author} {\bibfnamefont
  {AF}~\bibnamefont {Young}},\ }\bibfield  {title} {\enquote {\bibinfo {title}
  {Intrinsic quantized anomalous hall effect in a moir{\'e} heterostructure},}\
  }\href {\doibase 10.1126/science.aay5533} {\bibfield  {journal} {\bibinfo
  {journal} {Science}\ }\textbf {\bibinfo {volume} {376}},\ \bibinfo {pages}
  {900--903} (\bibinfo {year} {2019})}\BibitemShut {NoStop}%
\bibitem [{\citenamefont {Bandurin}\ \emph {et~al.}(2016)\citenamefont
  {Bandurin}, \citenamefont {Torre}, \citenamefont {Kumar}, \citenamefont
  {Shalom}, \citenamefont {Tomadin}, \citenamefont {Principi}, \citenamefont
  {Auton}, \citenamefont {Khestanova}, \citenamefont {Novoselov}, \citenamefont
  {Grigorieva} \emph {et~al.}}]{bandurin2016negative}%
  \BibitemOpen
  \bibfield  {author} {\bibinfo {author} {\bibfnamefont {DA}~\bibnamefont
  {Bandurin}}, \bibinfo {author} {\bibfnamefont {Iacopo}\ \bibnamefont
  {Torre}}, \bibinfo {author} {\bibfnamefont {R~Krishna}\ \bibnamefont
  {Kumar}}, \bibinfo {author} {\bibfnamefont {M~Ben}\ \bibnamefont {Shalom}},
  \bibinfo {author} {\bibfnamefont {Andrea}\ \bibnamefont {Tomadin}}, \bibinfo
  {author} {\bibfnamefont {A}~\bibnamefont {Principi}}, \bibinfo {author}
  {\bibfnamefont {GH}~\bibnamefont {Auton}}, \bibinfo {author} {\bibfnamefont
  {E}~\bibnamefont {Khestanova}}, \bibinfo {author} {\bibfnamefont
  {KS}~\bibnamefont {Novoselov}}, \bibinfo {author} {\bibfnamefont
  {IV}~\bibnamefont {Grigorieva}},  \emph {et~al.},\ }\bibfield  {title}
  {\enquote {\bibinfo {title} {Negative local resistance caused by viscous
  electron backflow in graphene},}\ }\href@noop {} {\bibfield  {journal}
  {\bibinfo  {journal} {Science}\ }\textbf {\bibinfo {volume} {351}},\ \bibinfo
  {pages} {1055--1058} (\bibinfo {year} {2016})}\BibitemShut {NoStop}%
\bibitem [{\citenamefont {An}\ \emph {et~al.}(2017)\citenamefont {An},
  \citenamefont {Meier},\ and\ \citenamefont {Gadway}}]{Ane1602685}%
  \BibitemOpen
  \bibfield  {author} {\bibinfo {author} {\bibfnamefont {Fangzhao~Alex}\
  \bibnamefont {An}}, \bibinfo {author} {\bibfnamefont {Eric~J.}\ \bibnamefont
  {Meier}}, \ and\ \bibinfo {author} {\bibfnamefont {Bryce}\ \bibnamefont
  {Gadway}},\ }\bibfield  {title} {\enquote {\bibinfo {title} {Direct
  observation of chiral currents and magnetic reflection in atomic flux
  lattices},}\ }\href {\doibase 10.1126/sciadv.1602685} {\bibfield  {journal}
  {\bibinfo  {journal} {Science Advances}\ }\textbf {\bibinfo {volume} {3}}
  (\bibinfo {year} {2017}),\ 10.1126/sciadv.1602685},\ \Eprint
  {http://arxiv.org/abs/https://advances.sciencemag.org/content/3/4/e1602685.full.pdf}
  {https://advances.sciencemag.org/content/3/4/e1602685.full.pdf} \BibitemShut
  {NoStop}%
\bibitem [{\citenamefont {Cooper}\ \emph {et~al.}(2019)\citenamefont {Cooper},
  \citenamefont {Dalibard},\ and\ \citenamefont
  {Spielman}}]{RevModPhys.91.015005}%
  \BibitemOpen
  \bibfield  {author} {\bibinfo {author} {\bibfnamefont {N.~R.}\ \bibnamefont
  {Cooper}}, \bibinfo {author} {\bibfnamefont {J.}~\bibnamefont {Dalibard}}, \
  and\ \bibinfo {author} {\bibfnamefont {I.~B.}\ \bibnamefont {Spielman}},\
  }\bibfield  {title} {\enquote {\bibinfo {title} {Topological bands for
  ultracold atoms},}\ }\href {\doibase 10.1103/RevModPhys.91.015005} {\bibfield
   {journal} {\bibinfo  {journal} {Rev. Mod. Phys.}\ }\textbf {\bibinfo
  {volume} {91}},\ \bibinfo {pages} {015005} (\bibinfo {year}
  {2019})}\BibitemShut {NoStop}%
\bibitem [{\citenamefont {Landsteiner}\ \emph {et~al.}(2016)\citenamefont
  {Landsteiner}, \citenamefont {Liu},\ and\ \citenamefont
  {Sun}}]{landsteiner-weyl-visc}%
  \BibitemOpen
  \bibfield  {author} {\bibinfo {author} {\bibfnamefont {Karl}\ \bibnamefont
  {Landsteiner}}, \bibinfo {author} {\bibfnamefont {Yan}\ \bibnamefont {Liu}},
  \ and\ \bibinfo {author} {\bibfnamefont {Ya-Wen}\ \bibnamefont {Sun}},\
  }\bibfield  {title} {\enquote {\bibinfo {title} {Odd viscosity in the quantum
  critical region of a holographic weyl semimetal},}\ }\href {\doibase
  10.1103/PhysRevLett.117.081604} {\bibfield  {journal} {\bibinfo  {journal}
  {Phys. Rev. Lett.}\ }\textbf {\bibinfo {volume} {117}},\ \bibinfo {pages}
  {081604} (\bibinfo {year} {2016})}\BibitemShut {NoStop}%
\bibitem [{\citenamefont {Arjona}\ and\ \citenamefont
  {Vozmediano}(2018)}]{vozmediano-rotational-strain}%
  \BibitemOpen
  \bibfield  {author} {\bibinfo {author} {\bibfnamefont {Vicente}\ \bibnamefont
  {Arjona}}\ and\ \bibinfo {author} {\bibfnamefont {Mar\'{\i}a A.~H.}\
  \bibnamefont {Vozmediano}},\ }\bibfield  {title} {\enquote {\bibinfo {title}
  {Rotational strain in weyl semimetals: A continuum approach},}\ }\href
  {\doibase 10.1103/PhysRevB.97.201404} {\bibfield  {journal} {\bibinfo
  {journal} {Phys. Rev. B}\ }\textbf {\bibinfo {volume} {97}},\ \bibinfo
  {pages} {201404} (\bibinfo {year} {2018})}\BibitemShut {NoStop}%
\bibitem [{\citenamefont {Copetti}\ and\ \citenamefont
  {Landsteiner}(2019)}]{copetti2019anomalous}%
  \BibitemOpen
  \bibfield  {author} {\bibinfo {author} {\bibfnamefont {Christian}\
  \bibnamefont {Copetti}}\ and\ \bibinfo {author} {\bibfnamefont {Karl}\
  \bibnamefont {Landsteiner}},\ }\bibfield  {title} {\enquote {\bibinfo {title}
  {Anomalous hall viscosity at the weyl-semimetal--insulator transition},}\
  }\href@noop {} {\bibfield  {journal} {\bibinfo  {journal} {Physical Review
  B}\ }\textbf {\bibinfo {volume} {99}},\ \bibinfo {pages} {195146} (\bibinfo
  {year} {2019})}\BibitemShut {NoStop}%
\bibitem [{\citenamefont {Becher}\ and\ \citenamefont
  {Joos}(1982)}]{becher1982dirac}%
  \BibitemOpen
  \bibfield  {author} {\bibinfo {author} {\bibfnamefont {Peter}\ \bibnamefont
  {Becher}}\ and\ \bibinfo {author} {\bibfnamefont {Hans}\ \bibnamefont
  {Joos}},\ }\bibfield  {title} {\enquote {\bibinfo {title} {The
  dirac-k{\"a}hler equation and fermions on the lattice},}\ }\href@noop {}
  {\bibfield  {journal} {\bibinfo  {journal} {Zeitschrift f{\"u}r Physik C
  Particles and Fields}\ }\textbf {\bibinfo {volume} {15}},\ \bibinfo {pages}
  {343--365} (\bibinfo {year} {1982})}\BibitemShut {NoStop}%
\end{thebibliography}%
\end{document}